\documentclass[aps,jmp,amsmath,amssymb,doublespace,reprint]{revtex4-1}

\usepackage{graphicx}
\usepackage{dcolumn}
\usepackage{bm}
\usepackage{lipsum}
\usepackage[colorlinks=true,citecolor=blue]{hyperref}
\usepackage{resizegather}
\usepackage{amsmath}

\DeclareMathOperator{\sgn}{sgn}

\begin{document}

\date{\today}

\title{Characterizing the mesh size of polymer solutions via the pore size distribution}

\author{Valerio Sorichetti}
\affiliation{Laboratoire Charles Coulomb (L2C), Univ. Montpellier, CNRS, F-34095, Montpellier, France and IATE, INRA, CIRAD, Montpellier SupAgro, Univ. Montpellier, F-34060, Montpellier, France}
\author{Virginie Hugouvieux}
\affiliation{IATE, INRA, CIRAD, Montpellier SupAgro, Univ. Montpellier, F-34060, Montpellier, France}
\author{Walter Kob}
\affiliation{Laboratoire Charles Coulomb (L2C), Univ. Montpellier, CNRS, F-34095, Montpellier, France}

\begin{abstract}

In order to characterize the geometrical mesh size $\xi$, we simulate a solution of coarse-grained polymers with densities ranging from the dilute to the concentrated regime and for different chain lengths. Conventional ways to estimate $\xi$ rely either on scaling assumptions which give $\xi$ only up to an unknown multiplicative factor, or on measurements of the monomer density fluctuation correlation length $\xi_c$. We determine $\xi_c$ from the monomer structure factor and from the radial distribution function, and find that the identification $\xi=\xi_c$ is not justified outside of the semidilute regime. In order to better characterize $\xi$, we compute the pore size distribution (PSD) following two different definitions, one by Torquato \emph{et al.} \cite{torquato1990nearest} and one by Gubbins \emph{et al.} \cite{gelb1999pore}. We show that the mean values of the two distributions, $\langle r \rangle_T$ and $\langle r \rangle_G$, both display the behavior predicted for $\xi$ by scaling theory, and argue that $\xi$ can be identified with either one of these quantities. This identification allows to interpret the PSD as the distribution of mesh sizes, a quantity which conventional methods cannot access. Finally, we show that it is possible to map a polymer solution on a system of hard or overlapping spheres, for which Torquato's PSD can be computed analytically and reproduces accurately the PSD of the solution. We give an expression that allows $\langle r \rangle_T$ to be estimated with great accuracy in the semidilute regime by knowing only the radius of gyration and the density of the polymers.

\end{abstract}

\maketitle

\section{Introduction}

Polymer solutions are outstanding liquids showing a rich behavior due to the exceptional variety of their structural and dynamical properties. While simple atomic and molecular liquids generally have only one relevant length scale --the size of the atoms/molecules-- a polymer solution has many: the monomer size, the Kuhn length, the entanglement length, all the way up to the chain size \cite{degennes1979scaling,rubinstein2003polymer}. In semidilute polymer solutions, i.e., solutions which are dense enough that different chains start to overlap, the only physically relevant length scale is (in the absence of entanglement) the \emph{mesh size}, $\xi$.  The general idea behind the concept of mesh size is that a dense polymer solution observed at a certain instant in time looks very similar to an intricate network with a certain mesh size $\xi$ \cite{degennes1979scaling}.

The relevance of the mesh size is most apparent when considering the problem of the diffusion of proteins or nanoparticles in polymer liquids, which has received a vast amount of attention in recent years because of its applications to biology \cite{zhou2008macromolecular,zhou2013influence} and medicine, e.g. for drug delivery \cite{soppimath2001biodegradable,lai2007rapid,schuster2013nanoparticle}.  Many theoretical approaches have been proposed to describe the diffusion of particles in polymer solutions and gels, such as geometric obstruction models \cite{ogston1973transport,altenberger1984theory,johansson1991diffusion}, hydrodynamic models \cite{phillies1985phenomenological,phillies1986universal,cukier1984diffusion,odijk2000depletion,fan2007motion,tuinier2008scaling} and mode-coupling theory \cite{egorov2011anomalous,yamamoto2011theory,dong2015diffusion} (for a review of some of these models, see Ref.~\citenum{masaro1999physical}). Although the predictions of different models can vary widely, most of them agree upon the fact that $\xi$ is central in controlling the diffusion of the particles. In the last few years, a model based on scaling theory has been proposed to describe the diffusion of nonsticky spherical nanoparticles in polymer liquids \cite{cai2011mobility}. This model, which seems to be the one in best agreement with recent experiments \cite{kohli2012diffusion,babaye2014mobility,poling2015size} and simulations \cite{chen2017coupling,chen2019influence}, predicts that particles with radius $R<\xi$ can slip through the mesh, experiencing the viscosity $\eta_s$ of the pure solvent, while larger particles will experience a higher effective viscosity. The mesh size represents therefore the length scale of the transition from free diffusion to obstructed diffusion in a polymer solution. Naturally, the mesh size is also a fundamental quantity when considering the diffusion of particles in polymer networks and gels \cite{lustig1988solute,masaro1999physical,dell2013theory,rose2013dynamics,cai2015hopping,parrish2017network,parrish2018temperature}.

When considering the diffusion of a probe particle in a polymer solution, it is clear that the relevant mesh size is, to a first approximation, just the average geometrical size of the pores, or holes, in the solution, i.e., those regions which are filled only with solvent. In view of applications to diffusion problems, we will in the present work adopt this intuitive concept of $\xi$ as the average pore size of the polymer solution. This also allows to extend the intuitive definition of $\xi$ proposed by de Gennes, as the average size of the transient mesh in a semidilute solution \cite{degennes1979scaling}, to the dilute and concentrated regimes: In the dilute regime the chains do not overlap, but a mesh size still exists as an average geometric distance between neighboring chains. In the concentrated regime, the chains overlap so strongly that there is no clear mesh structure, but we can still measure the size of the pores.

We are, however, immediately confronted with the question: How can one determine $\xi$? Normally, two ways are used to estimate the mesh size: scaling theory \cite{degennes1979scaling}, or measurements of the monomer density fluctuation correlation length, $\xi_c$ \cite{degennes1979scaling,rubinstein2003polymer,koshy2003density}. Scaling calculations are useful and allow to obtain easily an estimate of $\xi$. However, by construction they can only give the quantity of interest up to an unknown multiplicative factor, which is very inconvenient in cases where a precise knowledge of $\xi$ is required. Estimates based on the correlation length --which can be obtained from scattering experiments-- are also useful; however, as we will discuss below, the correlation length is not the same quantity as the mesh size, and taking them to be the same quantity can lead to apparently nonsensical results, such as $\xi$ increasing when the polymer concentration is increased \cite{koshy2003density}.

In the present paper, we use molecular dynamics simulations of coarse-grained polymers to perform a systematic comparison of the different conventional ways to measure $\xi$, i.e., those based on scaling theory and those based on static correlation functions in Fourier and real space. We show that these techniques have some limitations, can lead to apparently contradictory behaviors and provide at best an approximate value for the average mesh size in the semidilute regime. To overcome these problems, we propose a different method to estimate $\xi$, based on the concept of \emph{pore size distribution} \cite{scheidegger1957physics,prager1963interphase,torquato1990nearest,torquato1991diffusion,torquato1995nearest,torquato2013random,gelb1999pore,thomson2000modeling,pikunic2003structural,bhattacharya2006fast}. If the coordinates of the monomers are known, this method allows to measure $\xi$ reliably at any density, and to obtain not only an average value, but also the distribution of mesh sizes. This last feature can be particularly important, since it is known that particles diffusing in systems with the same average mesh size can display completely different dynamical behaviors \cite{parrish2017network}, i.e., the heterogeneity of the polymer medium plays a relevant role in particle diffusion. Although we consider here only the case of a solution of linear chains, the proposed method is very general and can be applied to systems of polymers with different topologies (e.g. rings, randomly branched, dendrimers) and also to networks and gels.

The paper is organized as follows: In Sec.~\ref{sec:theory} we review the theoretical background, discussing the concept of blob, of density fluctuation correlation length and of geometrical mesh size. In Sec.~\ref{sec:model}, the model and the details of the simulations are presented. Sec.~\ref{sec:results} deals with the main results: Measurements of the correlation length and scaling estimates of $\xi$ are compared with results from the pore size distribution. We conclude with a summary in Sec.~\ref{sec:summary}.

\section{Theoretical background} 
\label{sec:theory}

\begin{figure}
\centering
\includegraphics[width=0.45\textwidth]{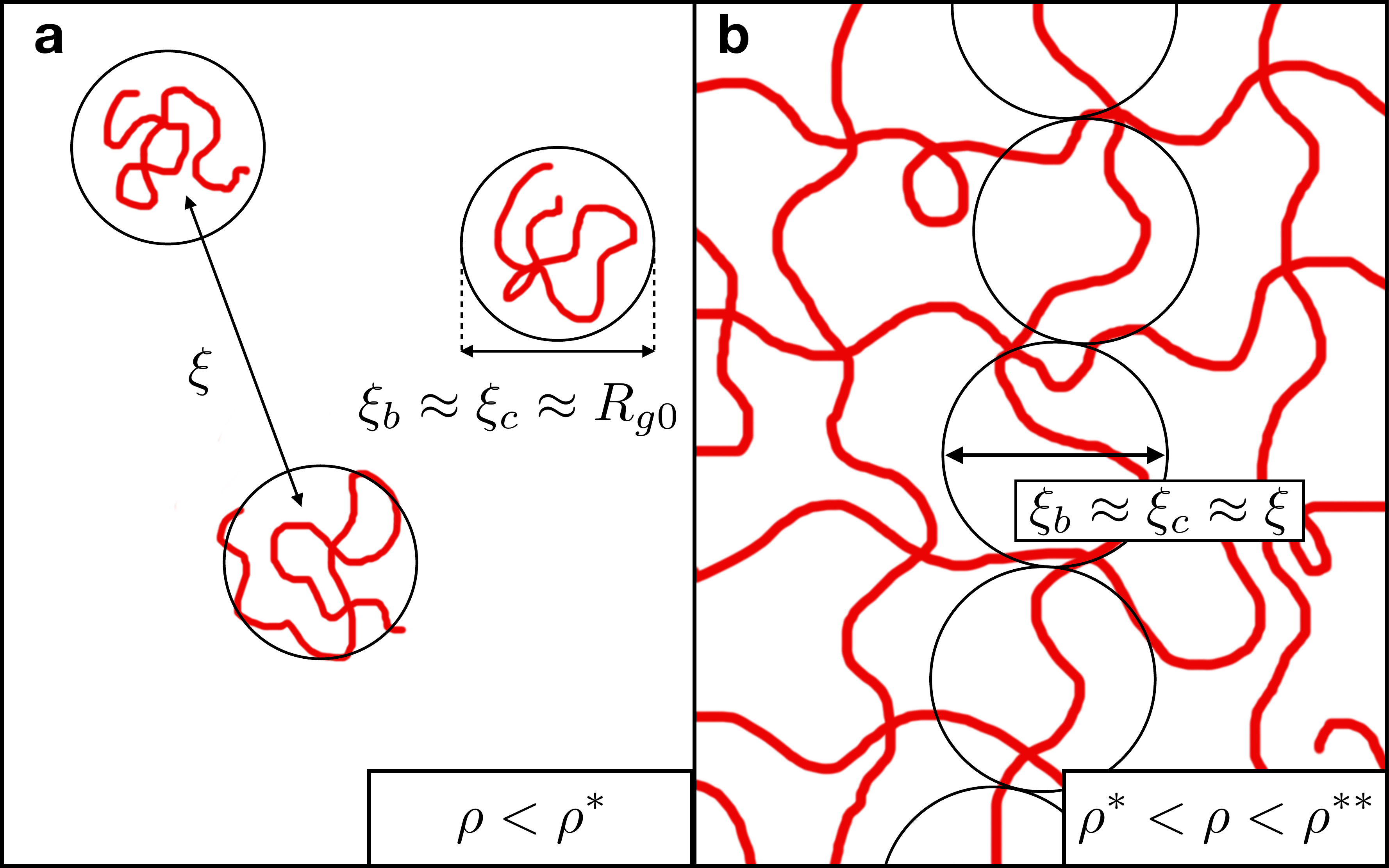}
\caption{Schematic representation of a polymer solution in the dilute (a) and semidilute (b) regime. Circles denote the blobs, of size $\xi_b$. The other relevant length scales are the geometrical mesh size, $\xi$, and the polymer correlation length, $\xi_c$. In the semidilute regime, these three length scales are proportional through a $\mathcal O (1)$ dimensionless factor. In the dilute regime, $\xi$ is very different from the other two length scales.}
\label{mesh_cartoon}
\end{figure}

Before discussing the approach that we propose to characterize the geometrical mesh size, we review the concepts of blob \cite{degennes1979scaling,rubinstein2003polymer}, monomer density fluctuation correlation length \cite{degennes1979scaling,rubinstein2003polymer,koshy2003density} and geometrical mesh size. Most of what is discussed in this section can be found in standard texts \cite{degennes1979scaling,doi1988theory,teraoka2002polymer,rubinstein2003polymer}. However, we feel it is useful to summarize these results, since they will be extensively used in what follows. Moreover, this section also serves to introduce the terminology and notations which will be used in the rest of this work.

To simplify the discussion, we limit ourselves to the case of polymers in athermal solvent \cite{rubinstein2003polymer}. This choice is also motivated by the fact that the model chosen for the simulations is athermal, in the sense that all the nonbonded interactions are purely repulsive, as discussed in Sec.~\ref{sec:model} below. Since our discussion is limited to the athermal solvent case, we use the symbol $\nu$ exclusively to denote the Flory exponent in good/athermal solvent, i.e., $\nu\simeq0.588$ \cite{rubinstein2003polymer,clisby2010accurate}.

\subsection{The concept of blob}
\label{sec:blob}

The structure of a polymer solution can be rather complicated, but its main features can easily be obtained from the \emph{blob model}, originally introduced by de Gennes  \cite{degennes1979scaling}. A blob is defined as a region of space of size $\xi_b$ which contains almost exclusively monomers from a single chain, or equivalently inside of which inter-chain interactions are negligible \cite{degennes1979scaling,rubinstein2003polymer}. A graphical representation of a blob is given in Fig.~\ref{mesh_cartoon}. The above definition implies that in a semidilute solution, where the polymer chains overlap, the blobs are \emph{space-filling}, i.e., if the number of monomers in a blob is $N_b$ we have \cite{rubinstein2003polymer}

\begin{equation}
\rho \approx \frac {N_b}{\xi_b^3},
\label{blob1}
\end{equation}

\noindent where $\rho$ is the bulk monomer density, see Fig.~\ref{mesh_cartoon}b. (In Eq.~\eqref{blob1}, and also in the following, we use the symbol $\approx$ to indicate that two quantities are proportional through a dimensionless factor of order $1$, while we will use the symbol $\simeq$ when two quantities can be considered approximately equal.)

On length scales $r \approx \xi_b$ or smaller, the chain is by definition mostly surrounded by solvent, and therefore it assumes a swollen configuration \cite{degennes1979scaling,rubinstein2003polymer}:

\begin{equation}
\xi_b \approx b N_b^\nu,
\label{blob2}
\end{equation}
\noindent where $b$ is the effective bond length. From Eqs.~\eqref{blob1} and \eqref{blob2} we obtain

\begin{equation}
\xi_b \approx  b (b^3 \rho)^{-\nu/(3\nu -1)}.
\label{blob_vs_rho}
\end{equation}

\noindent We note that, since $\nu\simeq0.588$, $-\nu/(3\nu -1)\simeq-0.770$.

On length scales $r>\xi_b$, the excluded volume interactions are screened by the surrounding chains and therefore the structure of the chain is ideal \cite{degennes1979scaling,rubinstein2003polymer}. If the chains are ideal at all length scales $r>b$, the solution is called \emph{concentrated} and $\xi_b\approx b$ \cite{rubinstein2003polymer,teraoka2002polymer} \footnote{We are assuming here that $b\approx \sigma$, with $\sigma$ the monomer size, and that the solvent is athermal, so that the size of the thermal blob is approximately equal to $b$.}. In the opposite limit we have a dilute solution, where chains do not overlap and therefore $\xi_b\approx R_{g0}$, where $R_{g0}\equiv \lim_{\rho \to 0} R_g$ (Fig.~\ref{mesh_cartoon}a). Between these two regimes, we find the semidilute behavior given by Eq.~\eqref{blob_vs_rho}. We can thus summarize the scaling behavior of $\xi_b$ as follows:

\begin{equation}
\xi_b \approx
\begin{cases}
R_{g0} & \rho < \rho^*\\
b (b^3 \rho)^{-\nu/(3\nu -1)} & \rho^* < \rho <\rho^{**}\\
b & \rho>\rho^{**},\\
\end{cases}
\label{blob_size}
\end{equation}

\noindent where $\rho^*$ (the \emph{overlap concentration}) marks the onset of the semidilute regime and $\rho^{**}$ the one of the concentrated regime. 

Imposing continuity between the three relations in Eq.~\eqref{blob_size}, and recalling that $R_{g0} \approx b N^{\nu}$, where $N$ is the degree of polymerization \cite{degennes1979scaling,rubinstein2003polymer}, we obtain

\begin{equation}
\rho^* \approx N R_{g0}^{-3} \approx N^{1-3\nu} b^{-3}
\label{rho_semidilute}
\end{equation}

\noindent and

\begin{equation}
\rho^{**} \approx b^{-3} \approx \rho_\text{melt}.
\label{rho_concentrated}
\end{equation}

We note that $\lim_{N\to\infty} \rho^* = 0$, i.e., infinitely long chains are never in the dilute regime.

Exploiting the relation $R_{g0} \approx b N^{\nu}$ and introducing the dimensionless scaling variable $\chi$, defined as \cite{paul1991crossover,gerroff1993new} \footnote{We choose to use $\chi$ and not $\rho/\rho^*$ as a scaling variable since $\rho^*$ is not a sharply defined quantity.}

\begin{equation}
\chi \equiv N (b^3 \rho)^{1/(3 \nu -1)} \approx \left(\frac{\rho}{\rho^*}\right)^{1/(3\nu-1)},
\label{chi_def}
\end{equation}

\noindent Eq.~\eqref{blob_size} can be rewritten in the following dimensionless form:

\begin{equation}
\frac{\xi_b}{R_{g0}} \approx
\begin{cases}
1 & \chi <  \chi^*\\
\chi^{-\nu} & \chi^* < \chi < \chi^{**}\\
N^{-\nu} & \chi > \chi^{**}.\\
\end{cases}
\label{blob_size_chi}
\end{equation}

\noindent It is easy to verify using Eqs.~\eqref{rho_semidilute}-\eqref{chi_def}, that $\chi^* \equiv \chi(\rho^*) = 1$ and $\chi^{**} \equiv \chi(\rho^{**}) = N$. We note that $\chi^*$ is independent of $N$, whereas $\chi^{**}$ depends on $N$. Conversely, $\rho^*$ depends on $N$, whereas $\rho^{**}$ is independent of $N$.

The concept of blob is extremely useful for scaling calculations; however, there is no quantity which can be measured in experiments or simulations which unambiguously corresponds to the blob size $\xi_b$. We will see below that it is possible to introduce two different quantities, the density fluctuation correlation length and the geometrical mesh size, which have the same scaling behavior as $\xi_b$ in the semidilute regime and are accessible from simulations or experimental data.

\subsection{Monomer density fluctuation correlation length}
\label{sec:corr}

Due to chain connectivity, the local monomer density in a polymer has significant spatial fluctuations. One possibility to quantify the spatial extent of these fluctuations is to study the radial distribution function $g(r)$ \cite{hansen1990theory}, defined in such a way that $4 \pi \rho r^2 g(r) dr$ represents the probability to find a monomer at distance $r<x<r+dr$ from a randomly chosen monomer. At low density, the large $r$ behavior of $g(r)$ is expected to have the form \cite{fisher1964correlation,degennes1979scaling,doi1988theory}

\begin{equation}
g(r) = 1+\frac A r e^{-r /\xi_c},
\label{rdf_oz}
\end{equation}

\noindent where $A>0$ and $\xi_c>0$ are constants. In Eq.~\eqref{rdf_oz}, $\xi_c$ has clearly the role of a correlation length, since it represents the typical decay length scale of the density fluctuations. Therefore $\xi_c$ is known as the monomer \emph{density fluctuation correlation length} (or simply \emph{correlation length}). At higher density, Eq.~\eqref{rdf_oz} must be modified by the introduction of a periodic modulation \cite{verlet1968computer,evans1994asymptotic,evans1993asymptotic,koshy2003density},

\begin{equation}
g(r)=1+\frac B r e^{-r /\xi_c} \sin \left( \frac{2 \pi r}{\lambda}+ \theta \right),
\label{rdf_oz_sin}
\end{equation}

\noindent where $B>0,\lambda >0$, and $\theta$ are constants.

The radial distribution function is easily accessible in simulations; in experiments, however, it is easier to study the static structure factor $S(q)$ \cite{hansen1990theory,rubinstein2003polymer}, which is related to $g(r)$ through a Fourier transform \cite{hansen1990theory}: 

\begin{equation}
S(q) = 1 + 4 \pi \rho \int_0^\infty [g(r)-1] \frac{\sin(qr)}{qr} r^2 dr.
\label{sq_transform}
\end{equation}

\noindent From Eqs.~\eqref{rdf_oz} and \eqref{sq_transform}, we obtain the low $q$ behavior of $S(q)$, which is given by the Ornstein-Zernike function \cite{fisher1964correlation,degennes1979scaling,doi1988theory} \footnote{To be more precise, Eq.~\eqref{sq_oz} follows formally from Eq.~\eqref{sq_transform} if $g(\mathbf r)=1+Ae^{-r/\xi_c}/r-\delta(\mathbf r)/\rho$. The presence of the delta distribution is however only an artifact stemming from the fact that we are connecting the large $r$ behavior of $g(r)$ with the small $q$ behavior of $S(q)$. In any case, since we are interested in the large $r$ behavior, this factor can be disregarded.},

\begin{equation}
S(q) =  \frac{S(0)}{1+(q\xi_c)^2},
\label{sq_oz}
\end{equation}

\noindent where $S(0)\equiv \lim_{q \to 0} S(q)$.

In the dilute regime $\rho<\rho^*$, the monomer structure factor coincides approximately with the structure factor of a single chain: $S(q) \simeq S_1(q)$ \cite{teraoka2002polymer,rubinstein2003polymer}. Since we know that for $q < 2\pi / R_g$ \cite{teraoka2002polymer,rubinstein2003polymer}

\begin{equation}
S_1(q) \simeq N\left[1-\left(\frac{q R_{g}}{\sqrt{3}}\right)^2\right] \simeq \frac{N}{1+(q R_{g}/\sqrt{3})^2},
\label{s1q}
\end{equation}

\noindent it follows from Eq.~\eqref{sq_oz} that for $\rho<\rho^*$, $\xi_c \simeq R_{g0}/\sqrt{3}$. Therefore, in a dilute solution $\xi_c\approx \xi_b \approx R_{g0}$.

Since the structure of a semidilute solution of short chains is identical to the one of long chains, we expect that in the semidilute regime $\xi_c$ is independent of $N$. Using scaling arguments, it is then straightforward to prove that \cite{rubinstein2003polymer,teraoka2002polymer}

\begin{equation}
\xi_c \approx  b (b^3 \rho)^{-\nu/(3\nu -1)} \ \ \  ( \rho^* < \rho <\rho^{**}),
\label{xic_semidilute}
\end{equation}

\noindent which is equivalent to \eqref{blob_vs_rho}. We can therefore conclude that $\xi_c\approx \xi_b$ in the whole range $\rho<\rho^{**}$.

The relation \eqref{xic_semidilute} loses its validity for $\rho>\rho^{**}$. The reason is that in this density regime chain connectivity becomes irrelevant, and the structure of the system is dominated by local packing constraints. The correlation length behaves therefore like that of a dense molecular liquid, i.e., it \emph{increases} with increasing density\cite{koshy2003density}. Therefore one can summarize the behavior of $\xi_c$ as follows:

\begin{equation}
\xi_c \approx
\begin{cases}
R_{g0} & \rho < \rho^*\\
b (b^3 \rho)^{-\nu/(3\nu -1)} & \rho^* < \rho <\rho^{**}\\
h(\rho) & \rho>\rho^{**},\\
\end{cases}
\label{corr_length}
\end{equation}

\noindent where $h(\rho)$ is an increasing function of $\rho$ for $\rho>\rho^{**}$. 

\subsection{Geometrical mesh size}
\label{sec:xi}

In the present work we are interested in measuring the geometrical mesh size (or simply ``mesh size") $\xi$, which we define as the average size of the pores in the system. Intuitively, in the semidilute regime $\xi$ will be very close to $\xi_b$, and therefore to $\xi_c$ (Eq.~\eqref{xic_semidilute} and Fig.~\ref{mesh_cartoon}b):

\begin{equation}
\xi \approx \xi_b \approx \xi_c \ \ \ ( \rho^* < \rho <\rho^{**}).
\label{blob_mesh}
\end{equation}

\noindent However, if we want to maintain our definition of $\xi$ as the average size of the pores, then it is clear that Eq.\eqref{blob_mesh} cannot be extended outside of the density range $\rho^* < \rho <\rho^{**}$. Indeed, in the dilute regime $\rho < \rho^*$ the mesh size is nothing else than the average distance between neighboring chains, i.e., $\xi \approx (\rho/{N})^{-1/3}$, and becomes infinite in the limit $\rho \to 0$, whereas $\xi_b \approx \xi_c \approx R_{g0}$. Moreover, in the concentrated regime $\rho>\rho^{**}$ the size of the pores becomes vanishingly small, $\xi \to 0$, while $\xi_b \approx b$, and $\xi_c$ \emph{increases}. To summarize, we expect the following behavior for the geometrical mesh size $\xi$ as a function of $\rho$:

\begin{equation}
\xi \approx
\begin{cases}
(\rho/N)^{-1/3} & \rho < \rho^*\\
b (b^3 \rho)^{-\nu/(3\nu -1)} & \rho^* < \rho <\rho^{**}\\
f(\rho) & \rho>\rho^{**},\\
\end{cases}
\label{xi_scaling_rho}
\end{equation}

\noindent where $f(\rho)$ is a decreasing function of $\rho$ for $\rho>\rho^{**}$.  In terms of the scaling variable $\chi$,

\begin{equation}
\frac{\xi}{R_{g0}} \approx
\begin{cases}
\chi^{-\nu+1/3}& \chi < \chi^*\\
\chi^{-\nu} & \chi^*<\chi<\chi^{**}\\
f[\rho(\chi)] & \chi > \chi^{**}.\\
\end{cases}
\label{xi_scaling}
\end{equation}

\begin{figure}
\centering
\includegraphics[width=0.43\textwidth]{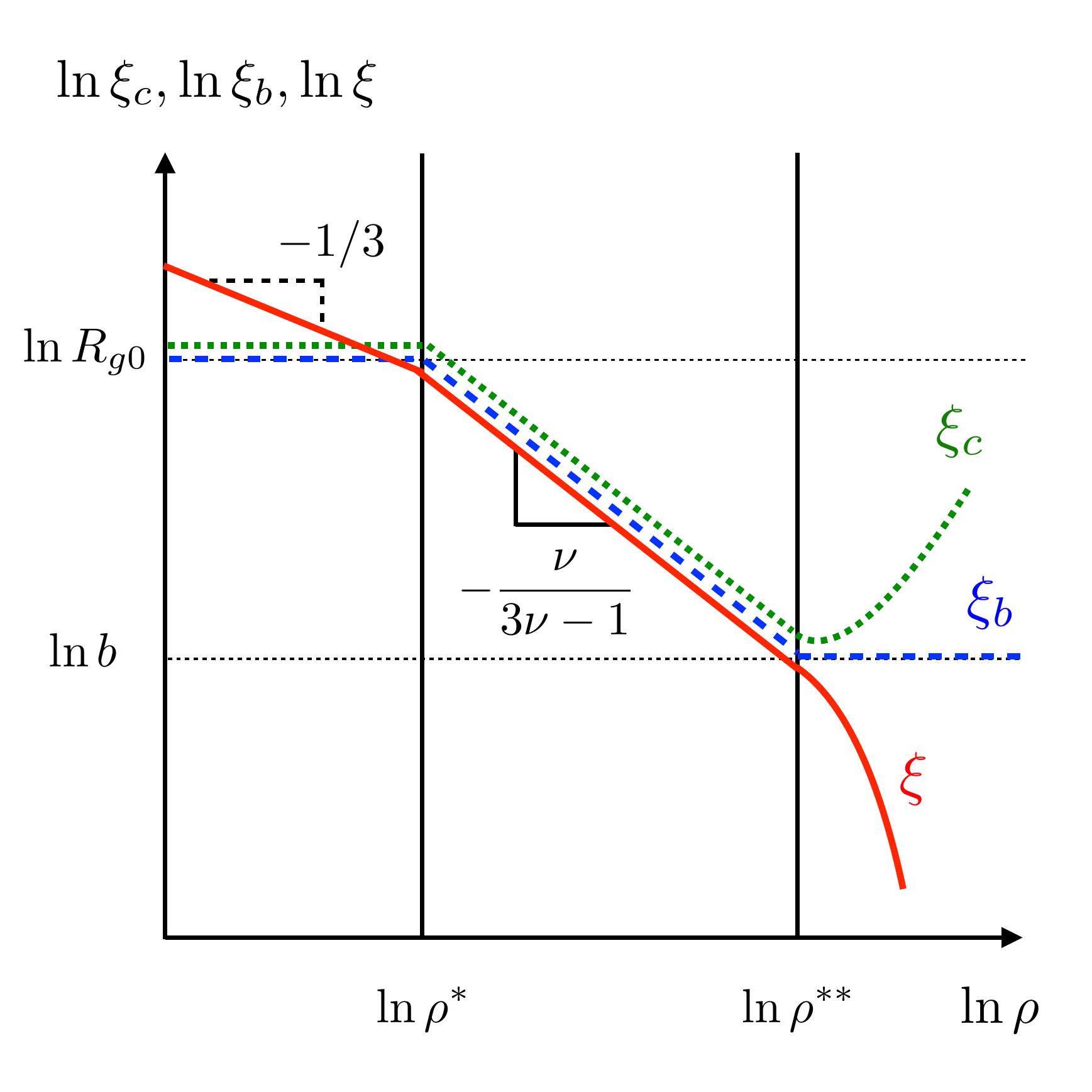}
\caption{Expected $\rho$-dependence of the blob size $\xi_b$, the correlation length $\xi_c$ and the geometrical mesh size $\xi$, Eqs.~\eqref{blob_size},\eqref{corr_length}, and \eqref{xi_scaling_rho}.}
\label{scaling_plot}
\end{figure}

The scaling behaviors of $\xi_b,\xi_c$ and $\xi$ as a function of $\rho$ are schematically represented in Fig.~\ref{scaling_plot}.

\noindent We mention that $\xi$ is often estimated using the relation (see Eqs.~\eqref{blob_vs_rho} and \eqref{rho_semidilute})

\begin{equation}
\xi \simeq R_{g0} \left(\frac \rho {\rho^*}  \right)^{-{\nu}/{(3\nu-1)}},
\label{xi_scaling_approx}
\end{equation}

\noindent where \cite{doi1988theory,teraoka2002polymer}

\begin{equation}
\rho^* \simeq \frac{3N} {4 \pi R_{g0}^3}.
\label{overlap_approx}
\end{equation}

\noindent Although Eq.~\eqref{overlap_approx} is a reasonable estimate, in reality the transition between the dilute and the semidilute regimes is far from being sharp (see Sec.~\ref{sec:results}), meaning that the uncertainty on $\xi$ estimated from Eqs.~\eqref{xi_scaling_approx}-\eqref{overlap_approx} can be quite large.

\section{Model and simulation method} \label{sec:model}

We have performed $NVT$ molecular dynamics simulations (MD) of a system of $N_c$ coarse-grained polymer chains of length (degree of polymerization) $N$ at different monomer densities $\rho=N N_c/V = M/V$. The model chosen to simulate the chains is the bead-spring model of Kremer and Grest \cite{kremer1990dynamics}. The excluded volume interactions between all monomers are modeled as a purely repulsive Weeks-Chandler-Andersen (WCA) potential \cite{weeks1971role}, 

\begin{equation}
\mathcal U (r) = 
\begin{cases}
4 \epsilon \left[ \left(\frac \sigma r\right)^{12} -\left(\frac \sigma r\right)^6+ \frac 1 4 \right] & r \leq 2^{1/6} \sigma\\
0 & \text{otherwise}.\\
\end{cases}
\label{lj}
\end{equation}

\noindent In addition, bonded monomers interact via a finite extensible nonlinear elastic (FENE) potential,

\begin{equation}
\mathcal U_{\text{bond}}(r)= -\frac {k r_0^2}2 \ln [1-(r/r_0)^2],
\label{fene}
\end{equation}

\noindent where $k=30 \epsilon/\sigma^2$ and $r_0 = 1.5 \sigma$. The combined effect of the FENE and the WCA potentials prevents the chains from crossing each other even at high density at temperature $T=1.0$ \cite{kremer1990dynamics}. Since non-bonded interactions are purely repulsive, this model mimics the behavior of polymers in an athermal solvent \cite{koshy2003density,rubinstein2003polymer}.

In the following, all quantities are given in reduced units. The units of energy, length and mass are respectively $\epsilon$, $\sigma$ and $m$, where $\epsilon$, and $\sigma$ are defined by Eq.~\eqref{lj} and $m$ is the mass of a monomer. The units of temperature, pressure, density and time are respectively $[T]=\epsilon/k_B, [P]=\epsilon\sigma^{-3},[\rho]= \sigma^{-3}$ and $[t]=\sqrt{m \sigma^2/\epsilon}$. 

We considered chain lengths $N=50,200$, and $1000$. For $N=50$ and $N=200$, we consider $N_c=200$ and $N_c=50$ chains, respectively, so that the total number of monomers is $M = N_c N=10^4$. For $N=1000$, we simulated two systems, one of $10$ chains ($M=10^4$) and one of $50$ chains ($M=5\cdot 10^4$), in order to check for the presence of finite-size effects. For the system of $10$ chains, five independent simulations were performed in order to improve the statistics. For none of the studied quantities any relevant difference was found between the systems with $M=10^4$ and those with $M=5\cdot 10^4$.

The range of monomer densities $\rho$ for the different systems are reported in Tab.~\ref{tab:details}. While many of these densities are meant to represent a polymer solution, we do not take into account hydrodynamic interactions between the monomers, since we focus on static quantities, which are unaffected by hydrodynamic interactions.

All the simulations were carried out using the LAMMPS software \cite{plimpton1995fast}. The temperature was kept constant at $T=1.0$ by means of a Langevin thermostat \cite{schneider1978molecular}, so that the force experienced by a monomer is 

\begin{equation}
m \ddot {\mathbf r}  = - \mathbf \nabla \mathcal U_\text{tot} (\{ \mathbf r\}) - m \Gamma \dot {\mathbf r} + \sqrt{2m\Gamma k_B T} \ \boldsymbol{\zeta}(t),
\label{langevin}
\end{equation}

\noindent where $\mathbf r$ is the position vector, and $\mathcal U_\text{tot}  (\{ \mathbf r\}) = \mathcal U_\text{tot} (\mathbf r_1, \dots, \mathbf r_M)$ is the total interaction potential acting on the monomer. The second term on the right side of Eq.~\eqref{langevin} represents a viscous damping, with $\Gamma=0.1$ the friction coefficient. The last term is a random, uncorrelated force mimicking the collisions with ``solvent'' particles. 

The simulation box is cubic and periodic boundary conditions are applied in all directions. In all the simulations, the MD integration time step is $\delta t = 3 \cdot 10^{-3}$. The initial configurations are prepared by randomly placing the polymers in the box; for every set of values $(N,N_c,\rho)$ a different initial configuration is created. Initially, overlaps between the monomers are allowed. The overlaps are then removed by using a soft potential whose strength is increased over a short amount of time (``fast push-off" method \cite{auhl2003equilibration}). After the overlaps have been removed, we switch to the WCA potential, Eq.~\eqref{lj}, and perform an equilibration run of duration $t_e$ (corresponding to $t_e/\delta t$ MD integration steps) before starting the production run. Since the relaxation time of a polymer chain scales as $N^2$ for unentangled chains and as $N^{3.4}$ for entangled chains \cite{rubinstein2003polymer}, longer equilibration times were chosen for the systems with longer chains. For each system, we checked that the equilibration time $t_e$ was larger than the longest relaxation time of the system (see Sec.~\ref{sec:msd} in the S.I.). The values of $t_e$ for the different systems are reported in Tab.~\ref{tab:details}. 

\begin{table}[h]
\centering
\caption{Details of the simulated systems: chain length $N$, number of chains $N_c$, number of monomers $M$, monomer density $\rho$, equilibration time $t_e$.}
\label{tab:details}
\begin{tabular}{@{\hspace{1em}} c @{\hspace{1em}} c @{\hspace{1em}} c @{\hspace{1em}} c @{\hspace{1em}} c @{\hspace{1em}}}
\toprule
$N$ & $N_c$ & $M=N_c N$ & $\rho$ (range) & $t_e$ \\
\colrule
$50$ & $200$ & $1 \cdot 10^4$ & $[0.001,1.00]$  & $1.5 \cdot 10^5$ \\
$200$ & $50$ & $1 \cdot 10^4$ & $[0.001,0.85]$  & $4.5 \cdot 10^5$ \\
$1000$ & $10$ & $1 \cdot 10^4$ & $[0.001,0.26]$  & $3.0 \cdot 10^6$ \\
$1000$ & $50$ & $5\cdot 10^4$ & $[0.001,0.26]$ & $3.0 \cdot 10^6$ \\
\botrule
\end{tabular}
\end{table}

\section{Results and discussion} \label{sec:results}

As a preliminary characterization of the structural properties of the simulated systems, we have analyzed the distribution of bond angles and the radius of gyration $R_g$ as a function of density. A complete discussion can be found in Sec.~\ref{sec:gyration} of the Appendix. The main result is that the transition from the semidilute to the concentrated regime occurs at density $\rho \approx 0.3-0.4$. In order to simplify the discussion, we will in the following operatively set the onset of the concentrated regime at $\rho^{**} = 0.3$, which implies that $\chi^{**} \equiv \chi(\rho^{**}) = 0.3N$ (Eq.~\eqref{chi_def}).

In the following subsection, we perform a systematic study of the correlation length, obtained alternatively from the structure factor and from the radial distribution function, as a function of density.

\subsection{Density fluctuation correlation length} \label{sec:xic}

\begin{figure}[h]
\centering
\includegraphics[width=0.45 \textwidth]{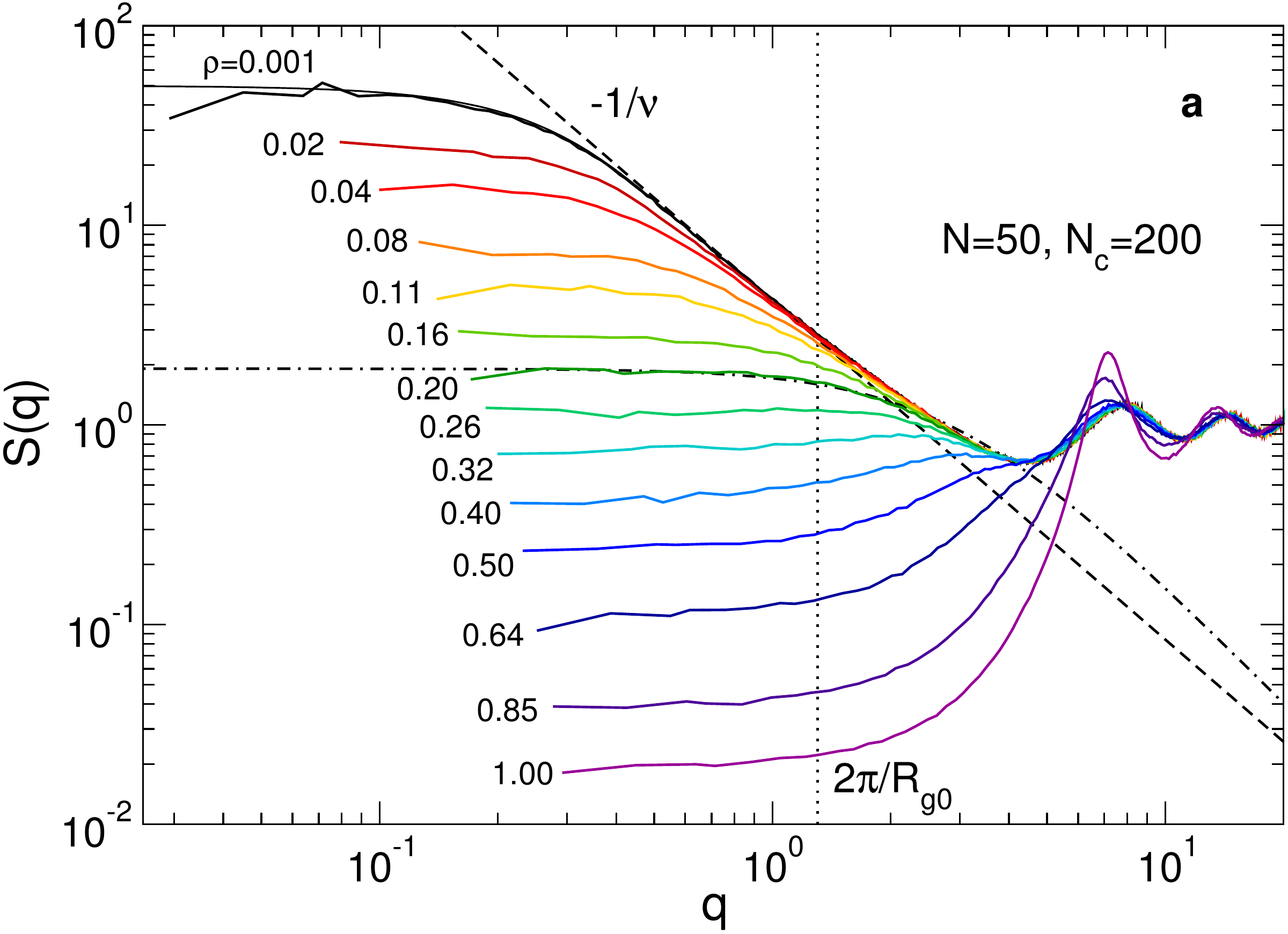}
\includegraphics[width=0.45 \textwidth]{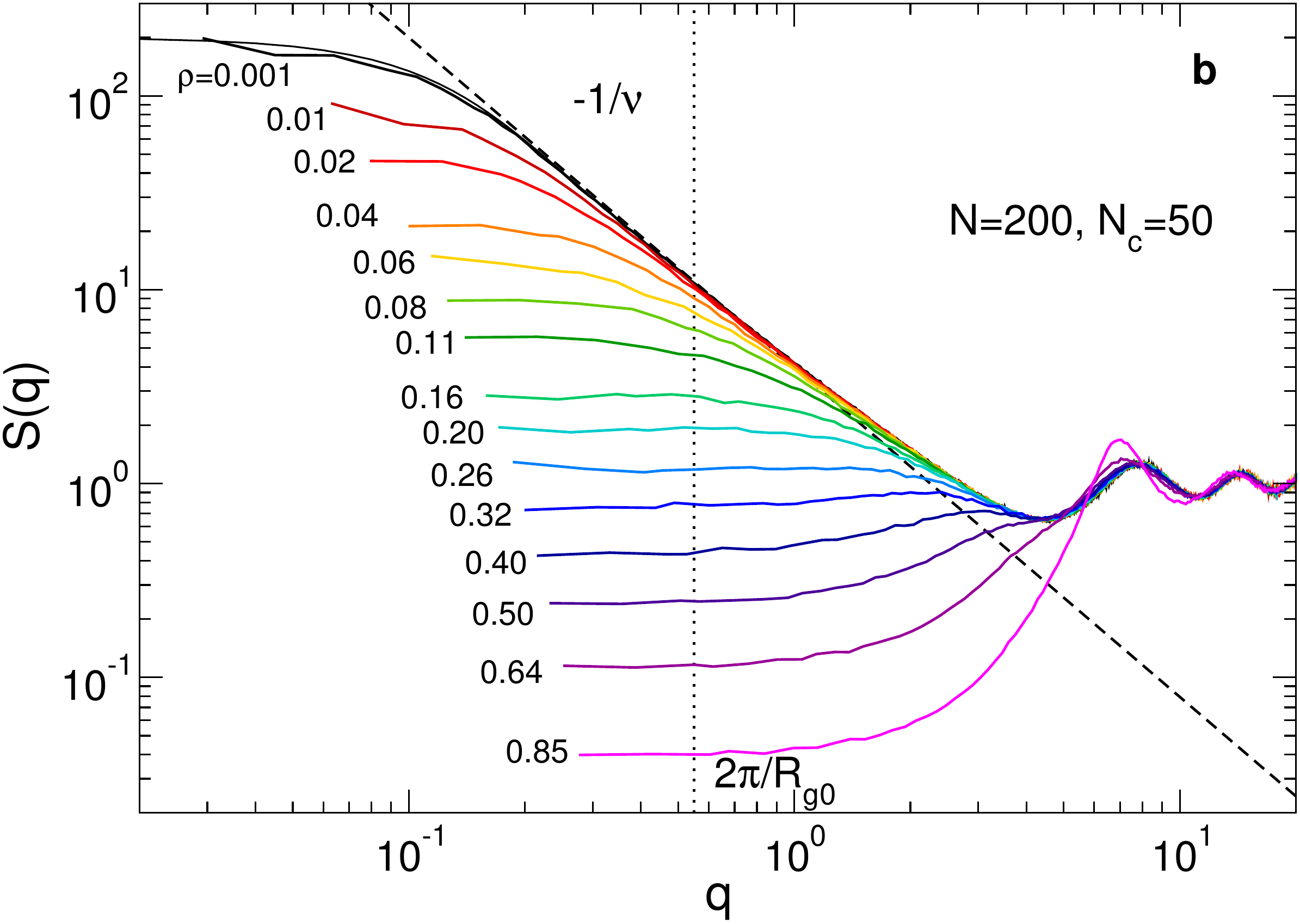}
\includegraphics[width=0.45 \textwidth]{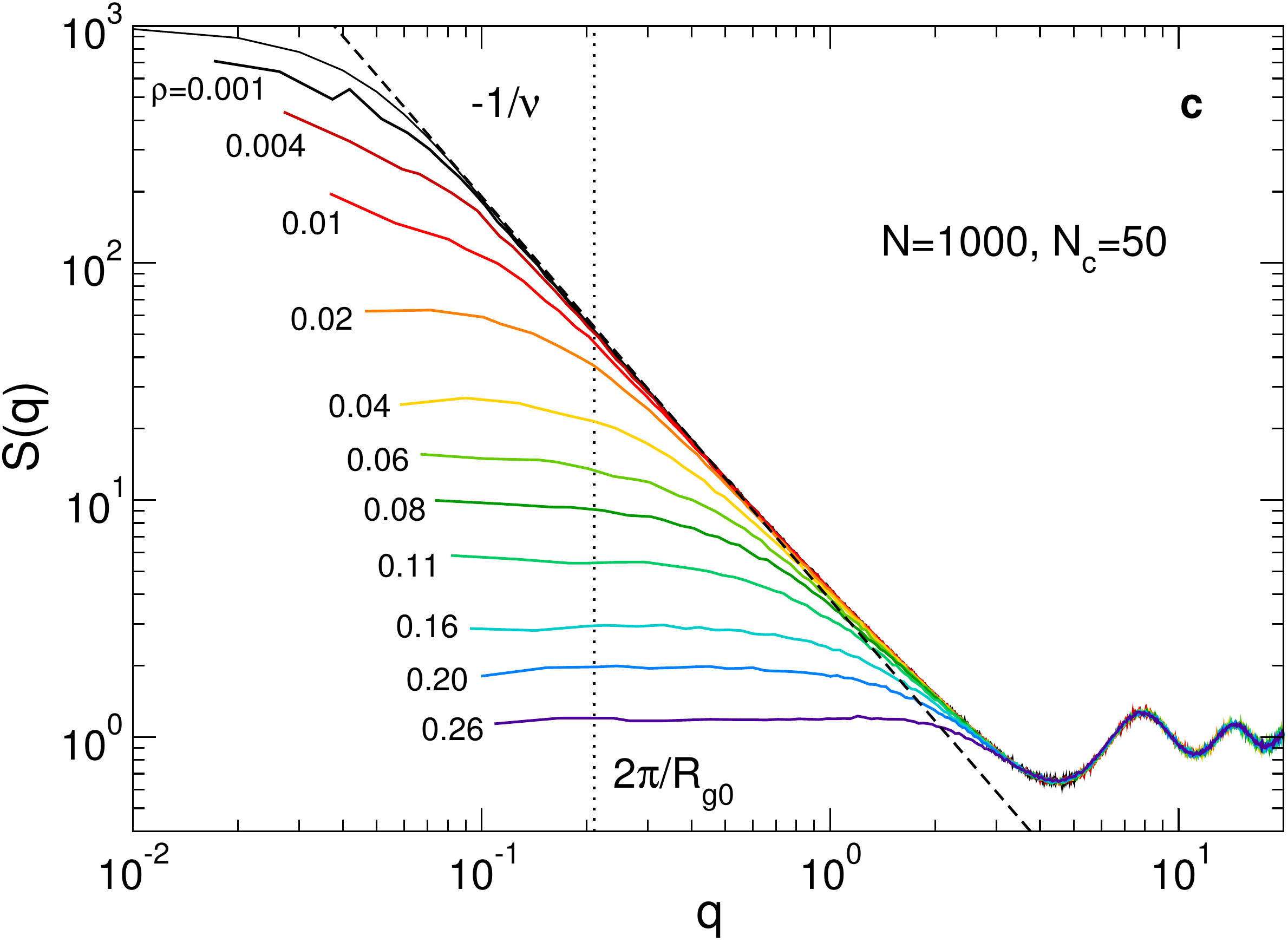}
\caption{Monomer structure factor $S(q)$, Eq.~\eqref{sq_def}, for $N=50$ (a), $N=200$ (b) and $N=1000$ (c), at different monomer densities. Thin continuous lines: $S_1(q)$ for $\rho=10^{-3}$. Dashed lines: slope $-1/\nu \simeq -1.70$ ($\nu \simeq 0.588$). Dash-dotted curve in (a): fit with the Ornstein-Zernike function, Eq.~\eqref{sq_oz}, for $\rho=0.20$.}
\label{sq_all}
\end{figure}

As discussed in Sec.~\ref{sec:theory}, the correlation length $\xi_c$ can be measured either from the monomer structure factor $S(q)$ or from the monomer radial distribution function $g(r)$. The monomer structure factor is defined as \cite{hansen1990theory}

\begin{equation} 
S(\mathbf q) \equiv \frac 1 M\sum_{k,j=1}^{M} \langle \exp [-i \mathbf q \cdot (\mathbf r_k - \mathbf r_j)]\rangle,
\label{sq_def}
\end{equation}

\noindent where $M$ is the number of monomers and $\langle \cdot \rangle$ denotes the thermodynamic average. Since the system is isotropic, we will consider the spherically averaged structure factor, $S(q)$, which depends only on $q\equiv|\mathbf q|$. 

In Fig.~\ref{sq_all} we show the monomer structure factor for $N=50,200$, and $1000$ and different densities.  In the dilute regime, we expect \cite{rubinstein2003polymer,teraoka2002polymer}

\begin{equation}
S(q) \simeq S_1(q) =
\begin{cases}
N /(1+q^2R_{g0}^2/3) & q < 2\pi/ R_{g0}\\
A q^{-1/\nu} &  2\pi/R_{g0}<q<2\pi/b\\
\mathcal O(1) &  q>2\pi/b,\\
\end{cases}
\end{equation}

\noindent where $S_1(q)$ is the single chain structure factor, which can be obtained by computing the expression \eqref{sq_def} for a single chain, and $A>0$ is a constant (the $\rho$-dependence of $S_1(q)$ is reported in Sec.~\ref{sec:form_factor} in the S.I.). In Fig.~\ref{sq_all}a-c, we show as thin continuous lines $S_1(q)$ for $\rho=10^{-3}$. For $N=50$ and $200$, we see that at this density $S(q)=S_1(q)$, indicating that the system is already in the infinite dilution limit, whereas for $N=1000$ this limit is still not yet reached.

In the semidilute regime, we expect \cite{rubinstein2003polymer}

\begin{equation}
S(q)\simeq
\begin{cases}
S(0)/[1+(q\xi_c)^{2}] & q<2\pi/\xi_c\\
B/[1+(q\xi_c)^{1/\nu}] & 2\pi/\xi_c<q<2\pi/b\\
\mathcal O(1) &  q>2\pi/b,\\
\end{cases}
\end{equation}

\noindent where $B>0$ is a constant. From our data at low $\rho$, we can observe at intermediate $q$ the slope $-1/\nu \simeq -1.70$; as expected, this regime is more clearly observable for the longest chains, $N=1000$ (Fig.~\ref{sq_all}c). 

As $\rho$ is increased, the isothermal compressibility $\kappa_T = S(0)/\rho k_B T$ decreases and eventually, for $\rho \geq 0.85$, the structure becomes virtually indistinguishable from that of a dense liquid \cite{hansen1990theory,verlet1968computer}. Comparing $S(q)$ for different chain lengths, we note that for densities $\rho > 0.11 \simeq \rho^*(N=50)$ (see Eq.~\eqref{overlap_approx}), $S(q)$ becomes independent of $N$ (see Fig.~\ref{sq_all_cfr} in the S.I.). This is in agreement with the prediction that in the semidilute and concentrated regimes, $\rho>\rho^*$, the global structure of the system does not depend on $N$ \cite{rubinstein2003polymer}.

\begin{figure}
\centering
\includegraphics[width=0.47 \textwidth]{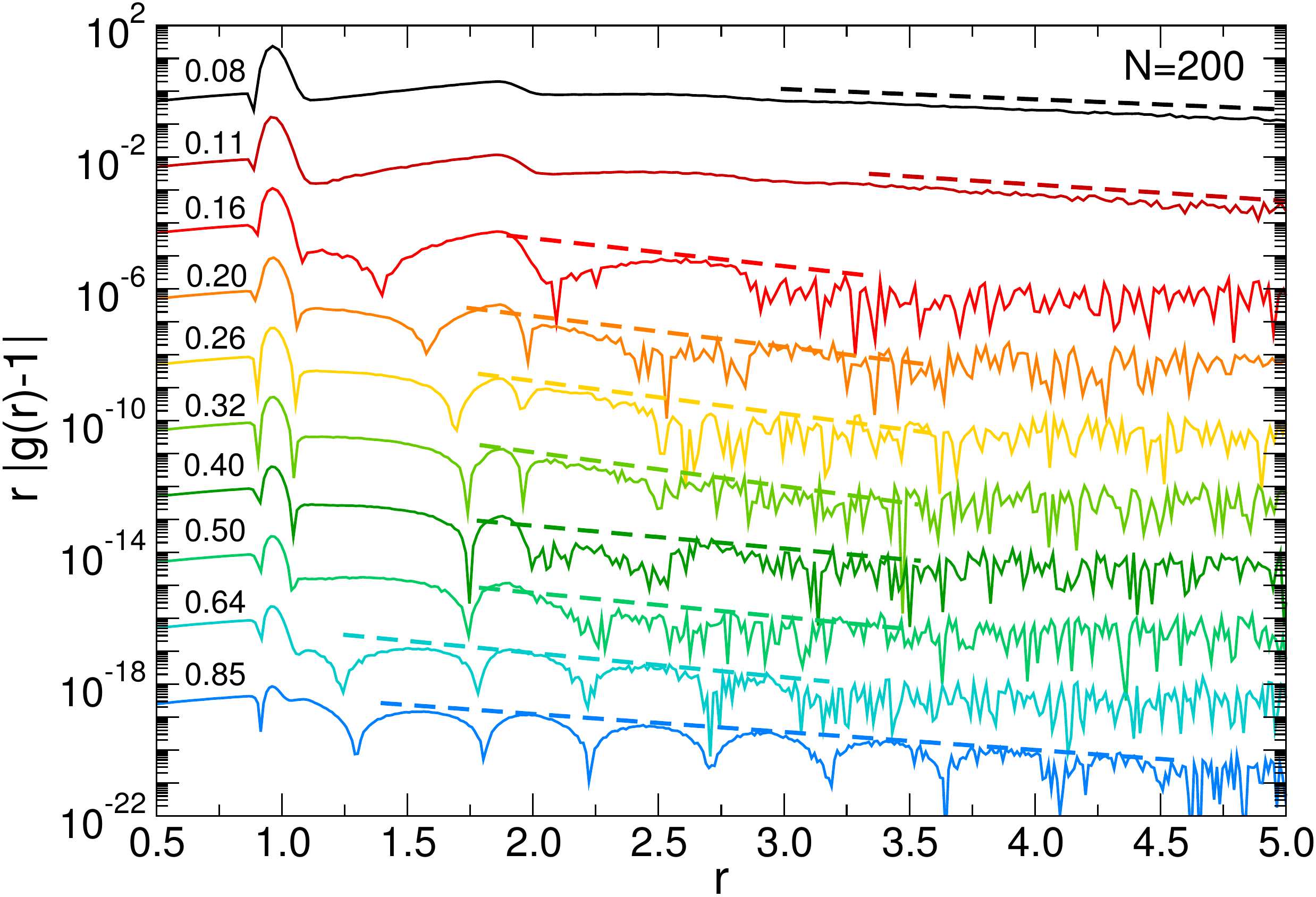}
\caption{ The function $r|g(r)-1| $, with $g(r)$ = radial distribution function, for $N=200$ and different monomer densities. To facilitate the visualization, every curve starting from $\rho=0.08$ was shifted by a factor $100$ with respect to the previous one, except for $\rho=0.85$, which is shifted by a factor $200$. The dashed lines are exponential fits.}
\label{rdf_n200}
\end{figure}

The correlation length $\xi_c$ can be measured by fitting $S(q)$ in the low-$q$ range with the Ornstein-Zernike expression, Eq.~\eqref{sq_oz}. It is clear, however, that this expression only gives a good description of the data for densities $\rho \lesssim 0.3 = \rho^{**}$. To obtain $\xi_c$ at higher densities we consider the radial distribution function of the monomers, $g(r)$, defined as \cite{hansen1990theory}

\begin{equation} 
g(r) \equiv \frac{1}{4 \pi M \rho r^2} \sum_{\substack{k=1\\ j\neq k}}^{M}  \langle \delta(|\mathbf r + \mathbf r_k- \mathbf r_j|)\rangle.
\label{rdf_def}
\end{equation}

\noindent 

\noindent The expected asymptotic expression of $g(r)$ at low and high $\rho$ is given by Eqs.~\eqref{rdf_oz} and \eqref{rdf_oz_sin}. At low $\rho$, we can therefore obtain $\xi_c$ from $g(r)$ by fitting it with an exponential, while at higher $\rho$ we fit the exponential envelope of the function \cite{koshy2003density}. This method is illustrated in Fig.~\ref{rdf_n200}, where we show $r|g(r)-1|$ for $N=200$ and different values of $\rho$, with the slopes resulting from the fit (dashed lines). One can see that the low-$\rho$ form works well up to density $\rho \simeq 0.11$, while the high-$\rho$ one gives a good description of the data starting from $\rho\simeq0.64$. For densities in-between, neither of these two expressions gives a really satisfactory description of the data. We decide nevertheless to fit the data with an exponential also in this intermediate range, as shown in Fig.~\ref{rdf_n200}, in order to have at least an estimate of $\xi_c$. 

\begin{figure}
\centering
\includegraphics[width=0.45 \textwidth]{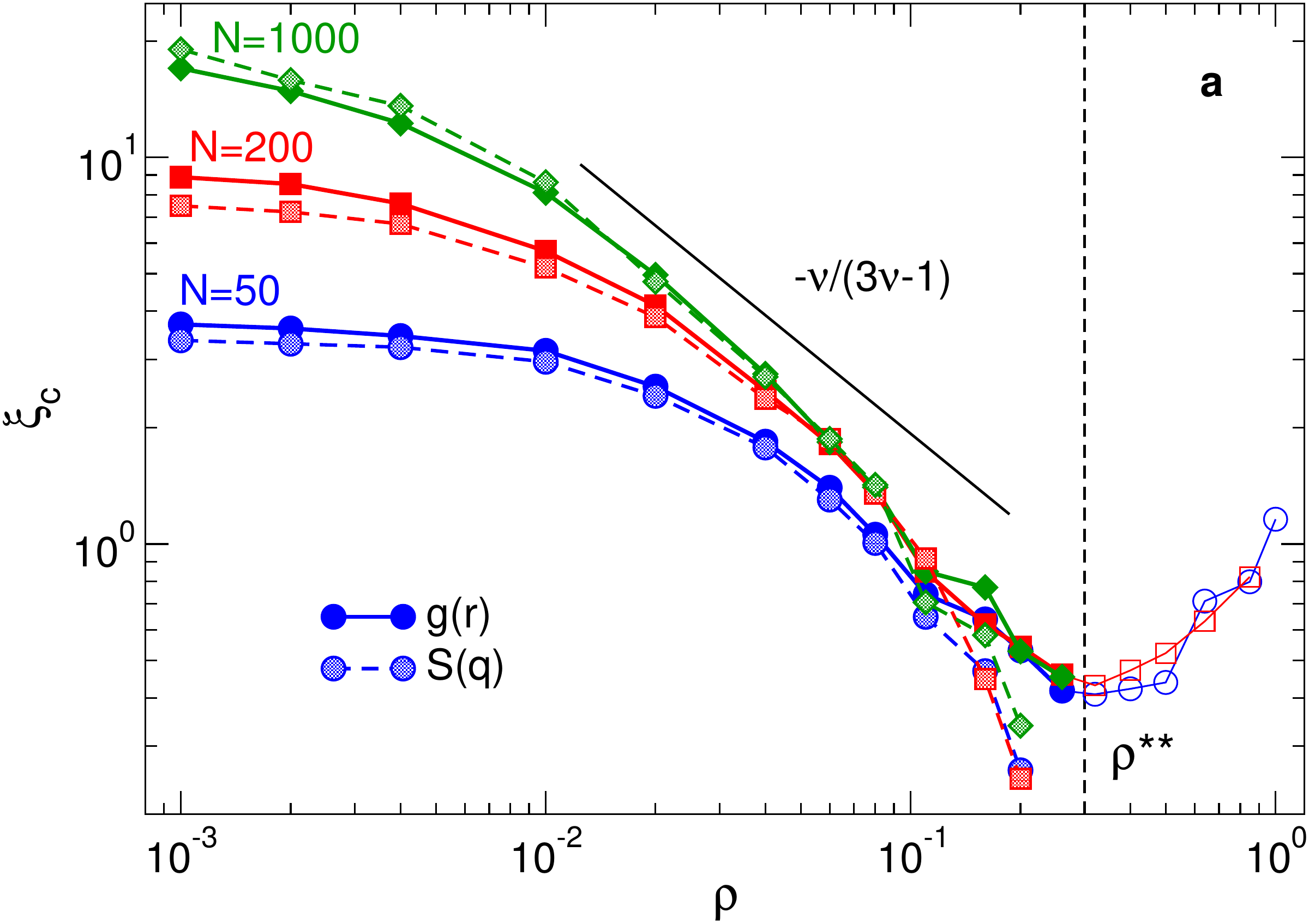}
\includegraphics[width=0.45 \textwidth]{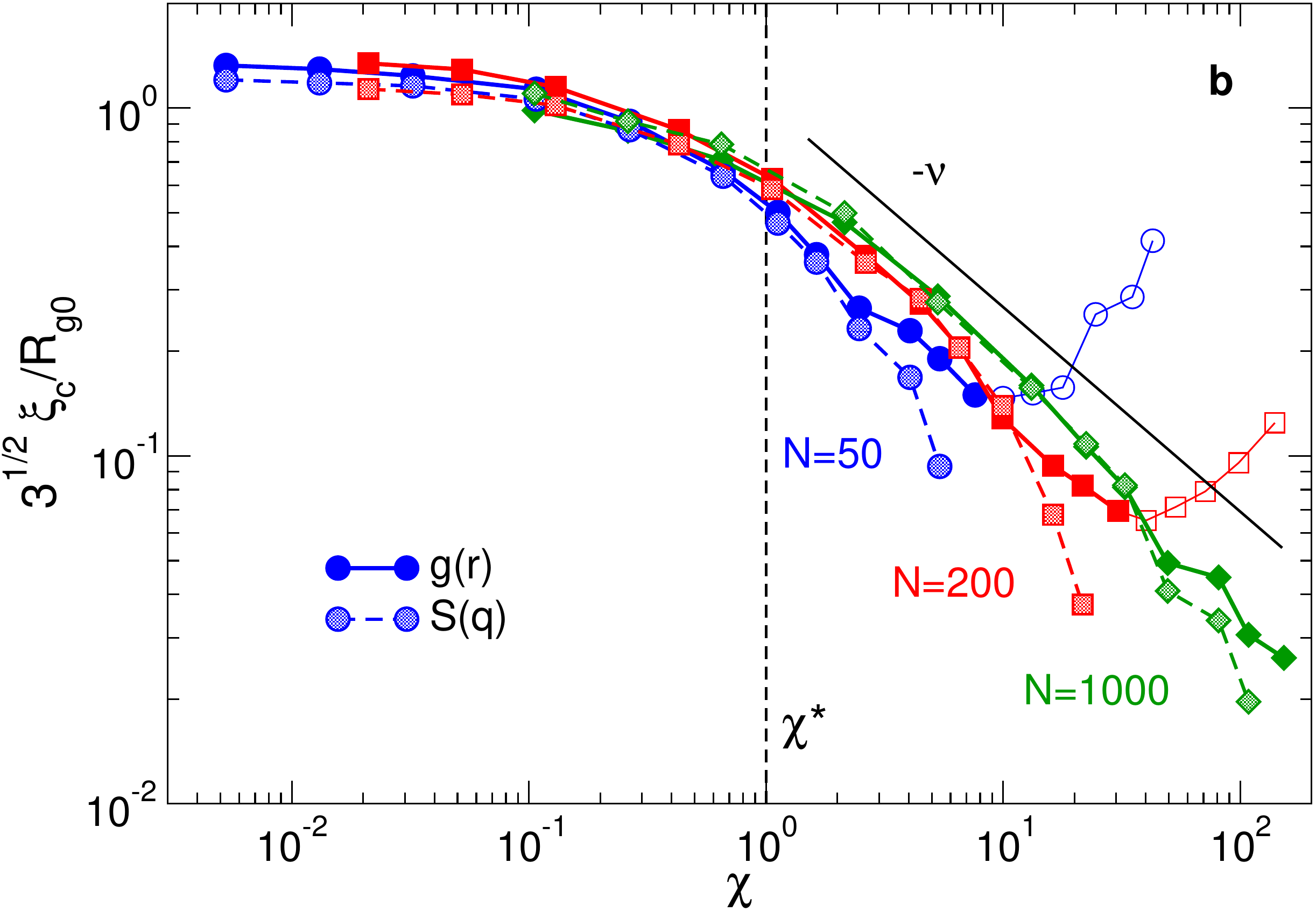}
\caption{The length $\xi_c$ measured from $g(r)$ (dark symbols, continuous lines) and from $S(q)$ (light symbols, dashed lines), as a function of monomer density (a) and the scaling variable $\chi$, Eq.~\eqref{chi_def} (b). Continuous lines represent the slopes predicted by scaling theory for the semidilute regime. Open symbols denote densities $\rho>\rho^{**}=0.3$}
\label{xic_rdf_sq}
\end{figure}

The values of $\xi_c$ obtained by fitting $S(q)$ and $g(r)$ are reported in Fig.~\ref{xic_rdf_sq}a as a function of $\rho$. We note that the two estimates give almost identical results for $\rho<0.2$. For $\rho\geq0.26$, i.e., when we start to approach $\rho^{**}=0.3$ we cannot fit reliably the structure factor with the Ornstein-Zernike function, Eq.~\eqref{sq_oz}, and we have to rely on the radial distribution function. We note that $\xi_c$ reaches a minimum at $\rho \simeq 0.3 = \rho^{**}$ and increases for larger $\rho$. The increase of $\xi_c$ with $\rho$ is due to local packing constraints, and it is a typical behavior for any dense liquid \cite{koshy2003density}.

In Fig.~\ref{xic_rdf_sq}b, we plot $\sqrt{3}\ \xi_c/R_{g0}$ as a function of the scaling variable $\chi$, Eq.~\eqref{chi_def}. At low densities, $\sqrt{3}\ \xi_c/R_{g0}\simeq 1$, in agreement with the theoretical predictions (Eqs.~\eqref{sq_oz} and \eqref{s1q}). We observe that in the semidilute regime, $\xi_c$ follows approximately the predictions of scaling theory Eq.~\eqref{corr_length}, i.e. $\xi_c \propto \rho^{-\nu/(3\nu-1)}$ (Fig.~\ref{xic_rdf_sq}a), or equivalently $\xi_c\propto \chi^{-\nu}$ (Fig.~\ref{xic_rdf_sq}b). The agreement with the theory becomes better with increasing $N$, as expected from the fact that $\lim_{N \to \infty} \rho^* = 0$, or analogously $\lim_{N \to \infty} \chi^{**} = \infty$.

\subsection{Pore size distribution} \label{sec:psd}

\begin{figure}
\centering
\includegraphics[width=0.40 \textwidth]{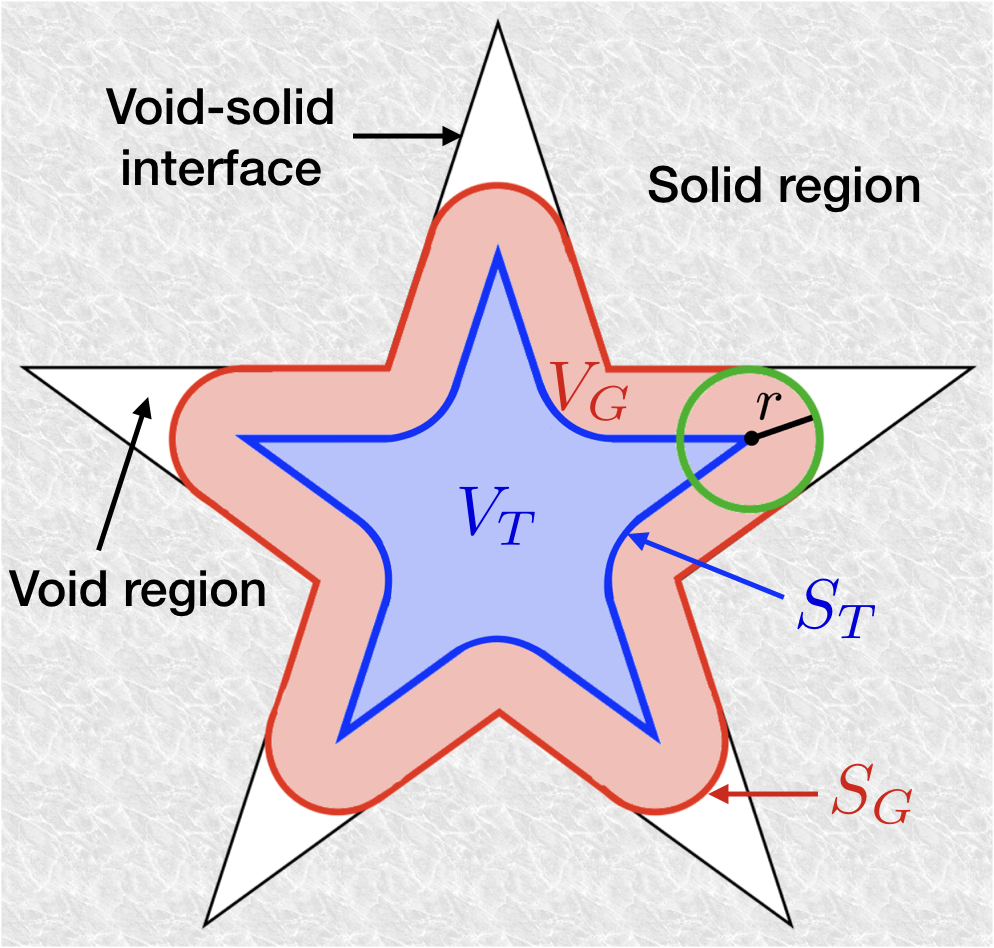}
\caption{Example of a porous medium with a single star-shaped pore. $V_T(r)$ is the volume which is accessible to the center of a spherical probe particle of radius $r$. This volume is enclosed in the surface $S_T$. $V_G(r)$ is the volume which is coverable by spheres of radius $r$ or smaller, and it is enclosed in the Connolly (or reentrant) surface $S_G$.}
\label{pore_cartoon}
\end{figure}

We have seen in the previous subsection how to estimate the monomer correlation length $\xi_c$. As discussed in Sec.~\ref{sec:theory}, only in the semidilute regime we have $\xi_c\approx\xi$. In the dilute regime $\xi_c\approx R_{g0}$ while $\xi\propto (\rho/N)^{-1/3}$. In the concentrated regime, $\xi_c$ increases, while $\xi$ is expected to decrease. In order to get the actual geometrical mesh size, we therefore need a way to measure directly the size of the pores in the system. With this objective in mind, we turn to  a concept that was developed to measure the distribution of pore sizes in solid porous media: the \emph{pore size distribution} (PSD) \cite{scheidegger1957physics,prager1963interphase,torquato2013random,torquato1990nearest,torquato1991diffusion,torquato1995nearest,gelb1999pore,thomson2000modeling,pikunic2003structural,bhattacharya2006fast}. 

We will consider two different definitions of the PSD: The first one, although introduced by others \cite{scheidegger1957physics,prager1963interphase}, was formalized and extensively studied by Torquato and coworkers \cite{torquato2013random,torquato1990nearest,torquato1991diffusion,torquato1995nearest}, while the second one was introduced by Gubbins and coworkers \cite{gelb1999pore,thomson2000modeling,pikunic2003structural,bhattacharya2006fast}. We will in the following consider a generic porous medium in three dimensions, but all the definitions can be extended to any two-phase system and any number of spatial dimensions. Our presentation closely follows the one given by Torquato \cite{torquato2013random,torquato1990nearest,torquato1991diffusion,torquato1995nearest}. A porous medium is a three-dimensional domain of volume $V$ which is composed of two sub-domains: a void (or pore) region, of volume $V_p$ and volume fraction $\phi_p \equiv V_p/V$ (also called the \emph{poro\-sity}), and a solid region of volume $V_s = V-V_p$ and volume fraction $\phi_s \equiv V_s/V=1-\phi_p$. In Fig.~\ref{pore_cartoon} we show a simple example of a porous medium with a single star-shaped pore.

Following Torquato, we start by defining the quantity $V_T(r)$ as the pore volume accessible to the center of a spherical particle with radius $r$. From Fig.~\ref{pore_cartoon} one recognizes that this is the volume inside the surface $S_T$ followed by the center of the particle when the particle slides over the void-solid interface. Naturally, $V_T(r) \leq V_p$. We can then define the \emph{fraction} of pore volume which is accessible to the particle as $F_T(r)\equiv V_T(r)/V_p$. It is clear from the definition that

\begin{equation}
F_T(0) = 1; \ \ \ \ \lim_{r \to \infty} F_T(r) = 0.
\label{ft_lim}
\end{equation}

\noindent Torquato's PSD is then defined as the probability density function \cite{torquato2013random,torquato1990nearest,torquato1991diffusion,torquato1995nearest}

\begin{equation}
P_T(r) \equiv - \frac{d F_T(r)}{dr}.
\label{psd_def}
\end{equation}

\noindent Thus, $P_T(r) dr$ represents the probability that a randomly chosen point in the pore region lies at a distance between $r$ and $r+dr$ from the nearest point on the pore-solid interface. From \eqref{ft_lim} and \eqref{psd_def} it follows that

\begin{equation}
F_T(r) = \int_r^\infty P_T(x) dx,
\label{ft_int}
\end{equation}

\noindent and that $P_T(r)$ is normalized to unity:

\begin{equation}
\int_0^\infty P_T(x) dx = 1.
\end{equation}

\noindent Moreover, $P_T(r)$ will vanish for $r \to \infty$.

The definition of Gubbins differs from that of Torquato by a simple, yet significant detail. Instead of considering the part of pore volume which is accessible to the \emph{center} of a spherical particle of radius $r$, we consider the volume which is accessible to \emph{any point} of the probe particle. In other words, one defines $V_G(r)$ as the pore volume coverable by spheres of radius $r$ or smaller \cite{gelb1999pore} (Fig.~\ref{pore_cartoon}). The surface $S_G$ in which $V_G$ is enclosed is sometimes called \emph{Connolly surface} or \emph{reentrant surface} \cite{connolly1983analytical,thomson2000modeling}.
Therefore $P_G(r) dr$ represents the probability that a randomly chosen point in the pore space is coverable by spheres of radius $r$ but \emph{not} by spheres of radius $r+dr$ \cite{gelb1999pore} \footnote{Other definitions of the PSD are possible. In the definition of Do \emph{et al.} (J. Colloid Interface Sci. \textbf{328}, 110 (2008)), for example, the void-solid interface is the boundary of the region $\mathcal D$ such that if the center of the probe particle is in $\mathcal D$, the solid-probe interaction energy is zero. If the solid-probe interaction potential is taken to be a hard-sphere interaction, this definition is equivalent to the one of Torquato. Despite the arbitrariness deriving from the choice of the solid-probe potential, this method can indeed be more suitable than the one of Torquato in situations where the solid-probe interaction cannot be approximated as a hard-sphere interaction (e.g., in the presence of very soft potentials).}. Once $V_G(r)$ is defined, all the other quantities can be defined exactly as done above for Torquato's PSD. We note that analogous definitions of $V_T$ and $V_G$ have been given by other authors \cite{lee1971interpretation,richards1977areas,connolly1983analytical,thomson2000modeling}.

The PSDs $P_T(r)$ and $P_G(r)$, obtained respectively from $V_T(r)$ and $V_G(r)$, can differ significantly from each other. To see this, let's consider the simple case of a solid material containing a spherical pore of radius $R$. It is easy to see that in this case \cite{prager1963interphase}:

\begin{equation}
F_T(r) = \left(\frac{R-r}{R}\right)^3;  \ \ \ P_T(r) = \frac{3(R-r)^2}{R^3},
\label{sphere_1}
\end{equation}

\noindent with $r\leq R$, while 

\begin{equation}
F_G(r) = \begin{cases}
1 & r \leq R\\
0 & r > R\\
\end{cases}
;\ \ \ \ P_G(r) = \delta(r-R),
\label{sphere_2}
\end{equation}

\noindent where $\delta$ is Dirac's delta distribution. It is clear that in this case the size $R$ of the pore is most readily identified by considering $P_G$, which is nothing else than a delta distribution centered at $R$, whereas $P_T$ goes to zero at $R$. We will see below that also when dealing with more realistic systems, $P_G$ often conveys the information regarding the typical size of the pore in a much more direct manner than $P_T$.

In order to compute $P_T$ and $P_G$ from the simulation data, we first need to divide the sample in pore and solid regions. For simplicity, we assume that the interaction potential between the probe particle of radius $r$ and the monomers is hard-sphere like, i.e.,

\begin{equation}
\mathcal U (d)=
\begin{cases}
\infty & d< r+r_m\\
0 & d \geq r+r_m,\\
\end{cases}
\end{equation}

\noindent where $d$ is the distance between the probe particle and the monomer and $r_m$ is the radius of the monomer when it is approximated as a hard sphere. In the present work, we choose $r_m=\sigma/2=0.5$ \footnote{We note that different choices of $r_m$ will lead to different PSDs. However, given two different values $r_m$ and $r'_m$, we expect the difference between the PSDs computed using these two values to be relevant only when the average pore size is comparable to, or smaller than, $|r_m-r'_m|$.}. Once this assumption is made, we can proceed to the calculation of the PSD, following the two definitions. The algorithms used to calculate $P_T$ and $P_G$ are described in Sec.~\ref{sec:app1}.

\begin{figure}
\centering
\includegraphics[width=0.45 \textwidth]{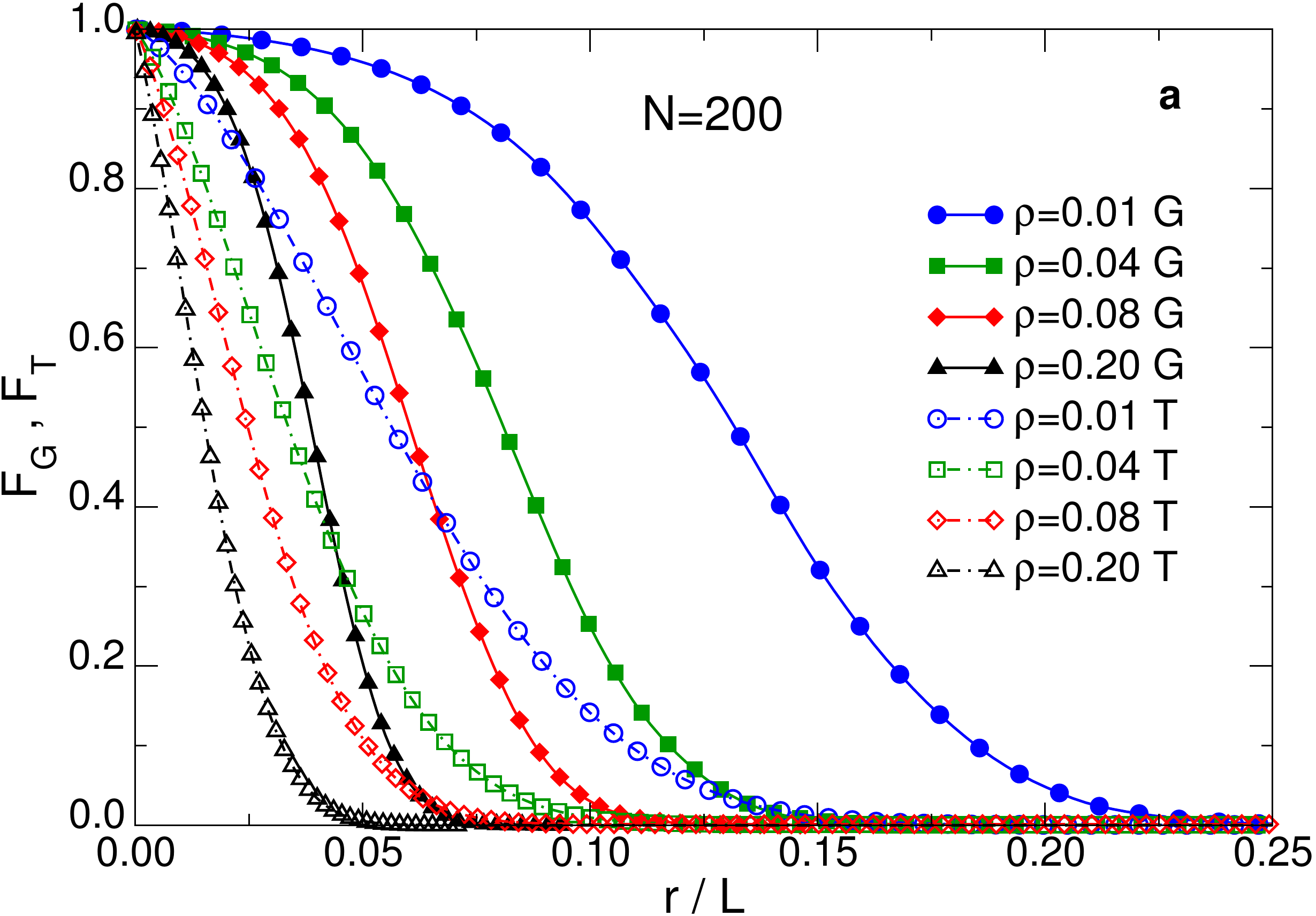}
\includegraphics[width=0.45 \textwidth]{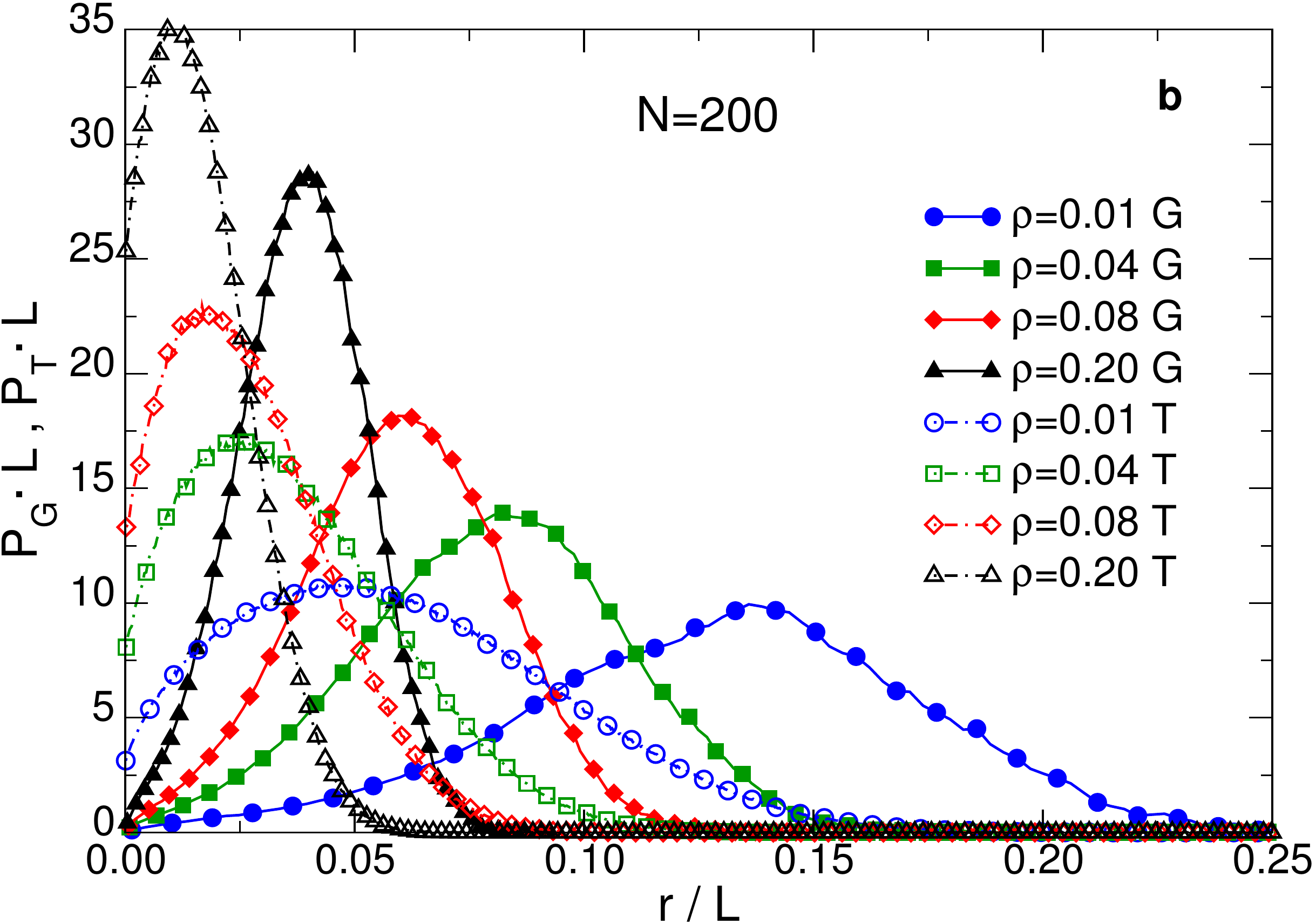}
\caption{Complementary cumulative distribution $F_\alpha(r)$ ($\alpha=T,G$) (a) and PSD $P_\alpha(r)$ (b), as obtained using the definitions of Gubbins (filled symbols, contineous lines) and of Torquato (open symbols, dashed lines).}
\label{vpore_psd}
\end{figure}

In Fig.~\ref{vpore_psd}, we show the complementary cumulative distribution functions $F_G$ and $F_T$ (Fig.~\ref{vpore_psd}a) and the  PSDs $P_G$ and $P_T$ (Fig.~\ref{vpore_psd}b) for $N=200$ and different densities. In both graphs, $r$ has been normalized by the length of the simulation box $L=(M/\rho)^{1/3}$ in order to make the plot more readable. We find that $F_T$ drops to zero much faster than $F_G$, and that the shapes of the two functions are quite different. As a consequence, also the distributions $P_T$ and $P_G$ are very different. 

\begin{figure}
\centering
\includegraphics[width=0.45 \textwidth]{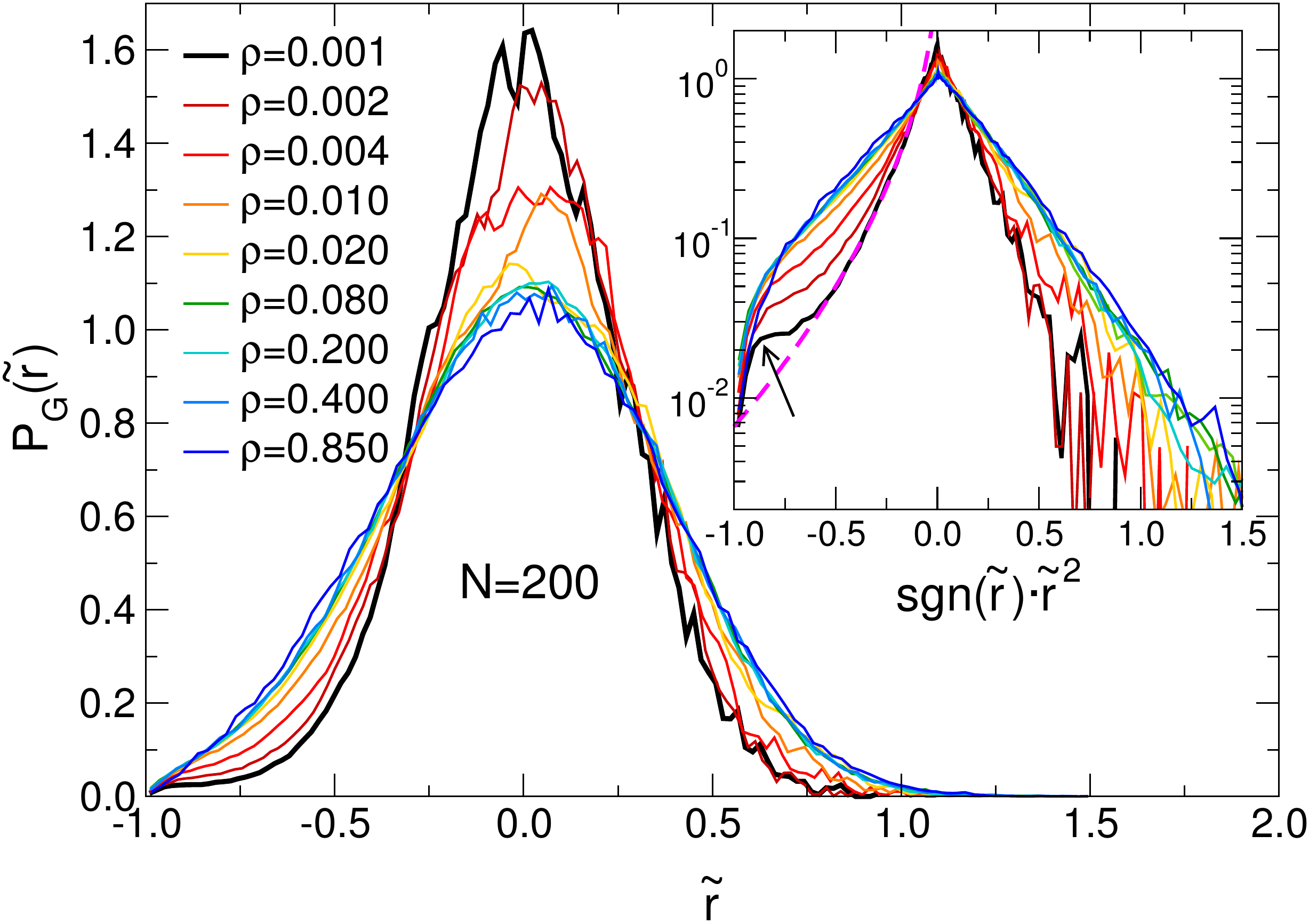}
\caption{Gubbins's PSD, $P_G(\tilde r)$, for $N=200$ and at different densities $\rho$, as a function of $\tilde r \equiv r/\langle r\rangle_G-1$. \emph{Inset}: Probability distribution of the variable $\sgn(\tilde r)\cdot {\tilde r}^2$ in semi-logarithmic scale. Dashed line: exponential fit. See main text for the explanation of the arrow.}
\label{psd_gubbins_rescaled}
\end{figure}

In Fig.~\ref{psd_gubbins_rescaled} we show $P_G(\tilde r)$, where $\tilde r \equiv r/\langle r\rangle_G-1$, for $N=200$ and different densities. One recognizes that at intermediate and high densities the distribution is very similar to a Gaussian. This becomes clearer when plotting the probability distribution of the variable $\sgn(\tilde r)\cdot {\tilde r}^2$ (with $\sgn$ the sign function) in semi-logarithmic scale, as shown in the inset of Fig.~\ref{psd_gubbins_rescaled}. Only at density $\rho<\rho^*\simeq 0.032$ we observe significant deviations from Gaussianity, with the left-side tail displaying a markedly exponential decay (dashed line in the inset of Fig.~\ref{psd_gubbins_rescaled}). The small shoulder observable at low $\rho$ for small values of $\tilde r$ (indicated by the arrow in the inset of Fig.~\ref{psd_gubbins_rescaled}) comes from distances inside the pervaded volume of the chain.

\begin{figure}
\centering
\includegraphics[width=0.45 \textwidth]{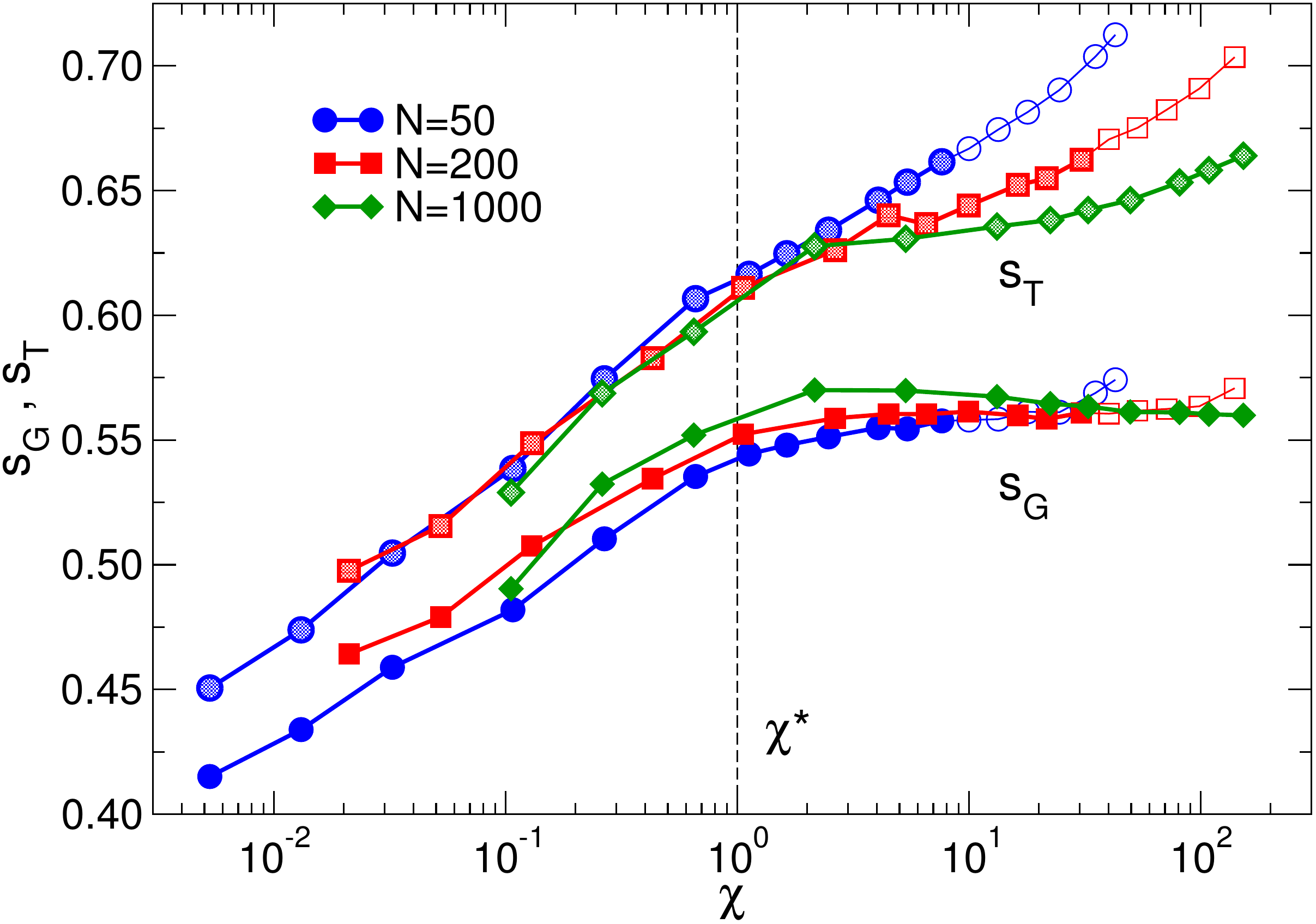}
\caption{Relative standard deviation $s_\alpha$ ($\alpha=G,T$), Eq.~\eqref{rel_std}, as a function of the scaling variable $\chi$.  $s_G$ is shifted up by $0.2$ to facilitate the comparison with $s_T$.  Open symbols denote densities $\rho>\rho^{**}=0.3$}
\label{rstd_over_rav}
\end{figure}

We note that the width of  $P_G(\tilde r)$ increases with increasing density, indicating that the pore space becomes more heterogeneous on the length scale of the mean pore size $\langle r\rangle_G$. Moreover, although the shape of $P_G(\tilde r)$ changes significantly when $\rho$ is increased in the dilute regime, it remains almost unchanged for $\rho>\rho^*$. This effect can be better appreciated by considering the relative standard deviation,

\begin{equation}
s_\alpha \equiv \sqrt{\frac{\langle r^2\rangle_\alpha}{\langle r \rangle_\alpha^2} -1 } \ \ \ \ (\alpha=T,G),
\label{rel_std}
\end{equation}

\noindent which is a measure of the width of the PSD relative to its mean value.

In Fig.~\ref{rstd_over_rav}, we show $s_G$ and $s_T$ as a function of the scaling variable $\chi$. One sees that $s_G$ increases with increasing $\chi$ up to the beginning of the semidilute regime at $\chi=\chi^*=1$, whereas it is a constant for $\chi > \chi^*$. This transition from dilute to semidilute behavior is quite sharp if compared with that of $\xi_c$ (Fig.~\ref{xic_rdf_sq}b) or $R_g$ (Fig.~\ref{gyr_norm} of Sec.~\ref{sec:gyration}), suggesting that $P_G$ can be used to determine $\rho^*$ in a more precise way. In addition, $s_G$ is almost independent of $N$ for a given value of $\chi$.

The relative standard deviation $s_T$ of Torquato's PSD behaves quite differently, showing a marked dependence on $N$ at fixed $\chi$ for $\chi>1$. Only for $N=1000$ a behavior similar to that of $s_G$, with an increase up to $\chi=1$ followed by a plateau, is recovered. These observations suggest that $P_G$ may be more suitable than $P_T$ to characterize the pore structure of the system. This will also be argued below based on different considerations. Additional details on $P_T$ will be given in Sec.~\ref{sec:analytical}.
Finally we note that the value of $s_G$ is significantly smaller than the one of $s_T$ (in Fig.~\ref{rstd_over_rav} $s_G$ is shifted upwards by 0.2) which shows that $P_G$ is much narrower than $P_T$ when both are normalized by the respective mean values. 

\subsection{Estimating the geometrical mesh size from the pore size distribution} \label{sec:estimating_xi}

Once the PSD has been determined, we need to extract from it a quantity that we can compare with $\xi_c$ and with the theoretical expectations for $\xi$. The most natural choice is to consider the mean pore radius,

\begin{equation}
\langle r \rangle_{\alpha} \equiv \int_0^\infty x \ P_{\alpha}(x) dx =  \int_0^\infty F_{\alpha}(x) dx \ \ \ \ (\alpha=T,G).
\end{equation}

\noindent Although other choices are possible, such as the position of the peak of $P(r)$,  we found that the main results do not depend significantly on this choice.

\begin{figure}
\centering
\includegraphics[width=0.43 \textwidth]{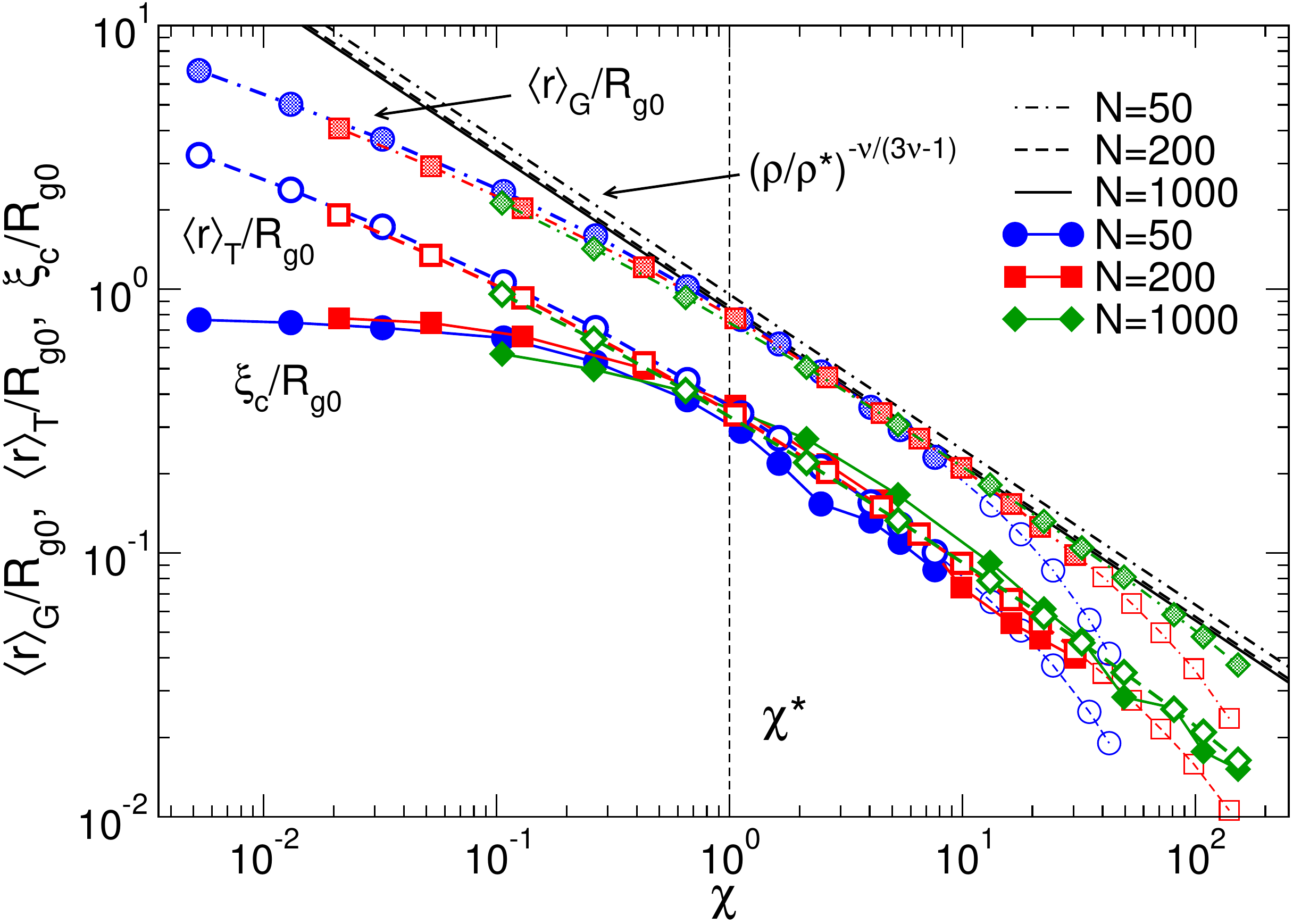}
\caption{Comparison between the average pore sizes, the density fluctuation correlation length $\xi_c$ and the scaling estimate of Eqs.~\eqref{xi_scaling_approx}-\eqref{overlap_approx} (with $\rho$ expressed as a function of $\chi$ using Eq.~\eqref{chi_def}), as a function of the scaling variable $\chi$. Dark filled symbols with continuous lines:  $\xi_c/R_{g0}$. Thick-bordered open symbols with dashed lines: $\langle r\rangle_T/R_{g0}$. Light filled symbols with dash-dotted lines: $\langle r\rangle_G/R_{g0}$. Thin-bordered open symbols denote densities $\rho>\rho^{**}=0.3$. Thick black lines: scaling estimate for $N=50$ (dash-dotted) $N=200$ (dashed) and $N=1000$ (continuous).}
\label{xi_all_cfr}
\end{figure}

\noindent In Fig.~\ref{xi_all_cfr} we show $\langle r \rangle_G/R_{g0}$ and $\langle r \rangle_T/R_{g0}$ as a function of the scaling variable $\chi$ for different values of $N$. These two quantities are compared with $\xi_c/R_{g0}$, where $\xi_c$ is measured from the radial distribution function (points for $\rho>0.3$ not shown), and with the scaling estimate $\xi/R_{g0} \simeq \left( \rho /\rho^{*}\right)^{-{\nu}/{(3\nu-1)}}$, Eq.~\eqref{xi_scaling_approx}, where $\rho^*$ was estimated using Eq.~\eqref{overlap_approx}. We observe that, despite the fact that $\langle r \rangle_G > \langle r \rangle_T$, both quantities show a remarkably good agreement with the scaling prediction, Eq.~\eqref{xi_scaling}. Moreover, above the dilute regime, i.e., $\chi \gtrsim 1$, $\langle r\rangle_T \simeq \xi_c$, likely because the calculation of both quantities involves averaging over distances which are of similar magnitude.

\begin{figure}
\centering
\includegraphics[width=0.45 \textwidth]{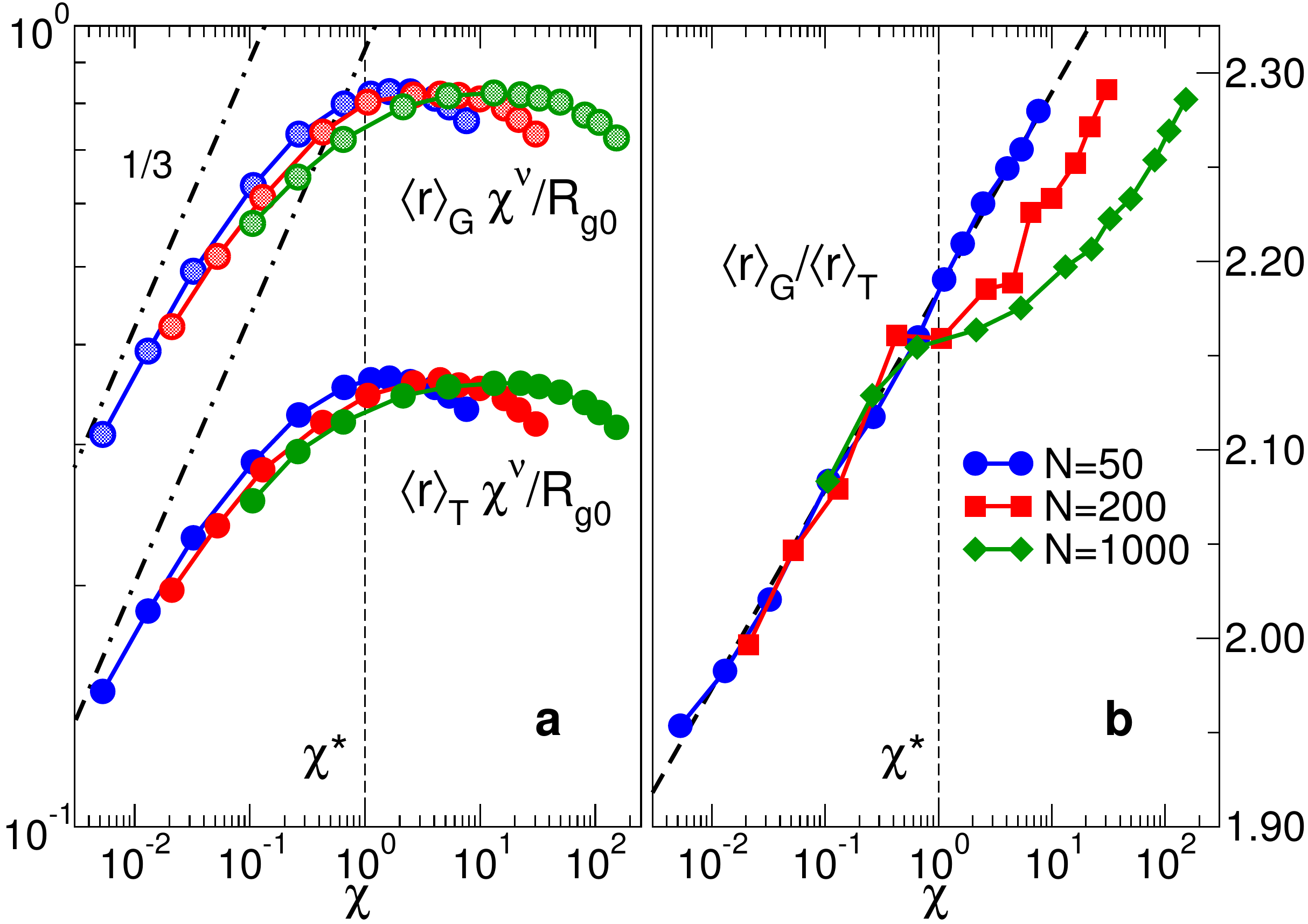}
\caption{(a) $\chi^\nu \cdot \langle r\rangle_\alpha/ R_{g0}$ ($\alpha=G,T$) as a function of the scaling variable $\chi$. Dash-dotted lines: Predicted slopes in the dilute limit, Eq.~\eqref{xi_scaling}. (b) Ratio $\langle r \rangle_G/\langle r\rangle_T$ as a function of $\chi$.  Thick dashed line: $2.18+0.0458 \ln(\chi)$. Points corresponding to $\rho>\rho^{**}=0.3$ are not shown.}
\label{rav_rescaled}
\end{figure}

To better see that $\langle r \rangle_\alpha$ ($\alpha=T,G$) follow the scaling prediction, we show in Fig.~\ref{rav_rescaled}a $\langle r\rangle_\alpha$ multiplied by $\chi^\nu$ (points corresponding to $\rho>0.3$ not shown): For $\chi \gtrsim 1$, they reach a plateau, the length of which increases with increasing $N$ implying that $\langle r\rangle_\alpha$ is indeed proportional to $\chi^{-\nu}$. Moreover, we also see that for low $\chi$ the function tends towards the predicted behavior $\chi^{-\nu+1/3}$, corresponding to the average distance between neighboring chains, Eq.~\eqref{xi_scaling}. In Fig.~\ref{rav_rescaled}b we show the ratio $\langle r\rangle_G/\langle r \rangle_T$ as a function of $\chi$ (points corresponding to $\rho>0.3$ not shown). For $\chi<1$, the ratio increases logarithmically with increasing $\chi$ and is independent of $N$. This demonstrates that in the dilute regime $\langle r\rangle_G$ and $\langle r\rangle_T$ are directly related to each other via an $N$-independent function. For $\chi>1$ the $N=50$ data follows this logarithmic dependence whereas the larger systems show a plateau the width of which increases with $N$, and only for larger $\chi$ the ratio increases again.
In summary one can conclude that this ratio displays a surprisingly weak dependence on $\chi$, changing only by $\simeq 15\%$ over several decades in $\chi$.

To summarize, we have two possible estimators for the mesh size, which are both in excellent agreement with the scaling prediction and differ from each other by a multiplicative factor $\simeq 2$. This factor is of course irrelevant if we are only interested in order-of-magnitude estimates of the mesh size, but it is relevant when more precise information about the size of the pores are needed. In other words, it is important to understand which PSD, $P_G$ or $P_T$, gives us a more precise information about the ``real'' size of the pores. In the following we will therefore  discuss how to interpret the PSD, and which estimator to choose between $\langle r \rangle_G$ and $\langle r \rangle_T$ when more quantitative information about the mesh size is required.

\begin{figure}
\centering
\includegraphics[width=0.45 \textwidth]{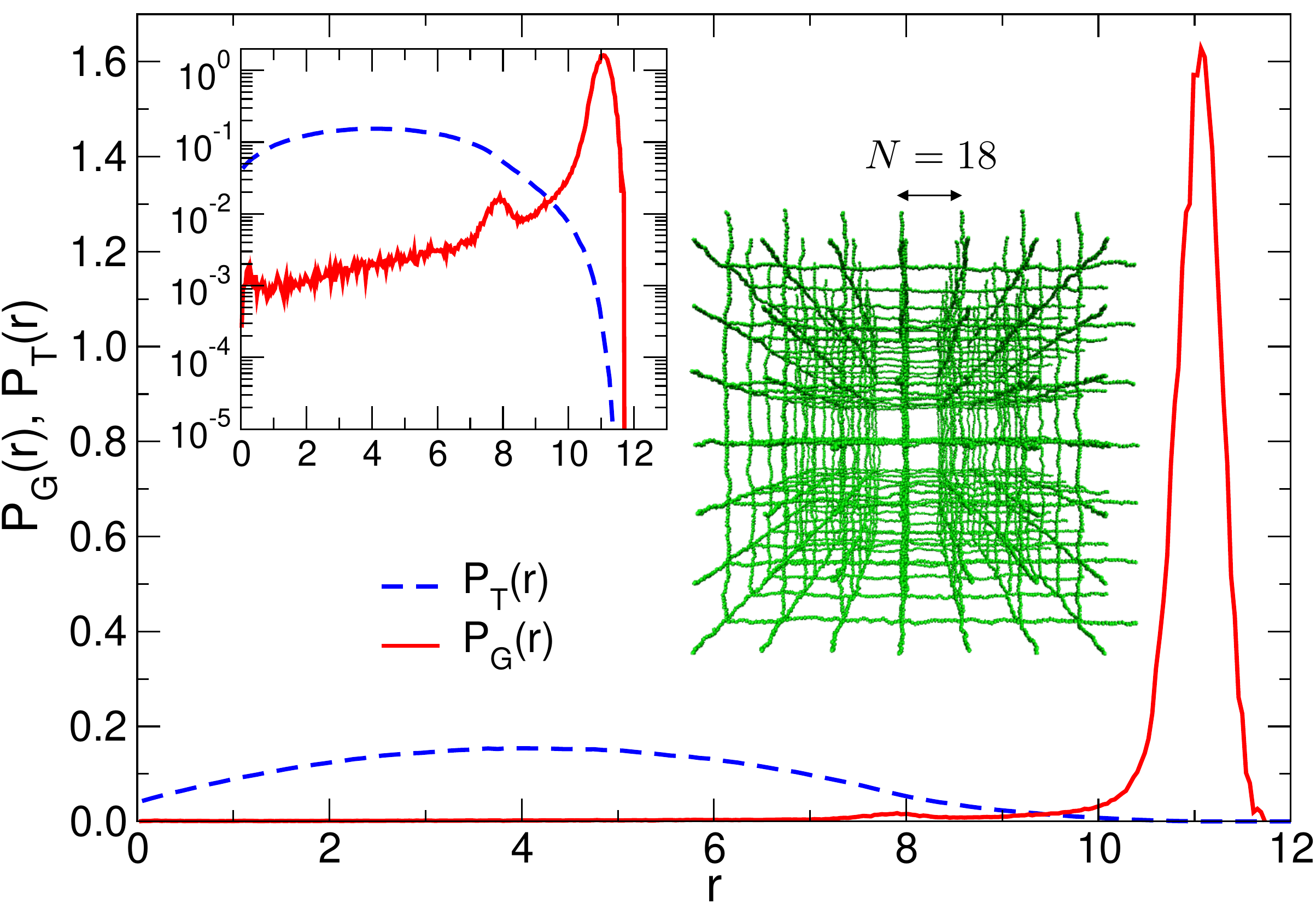}
\caption{PSD for a model polymer gel (cubic lattice with strand length $N=18$; a snapshot of the system is reported in the main figure). Dashed blue line: Torquato's PSD. Continuous red line: Gubbins's PSD. \emph{Inset:} double logarithmic plot of the same PSDs.}
\label{psd_lattice}
\end{figure}

When considering the case of a solid material containing disconnected spherical pores, Eqs.~\eqref{sphere_1} and \eqref{sphere_2}, one can see that it is much more straightforward to infer the value of $R$ from $P_G(r)$ than from $P_T(r)$. Indeed, using Eq.~\eqref{sphere_1} we find $\langle r\rangle_T = R/4$, whereas from Eq.~ \eqref{sphere_2} one obtains $\langle r\rangle_G = R$. Thus already this simple example hints that $\langle r \rangle_G$ is a better indicator for the ``real'' pore size than $\langle r \rangle_T$. In order to consider a somewhat less artificial example, we simulated a simple model of a polymer gel: Polymer strands of length $N=18$ are placed on the edges of a cubic lattice and connected to each other at the vertices (see snapshot in Fig.~\ref{psd_lattice}). Initially, the distance between each pair of bonded monomers is $r=1$. The network is then allowed to relax at constant volume. In order to grant additional flexibility to the chains, we set the $k$ parameter in the FENE potential, Eq.~\eqref{fene}, to $k=20$. By construction, the mesh size of this system is $\xi \simeq N$, and therefore the ``ideal'' PSD should have a strong signal at $r \simeq N/2 = 9$. More precisely, taking into account the tri-dimensionality of the system and the finite diameter of the monomers, $2r_m=1$, we expect a strong signal between $(N-1)/2=8.5$ (edge of the cubic cell) and $(\sqrt{2}N-1)/2\simeq 12.2$ (face diagonal of the cubic cell).

In Fig.~\ref{psd_lattice}, we show $P_T(r)$ and $P_G(r)$ for this model gel. We observe that, while $P_T(r)$ displays a broad profile which peaks at $r\simeq 4$, $P_G(r)$ has a negligible value over the whole $r$ range, except for a very sharp peak at $r\simeq 11\simeq(\sqrt{2}N-1)/2$. The information about the mesh size is not, however, completely absent from $P_T(r)$, but it is ``hidden'' in the value of $r_\text{max}$ at which $P(r_\text{max}) = 0$ (the largest pore size), as it becomes clear when plotting $P_T(r)$ in semi-logarithmic scale (inset of Fig.~\ref{psd_lattice}). From this semi-logarithmic plot, one can also see that $P_G(r)$ has a secondary peak (much smaller than the main one) at $r\simeq8\simeq(N-1)/2$, which originates from the portion of pore space in the vicinity of the surface of the lattice cells. Furthermore, we note that also for this idealized system the ratio $\langle r\rangle_G/\langle r\rangle_T$ is very close to $2$ ($\langle r\rangle_G/\langle r\rangle_T\simeq2.6$), and this remains true also when other densities are considered (not shown). This is a consequence of the fact that for all the systems studied $P_T(r)$ is fairly symmetrical, and $P_G(r)$ peaks close to $r_\text{max}$. The precise value of  $\langle r\rangle_G/\langle r\rangle_T$ will however depend on the relative forms of $P_T$ and $P_G$.

From these two examples, we conclude that $P_G(r)$ is the quantity that gives a more immediate and easy to interpret information on the pores in the system. This does not mean that $P_T(r)$ should not be used, but only that care should be taken in its interpretation.

\subsection{$P_T(r)$: Comparison with analytical models} \label{sec:analytical}

In the previous section we have shown how the PSD can be connected to the geometrical mesh size and have compared the PSD of Torquato, $P_T(r)$, to that of Gubbins, $P_G(r)$. In this section, we will illustrate some analytical results which can be used to gain insight into $P_T(r)$. We focus on this distribution because quasi-exact results are available for it \cite{torquato2013random,torquato1990nearest,torquato1991diffusion,torquato1995nearest}, whereas this is not the case for $P_G(r)$.

Torquato and coworkers have studied extensively the properties of porous media, and in particular the PSD of systems of identical particles interacting via an arbitrary potential. For a system of hard spheres (HS) of radius $R$ at density $\rho$, they demonstrated that \cite{torquato2013random,torquato1990nearest,torquato1991diffusion,torquato1995nearest}

\begin{equation}
P_T^\text{\tiny{HS}}(x)= \frac{3 \eta}{R} F^\text{\tiny{HS}}_T(x) ( a_0 x^2 + 2 a_1 x + 4 a_3),
\label{psd_hs}
\end{equation}

\noindent with

\begin{equation}
F^\text{\tiny{HS}}_T(x)  = \exp [ - \eta ( a_0 x^3 + 3 a_1 x^2 +12 a_2 x + a_3 ) ].
\end{equation}

\noindent Here $x\equiv (r+R)/R$, $\eta\equiv 4 \pi R^3 \rho/3$ is a dimensionless density which for hard spheres is equivalent to the solid volume fraction $\phi_s$, and $a_i$, $i=0,1,2,3$ are functions of $\eta$ only, whose explicit expression can be found in Ref.~\citenum{torquato1995nearest}.

For overlapping spheres (OS), i.e., spheres which can overlap with no energy penalty, they found \cite{torquato1991diffusion,torquato2013random}

\begin{equation}
P_T^\text{\tiny{OS}} (x) = \frac{3 \eta}{R} x^2 \exp \left[-\eta (x^3 - 1) \right].
\label{psd_os}
\end{equation}

\noindent Note that in this case, $\eta$ is not equivalent to the particle volume fraction $\phi_s$, and indeed it can be shown that $\phi_s=1-e^{-\eta}$ \cite{torquato1991diffusion,torquato2013random}. For small values of $\eta$, Eq.~\eqref{psd_hs} reduces to Eq.~\eqref{psd_os}, i.e., the HS model is equivalent to the OS model \cite{torquato1991diffusion,torquato2013random}.

We make the assumption that the measured PSD $P_T(r)$ can be fitted by one of the two functional forms \eqref{psd_hs}-\eqref{psd_os}, with $R$ and $\rho$ as fit parameters. In other words, we assume that it is possible to map our system on a system of hard (resp. overlapping) spheres with radius $R_\text{\tiny{HS}}(\rho,N)$ (resp. $R_\text{\tiny{OS}}(\rho,N)$) and density $\rho_\text{\tiny{HS}}(\rho,N)$ (resp. $\rho_\text{\tiny{OS}}(\rho,N)$).

\begin{figure}
\centering
\includegraphics[width=0.45 \textwidth]{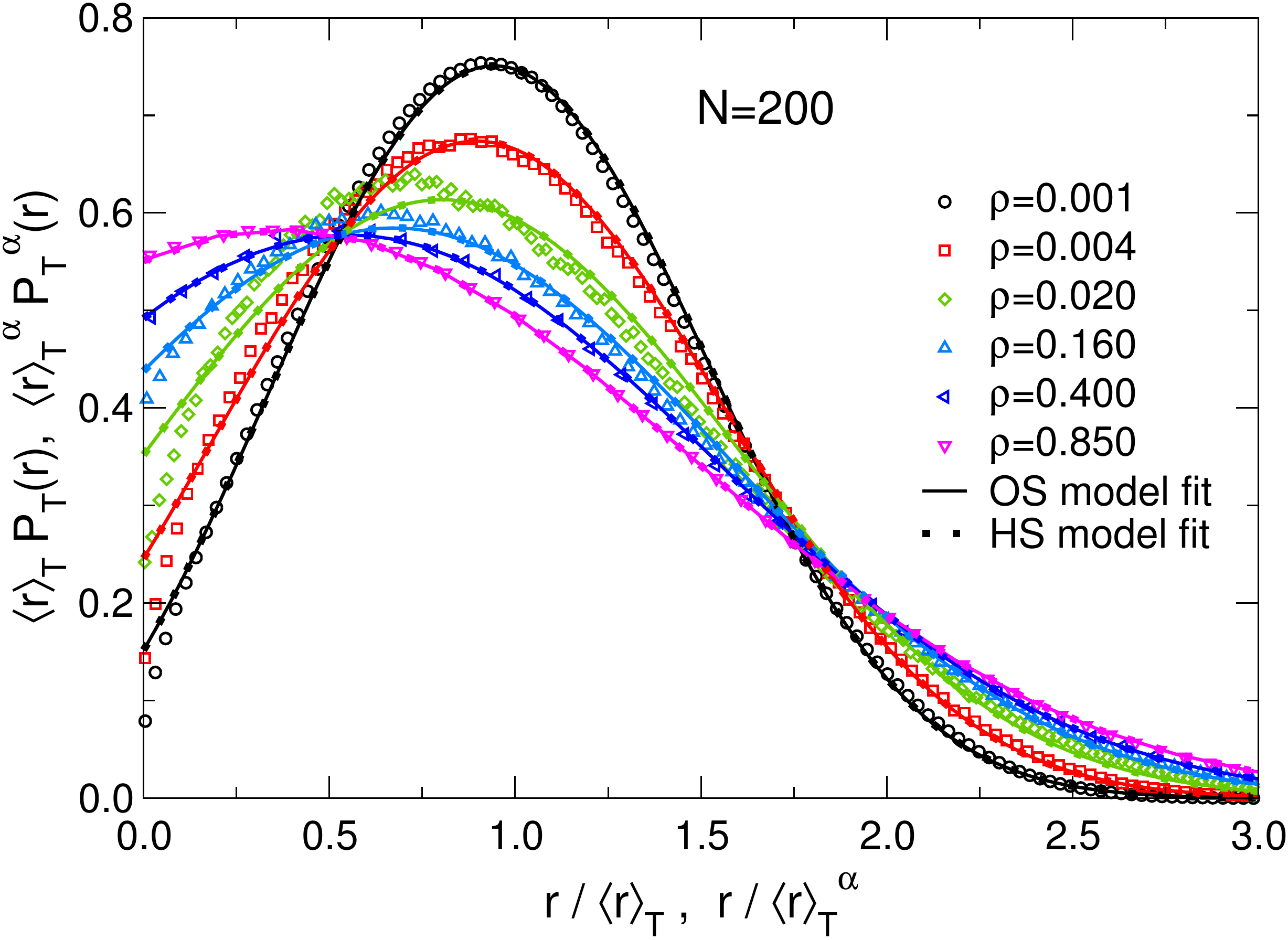}
\caption{Torquato's PDS, $P_T(r)$, for $N=200$ and at different densities (open symbols), compared with the fit results $P_T^{\alpha}(r)$ (with $\alpha=$ OS or HS) employing the HS (Eq.~\eqref{psd_hs}, points) and OS (Eq.~\eqref{psd_os}, lines) models. The HS and OS fits superimpose almost perfectly.}
\label{psd_fit_cfr}
\end{figure}

In Fig.~\ref{psd_fit_cfr} we compare for $N=200$ and different monomer densities the calculated PSD, with the result of the fits using the HS model, $P_T^\text{\tiny{HS}}(r;R_\text{\tiny{HS}},\rho_\text{\tiny{HS}})$, and with the OS model $P_T^\text{\tiny{OS}}(r;R_\text{\tiny{OS}},\rho_\text{\tiny{OS}})$. It is virtually impossible to distinguish $P_T^\text{\tiny{HS}}(r;R_\text{\tiny{HS}},\rho_\text{\tiny{HS}})$ from $P_T^\text{\tiny{OS}}(r;R_\text{\tiny{OS}},\rho_\text{\tiny{OS}})$ via a simple visual inspection, since the two curves superimpose almost perfectly. An analysis of the squared difference between the calculated and the fitted function (not shown) reveals however that the HS model fits the data slightly better for all densities, except $\rho=0.85,1.00$.

\begin{figure}
\centering
\includegraphics[width=0.45 \textwidth]{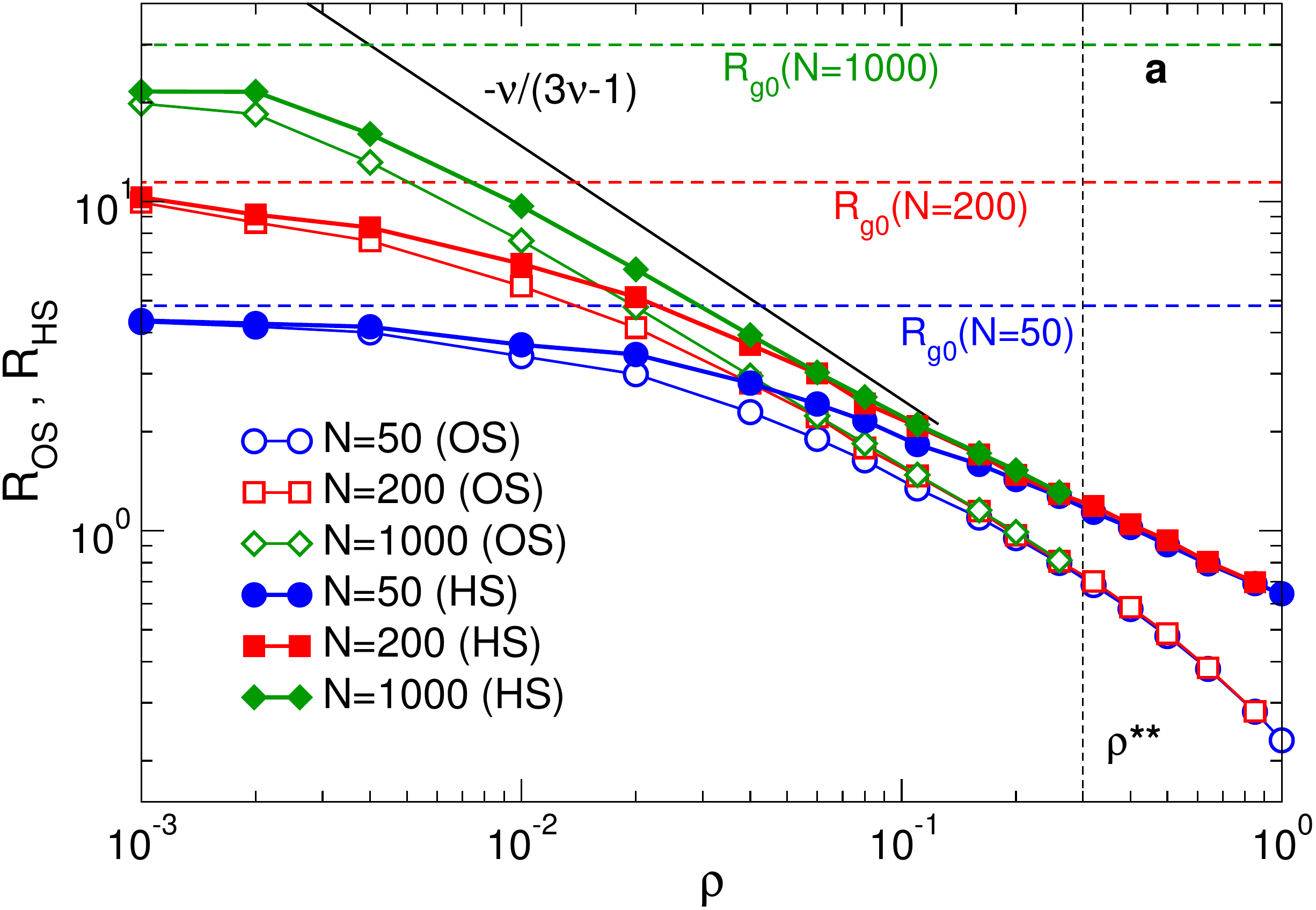}
\includegraphics[width=0.45 \textwidth]{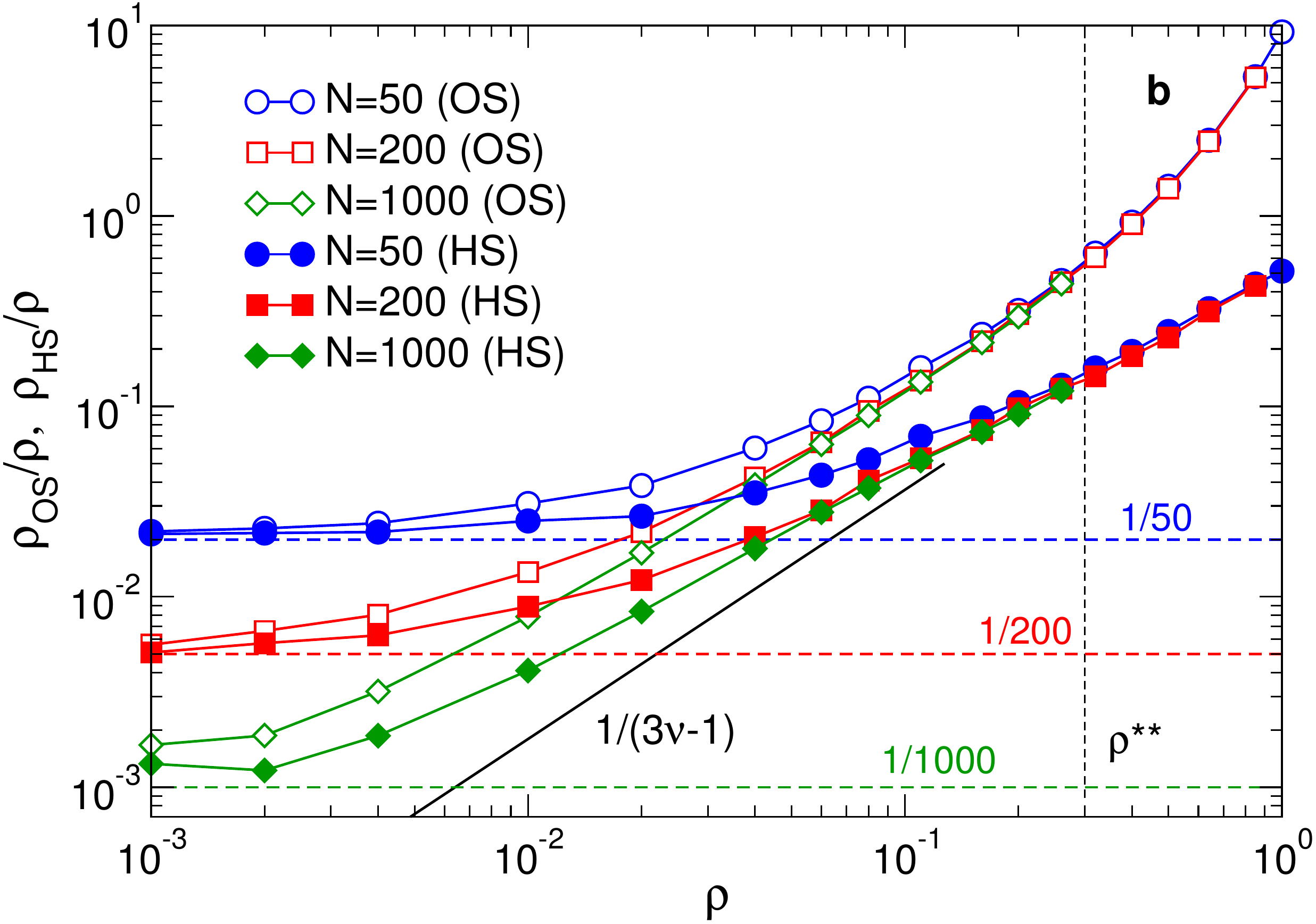}
\caption{Fit parameters as a function of monomer density $\rho$ for the fit of the PSD $P_T(r)$ with the HS (Eq.~\eqref{psd_hs}, filled symbols) and OS (Eq.~\eqref{psd_os}, open symbols) models. (a): Effective radius. (b): Effective density divided by $\rho$.}
\label{fit_params}
\end{figure}

In Figs.~\ref{fit_params}a and b we show, respectively, $R_\text{\tiny{HS}},R_\text{\tiny{OS}}$ and $\rho_\text{\tiny{HS}}/\rho,\rho_\text{\tiny{OS}}/\rho$ as function of the monomer density. Through an extrapolation of the low density behavior, we deduce that in the limit $\rho\to 0$, $R_\text{\tiny{HS}}, R_\text{\tiny{OS}} \to R_{g0}(N)$, and $\rho_\text{\tiny{HS}},\rho_\text{\tiny{OS}} \to \rho/N$ (dashed horizontal lines in Figs~\ref{fit_params}a and \ref{fit_params}b). In other words, $P_T(r)$ is described well by the PSD of a system of spheres with radius $R_{g0}$, and density equal to the chain density $\rho/N$. This is valid independently of the model used (OS or HS), since, as mentioned above, both models give the same result in the limit $\eta \to 0$, or equivalently $\rho \to 0, R=\text{const.}$ 

In the semidilute regime $\rho^*(N)<\rho<\rho^{**}$, both $R_\alpha$ and $\rho_\alpha$ ($\alpha=\text{OS,HS}$) tend towards a power-law behavior with increasing $N$. Since in the semidilute regime the only relevant length scale is the mesh size $\xi$, we make the hypothesis that for large $N$ these quantities will show the following scaling behavior:

\begin{equation}
\begin{split}
R_\alpha & \propto \xi \approx \rho^{-\nu/(3 \nu-1)}\\
\rho_\alpha & \propto \xi^{-3} \approx \rho^{3\nu/(3 \nu-1)}.\\
\end{split}
\ \ \ \ \ (\alpha=\text{OS,HS})
\label{fit_scaling}
\end{equation}

This \emph{ansatz} is motivated by the fact that the PSD is a purely geometrical object and hence has to scale with the intrinsic length scales of the system.

\begin{figure}
\centering
\includegraphics[width=0.45 \textwidth]{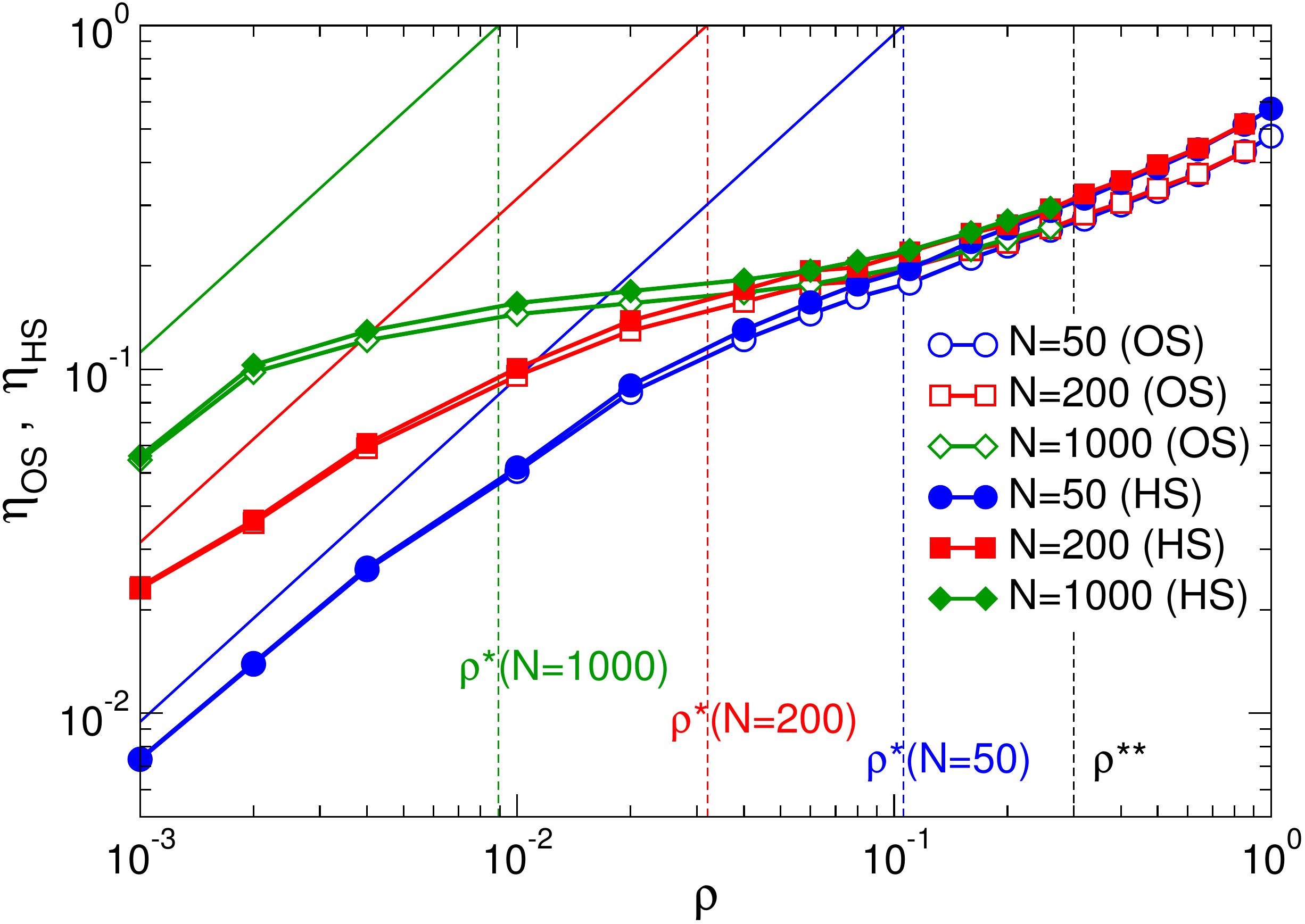}
\caption{Dimensionless densities $\eta_\text{\tiny{OS}}$ (open symbols) and $\eta_\text{\tiny{HS}}$ (filled symbols), Eq.~\eqref{eta_alpha}, calculated using the fit parameters from the OS and HS model fits, as a function of monomer density. Continuous lines: $\eta_c(0,N) = 4\pi R_{g0}^3 \rho / 3N$ (see Eq.~\eqref{eta_c}). The approximate values of the overlap densities $\rho^*(N)$ have been estimated using Eq.~\eqref{overlap_approx}.}
\label{eta_fit}
\end{figure}

\noindent In Fig.~\ref{eta_fit} we show the dimensionless density

\begin{equation}
\eta_\alpha \equiv 4 \pi R_\alpha^3 \rho_\alpha/3 \ \ \ \ (\alpha=\text{OS,HS})
\label{eta_alpha}
\end{equation}

\noindent as a function of $\rho$. We note that both $\eta_\text{\tiny{HS}}$ and $\eta_\text{\tiny{OS}}$ depend basically only on $(\rho,N)$, with $\eta_\text{\tiny{OS}}$ being slightly smaller than $\eta_\text{\tiny{HS}}$.  At low $\rho$, both $\eta_\text{\tiny{OS}}$ and $\eta_\text{\tiny{HS}}$ approach the asymptotic expression $\eta_c(0,N)=4\pi R_{g0}^3 \rho / 3N$  (continuous lines), where $\eta_c$ is in general defined as

\begin{equation}
\eta_c (\rho, N) \equiv 4\pi R_g^3 \rho / 3N, 
\label{eta_c}
\end{equation}

\noindent i.e., the dimensionless density the chains would have were they spheres of radius $R_g$. Importantly, we observe that when $N$ increases, $\eta_\text{\tiny{OS}}$ and $\eta_\text{\tiny{HS}}$  develop a plateau in the semidilute regime: This corroborates the hypothesis that for large $N$ Eq.~\eqref{fit_scaling} is valid, since this equation implies $\eta_\alpha\approx \text{const.}$

\begin{figure}
\centering
\includegraphics[width=0.45 \textwidth]{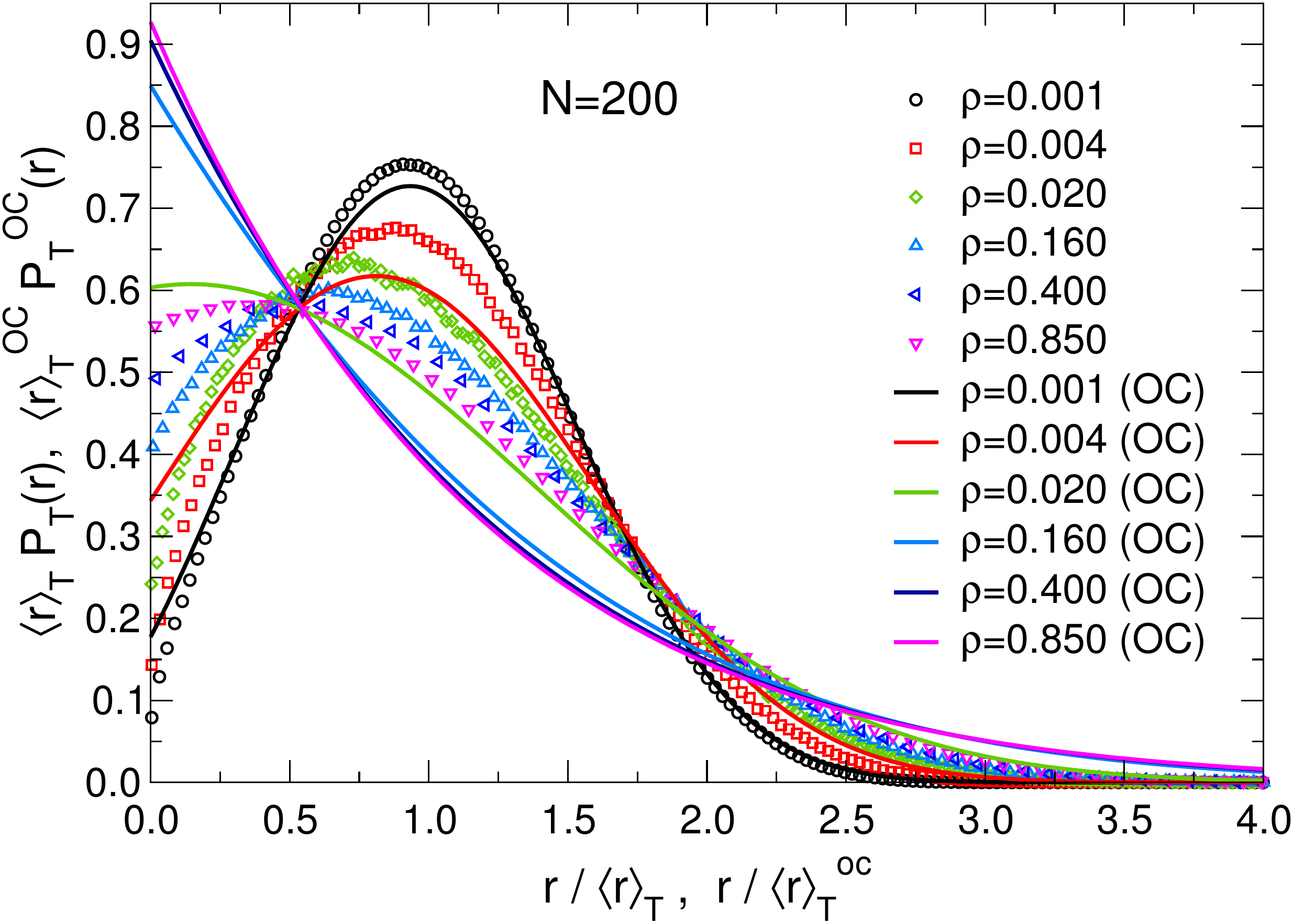}
\caption{Torquato's PSD for $N=200$ and at different densities (open symbols), compared with the OC model prediction, Eqs.~\eqref{psd_os}-\eqref{oc_model}.}
\label{psd_oc_cfr}
\end{figure}

We have seen in the preceding paragraph that at low $\rho$ the polymer solution can be mapped on a system of overlapping spheres of radius $R_g$ and density equal to the chain density, $\rho/N$. Inspired by this result, we introduce the \emph{overlapping chains} (OC) model, by assuming that the PSD of the system is approximated \emph{for all} $\rho$ by Eq.~\eqref{psd_os}, with

\begin{equation}
\begin{cases}
R_\text{\tiny{OS}} = R_g(\rho, N) &\\
\rho_\text{\tiny{OS}} =\rho/N & \\
\eta_\text{\tiny{OS}} = \eta_c. &\\
\end{cases}
\label{oc_model}
\end{equation}

\noindent This approximation fails at intermediate and high $\rho$, as it is clear from Fig.~\ref{fit_params}a-b. This is also illustrated in Fig.~\ref{psd_oc_cfr}, where the calculated PSD for $N=200$ and different densities is compared with the one predicted by the OC model and one sees that already for $\rho$ as low as $0.02$ the two PSDs are qualitatively very different. However, given the simplicity of the OC model, we decide nevertheless to compare the computed average pore size, $\langle r \rangle_T$, with the one that the model predicts.

From the PSD \eqref{psd_os} one can compute the average pore size with the result

\begin{equation}
\langle r \rangle_T^\text{\tiny{OS}} = \frac {R} 3 \exp(\eta) E_{2/3}(\eta),
\label{rav_os}
\end{equation}

\noindent where $E_n(\eta)$ is an exponential integral \cite{abramowitz1965handbook},

\begin{equation}
E_{n}(\eta) = \int_1^\infty \frac{e^{-\eta x}}{x^{n}} dx.
\end{equation}

\noindent The OC model result is obtained simply by applying Eqs.~\eqref{oc_model} to Eq.~\eqref{rav_os}, i.e.,

\begin{equation}
\langle r \rangle_T^\text{\tiny{OC}} = \frac {R_g(\rho,N)} 3 \exp(\eta_c) E_{2/3}(\eta_c). 
\label{rav_oc}
\end{equation}

\begin{figure}
\centering
\includegraphics[width=0.45 \textwidth]{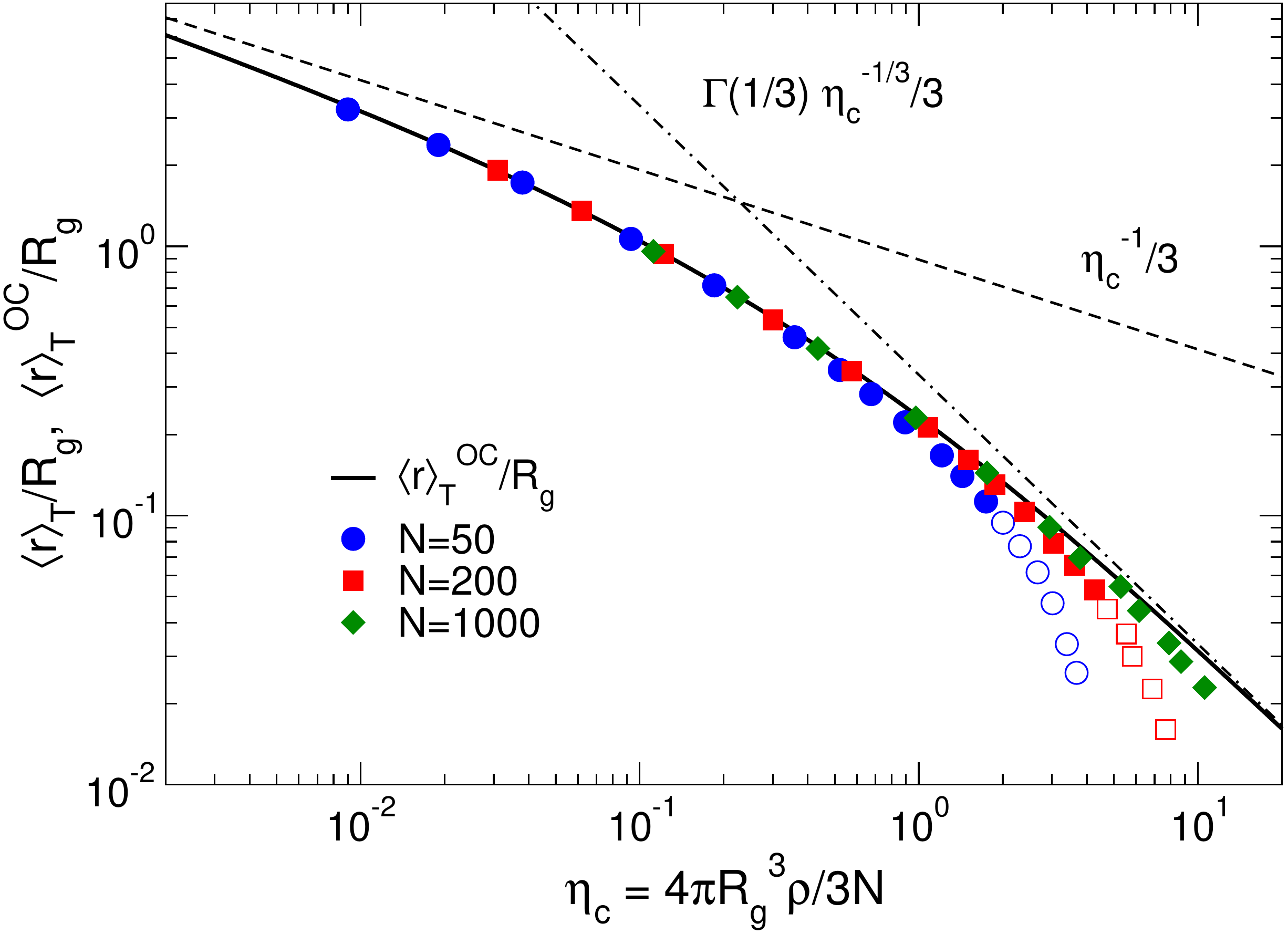}
\caption{Reduced average pore size $\langle r \rangle_T/R_g$ (symbols) compared with the OC model prediction $\langle r \rangle_T^\text{\tiny{OC}}/R_g$ (thick line), Eq.~\eqref{oc_model}, as a function of $\eta_c$. Open symbols correspond to densities $\rho>0.3$ (concentrated regime). Dashed (dash-dotted) line: Low (high) density behavior predicted by the OC model, Eq.~\eqref{expansion}.}
\label{rav_cfr_os}
\end{figure}

In Fig.~\ref{rav_cfr_os}, we compare the measured average pore size $\langle r\rangle_T$ with the OC model result $\langle r \rangle_T^\text{\tiny{OC}}$. One observes that, except for the points corresponding to the concentrated regime ($\rho>0.3$), the prediction of the OC model is in surprisingly good agreement with the data, without using any fit parameter. This is a striking result, since we have seen that the calculated PSD starts to diverge from the theoretical prediction already at relatively low density (Fig.~\ref{psd_oc_cfr}). 

Another related surprising result is the following. The scaling behavior of $R_g$ is \cite{degennes1979scaling,rubinstein2003polymer,teraoka2002polymer}

\begin{equation}
R_g\approx
\begin{cases}
R_{g0} & \chi<\chi^*\\
R_{g0} \chi^{-\nu+1/2}&  \chi^*<\chi<\chi^{**}.\\
\end{cases}
\label{rg_scaling}
\end{equation}

\noindent This implies, given the definition of $\eta_c$, that

\begin{equation}
\eta_c\approx
\begin{cases}
\chi^{3\nu-1} & \chi<\chi^*\\
\chi^{1/2}&  \chi^*<\chi<\chi^{**}.\\
\end{cases}
\label{eta_scaling}
\end{equation}

\noindent The function appearing on the right hand side of  Eq.~\eqref{rav_oc} can be expanded for $\eta_c \to 0$ and $\eta_c \to \infty$ via a Puiseux series \cite{davenport1993computer}; the leading terms are

\begin{equation}
\exp(\eta_c) E_{2/3}(\eta_c)=
\begin{cases}
\Gamma\left(\frac 13\right)\eta_c^{-1/3} + O(1) &\eta_c \to 0\\
\eta_c^{-1} + O(\eta_c^{-2}) &\eta_c \to \infty,
\end{cases}
\label{expansion}
\end{equation}

\noindent where $\Gamma$ is the Gamma function, and therefore

\begin{equation}
\frac{\langle r \rangle_T^\text{\tiny{OC}}}{R_g} \approx
\begin{cases}
\eta_c^{-1/3}  &\eta_c \to 0\\
\eta_c^{-1} &\eta_c \to \infty.
\end{cases}
\label{rav_eta}
\end{equation}

\noindent From Eqs~\eqref{rg_scaling}, \eqref{eta_scaling}, \eqref{rav_eta}, along with the fact that $\lim_{N\to \infty} \chi^{**}(N) = \infty$, we obtain for infinitely long chains

\begin{equation}
\frac{\langle r \rangle_T^\text{\tiny{OC}}}{R_{g0}} = \frac{\langle r \rangle_T^\text{\tiny{OC}}}{R_g} \frac {R_g}{R_{g0}}  \approx
\begin{cases}
\chi^{-\nu+1/3} & \chi<\chi^*\\
\chi^{-\nu}&  \chi^*<\chi.\\
\end{cases}
\label{rav_oc_chi}
\end{equation}

\noindent We thus find that for very long chains, ${\langle r \rangle_T^\text{\tiny{OC}}}/{R_{g0}}$ behaves exactly as $\xi/R_{g0}$, Eq.~\eqref{xi_scaling}, for all values of $\chi$. We stress, however, that while  $\lim_{N\to \infty} \chi^{**}(N) = \infty$, $\rho^{**}$ is independent of $N$: Therefore, when considering functions of density instead of functions of $\chi$, we must be aware that the approximation $\langle r\rangle_T \simeq \langle r\rangle_T^\text{\tiny{OC}}$ will always break down for $\rho \geq \rho^{**}=0.3$.

It is not evident whether there is a deep reason behind the fact that $\langle r\rangle_T$ and $\langle r \rangle_T^\text{\tiny{OC}}$ are so similar even if $P_T(r)$ and $P_T^\text{\tiny{OC}}(r)$ are qualitatively very different. Nonetheless, this remarkable fact allows to estimate $\langle r\rangle_T$ for $\rho<\rho^{**}$ with good accuracy by knowing only $R_g$ and $\rho$. Since we have seen that $\langle r\rangle_G \simeq 2 \langle r \rangle_T$ (Sec.~\ref{sec:estimating_xi}), this means that we are also able to estimate with reasonable accuracy the average value of the PSD of Gubbins. This is a useful result for estimating $\xi$ in real systems, since measuring the PSD experimentally is not an easy task \cite{rintoul1996structure}.

\section{Summary and conclusions} \label{sec:summary}

The geometrical mesh size $\xi$ is a key quantity in polymer solutions and networks, since it governs the diffusion of particles or molecules in these systems. Conventional methods to measure the mesh size rely on scaling estimates or on measurements of the monomer density fluctuation correlation length $\xi_c$. Scaling estimates, however, only give $\xi$ up to an unknown multiplicative factor, and identifying $\xi_c$ with $\xi$ works only in the semidilute regime, whereas it gives nonsensical results in dense systems.

We propose a method to directly probe the geometrical mesh size $\xi$, which we identify with the average size of the pores of the polymer solution, i.e., the space filled by the solvent. To obtain the distribution of the pore sizes, we make use of the definitions of pore size distribution (PSD) formalized by Torquato \cite{torquato2013random} and by Gubbins \cite{gelb1999pore}. We find that in both cases, the average of the PSD, $\langle r \rangle$, follows the expected scaling behavior for $\xi$ with high accuracy, validating the hypothesis that $\langle r \rangle$ can be identified with the geometrical mesh size. An important feature of our approach is that it naturally gives access to the {\it distribution} of mesh (pore) sizes in a polymer medium: This aspect can be quite relevant, since we know from experiments on polymer networks that the degree of heterogeneity of the medium can strongly influence the diffusion of particles \cite{parrish2017network}.

The range of applicability of the PSD method to evaluate the mesh size is very broad, and by no means limited to polymer solutions. Moreover, it can also be extended to polymer networks and gels. We point out, however, that in gels there exist another relevant  ``mesh size'', which is related to the average distance between neighboring crosslinks and is in general quite different from $\xi$ \cite{tsuji2018evaluation}. If, for example, a polymer melt is crosslinked, one expects the structural properties, and therefore $\xi$, to remain basically unchanged with respect to the un-crosslinked system \cite{duering1994structure}. However, the diffusivity of a spherical probe particle will change significantly if the particle diameter is comparable to the average distance between the crosslinks. The same is of course true for entangled polymer melts, where $\xi\simeq 0$ and the relevant parameter controlling the diffusion of particles and the mechanical properties is nothing else than the diameter of the reptation tube \cite{doi1988theory}, since on timescales shorter than the entanglement relaxation time the entanglements effectively act as crosslinks  \cite{dell2013theory,cai2015hopping}. 

Another factor which should be taken into account is that the PSD does not depend on the flexibility of the polymers, although in reality this can have a very strong influence on the diffusion of nanoparticles in polymeric systems. For example, it has recently been shown in simulations that increasing the chain rigidity will decrease the diffusivity of nanoparticles in polymer solutions \cite{chen2019influence}. This rigidity will, however, also affect estimates based on scaling predictions and on measurements of the correlation length, which in addition only give access to an average mesh size, and not to the whole distribution.

To summarize, the method which we have proposed to measure the mesh size should represent a substantial improvement over conventional methods such as scaling estimates and methods which identify the mesh size with the correlation length. It will be interesting, in further studies, to address the problem of characterizing the relevant mesh size in networks and gels and understanding the role played by polymer flexibility.

\section*{Appendix} 

\setcounter{section}{0}
\renewcommand{\thesection}{A\Roman{section}}

\section{Radius of gyration and bond-bond correlation function} \label{sec:gyration}

\begin{figure}
\centering
\includegraphics[width=0.45 \textwidth]{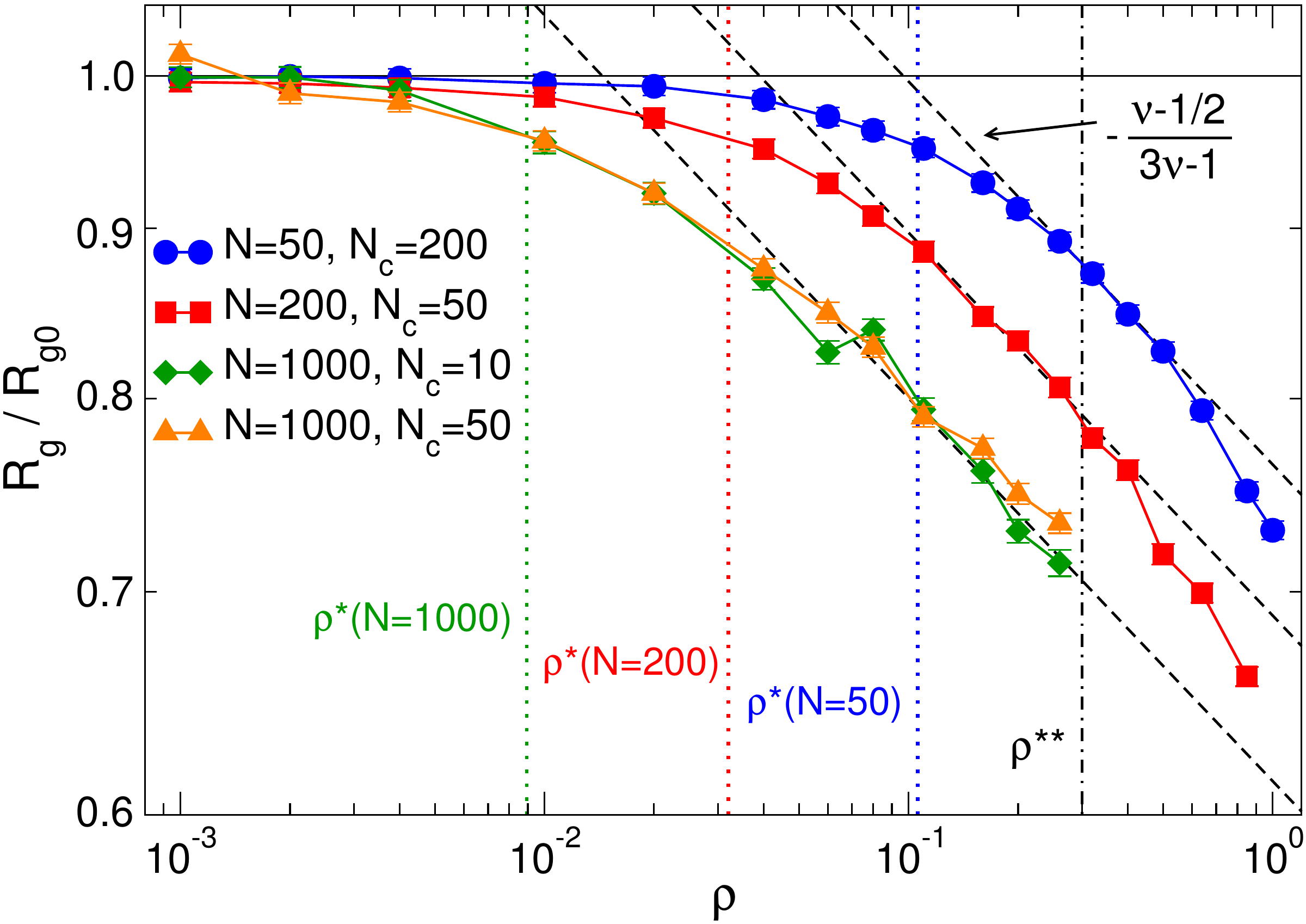}
\caption{Normalized polymer radius of gyration, Eq.~\eqref{rg_def}, as a function of monomer density $\rho$. The dashed lines have slope $-\frac{\nu-1/2}{3\nu-1} \simeq -0.115$ ($\nu \simeq 0.588$). The overlap densities have been estimated using Eq.~\eqref{overlap_approx}.}
\label{gyr_norm}
\end{figure}

As a preliminary analysis of the structural properties of the simulated systems, we have studied the behavior of the radius of gyration $R_g$ as a function of monomer density $\rho$. The radius of gyration is defined as

\begin{equation}
R_g^2 \equiv \frac 1 N \sum_{i=1}^{N} \langle (\mathbf r_i - \mathbf r_\text{cm})^2\rangle,
\label{rg_def}
\end{equation}

\noindent where $\mathbf r_\text{cm}$ is the position of the center of mass of the  chain and $\langle \cdot \rangle$ represents the thermodynamic average. Scaling theory predicts the dependence of $R_g$ on $\rho$ to be  \cite{degennes1979scaling,rubinstein2003polymer,teraoka2002polymer} 

\begin{equation}
\frac{R_g}{R_{g0}}\approx
\begin{cases}
1 & \rho < \rho^*\\
(\rho/{\rho^*})^{-\frac{\nu-1/2}{3\nu-1}} &  \rho^* < \rho < \rho^{**}\\
N^{1/2-\nu} & \rho>\rho^{**},\\
\end{cases}
\label{rg_density}
\end{equation}

\noindent where $R_{g0}\equiv \lim_{\rho \to 0} R_g(\rho)$ and $\rho^*$ is the overlap density. 

In Fig.~\ref{gyr_norm}, we report $R_g/R_{g0}$ as a function of $\rho$, with the dashed lines representing power laws with exponent ${-\frac{\nu-1/2}{3\nu-1}}\simeq -0.115$. The radius of gyration in the dilute limit, $R_{g0}$, was computed by simulating a single polymer chain at very low density ($\rho=10^{-3}$ for $N=50,200$ and $\rho=10^{-4}$ for $N=1000$): The results are reported in Tab.~\ref{tab:properties}, alongside with the resulting estimate for the overlap density, given by Eq.~\eqref{overlap_approx}. Figure~\ref{gyr_norm} shows that $R_g/R_{g0}$ follows the scaling predictions for all values of $N$ up to the density $\rho\simeq 0.3$. For $\rho \gtrsim 0.3$, the decrease is steeper than what is predicted by scaling theory. This is due to the fact that in this density range the persistence length is density-dependent, an effect which scaling theory does not take into account. 

The $\rho$-dependence of the persistence length can be understood by studying the bond-bond correlation function $\langle \cos(\theta_s)\rangle$, where $\cos(\theta_s)$ is defined as \cite{wittmer2004long}

\begin{equation}
\cos(\theta_s) \equiv  \frac{\mathbf b_n \cdot \mathbf b_{n+s}}{|\mathbf b_n| \ |\mathbf b_{n+s}|} \ \ (s=1,\dots, N-n),
\label{bond_corr_def}
\end{equation}

\noindent where $\mathbf b_n \equiv\mathbf r_{n+1} - \mathbf r_n$ is the $n$-th bond vector, with $\mathbf r_n$ the monomer's position vector. For very long chains, $\langle \cos(\theta_s)\rangle$ is expected to have the following behavior \cite{schafer2004calculation,wittmer2004long,hsu2010standard}:

\begin{equation}
\langle \cos(\theta_s)\rangle \propto 
\begin{cases}
s^{-2(1-\nu)} & l_p/l_b < s \ll N_b(\rho)\\
s^{-3/2} & N_b(\rho) \ll s \ll N,\\
\end{cases}
\label{bond_corr_theo}
\end{equation}

\noindent where $l_p$ is the persistence length \cite{rubinstein2003polymer}, $l_b = \sqrt{\langle \mathbf b_n \rangle}$ is the RMS bond length and $N_b$ is the number of monomers in a blob (see Eq.~\eqref{blob1}). For the fully flexible Kremer-Grest model considered here, $l_p \approx l_b$ \cite{hsu2016static}. Since in the dilute regime $N_b \approx N$, we expect in this regime $\langle \cos(\theta_s)\rangle \propto s^{-2(1-\nu)} \ \forall s > l_p/l_b \approx 1$. Analogously, since in the concentrated regime $N_b \approx 1$, we expect $\langle \cos(\theta_s)\rangle \propto s^{-3/2} \ \forall s \ll N$.

\begin{figure}
\centering
\includegraphics[width=0.45 \textwidth]{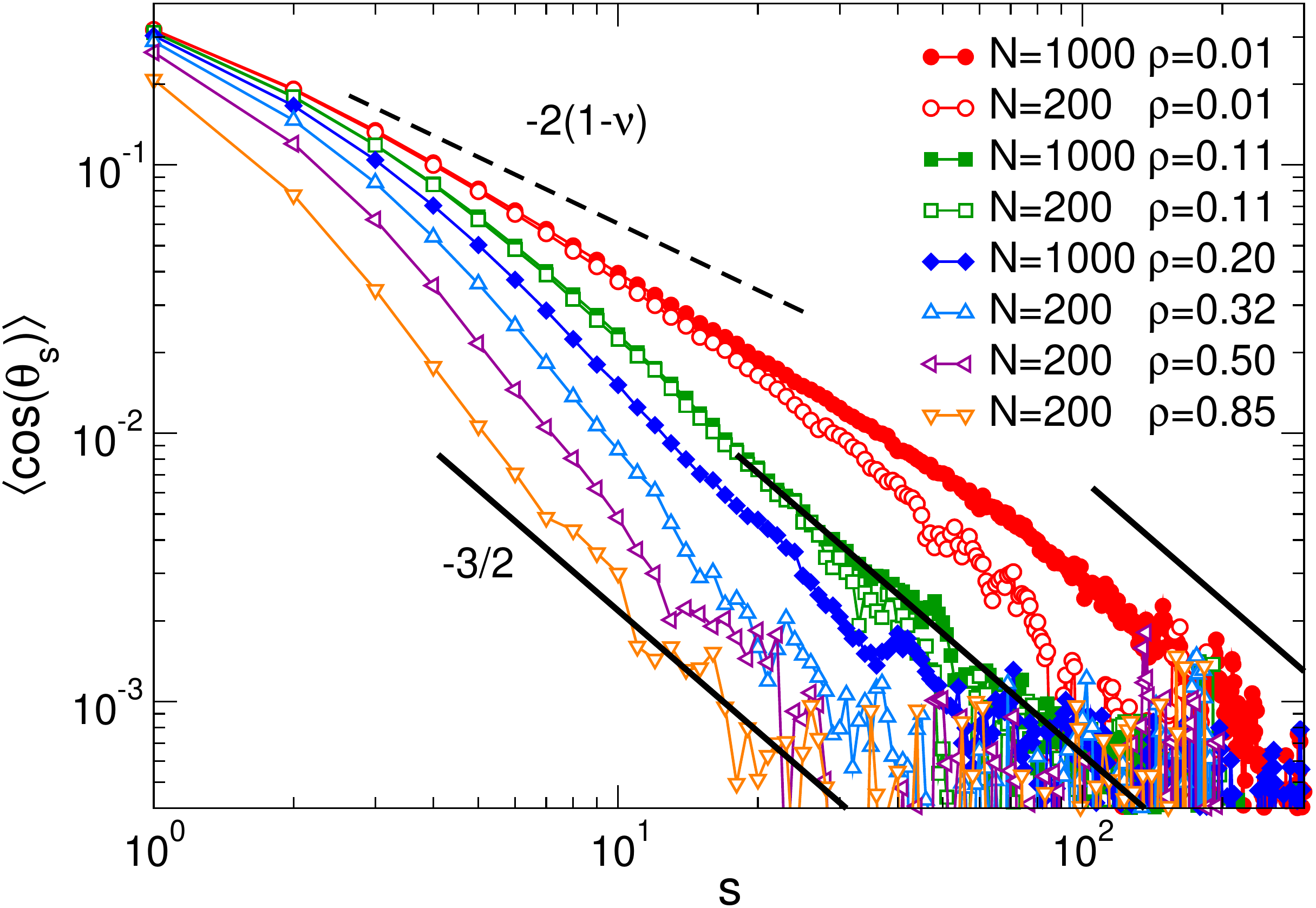}
\caption{Bond-bond correlation function $\langle \cos(\theta_s)\rangle$, Eq.~\eqref{bond_corr_def}, for $N=200$ and $1000$ and for different values of $\rho$. The dashed line indicates the predicted theoretical behavior for small $s$, whereas the continuous lines indicate the predicted high-$s$ behavior (see Eq.~\eqref{bond_corr_theo}). We note that $2(1-\nu)\simeq 0.824$.}
\label{bond_corr}
\end{figure}

In Fig.~\ref{bond_corr} we report $\langle \cos(\theta_s)\rangle$ for $N=200$ and $1000$ and different densities (for $N=50$ we find similar results). The chains considered here are too short to clearly observe the small-$s$ regime $s^{-2(1-\nu)}$ (dashed line). On the other hand, the large-$s$ behavior is compatible with the theoretical prediction $\langle \cos(\theta_s)\rangle \propto s^{-3/2}$, as illustrated by the continuous lines. Although it is difficult to precisely quantify the persistence length outside of the melt regime for highly flexible and rather short chains as those considered here \cite{hsu2010standard}, the marked reduction of $\langle \cos(\theta_s)\rangle$ as a function of $\rho$ observed for small value of $s$ is compatible with a reduction of the persistence length $l_p$. A reduction of $l_p$ with increasing density can also be inferred by considering the distribution of the bond angle $\theta = \pi-\theta_1$ (Sec.~\ref{bond_angle} in the S.I.) These observations explain qualitatively the high-$\rho$ behavior of $R_g$ observed in Fig.~\ref{gyr_norm}. 

\begin{table}
\centering
\caption{$N$-dependent properties of the simulated systems: Radius of gyration in the dilute limit, $R_{g0}$ and approximate overlap concentration (Eq.~\eqref{overlap_approx}).}
\label{tab:properties}
\begin{tabular}{@{\hspace{1em}} c @{\hspace{1em}} c @{\hspace{1em}} c @{\hspace{1em}}}
\toprule
$N$ & $R_{g0}$ & $3N/4\pi R_{g0}^3$  \\
\colrule
 $50$ & $4.83\pm0.02$ & $1.1 \cdot 10^{-1}$\\
$200$ & $11.44 \pm 0.07$ & $3.2 \cdot 10^{-2}$\\
$1000$ & $29.9 \pm 0.2$ & $8.9 \cdot 10^{-3}$\\
\botrule
\end{tabular}
\end{table}

\section{Calculation of the pore size distribution} \label{sec:app1}

\subsection{Torquato's PSD}

The algorithm to compute Torquato's PSD is described in Ref.~\citenum{torquato2013random}.
First of all, as mentioned in Sec.~\ref{sec:psd}, we have to divide the sample in a ``pore'' region and a ``solid'' region. In order to do so, we approximate the monomers as hard spheres of diameter $\sigma$. The procedure to calculate $P_T(r)$ is then as follows:

\begin{enumerate}
\item{A random point $\mathbf r$ is chosen in the pore phase.}
\item{The smallest distance $r_\text{min}$ between $\mathbf r$ and the center of a monomer is calculated. The distance between $\mathbf r$ and the nearest pore-solid interface is calculated as $r=r_\text{min}-\sigma/2$.}
\end{enumerate}

This procedure is repeated many times, until a large number $\mathcal N$ of $r$ values is recorded.  In the limit of large $\mathcal N$, the normalized histogram of these values is equivalent to $P_T(r)$.

\subsection{Gubbins's PSD}

In order to calculate $P_G(r)$, we have used the algorithm proposed by Bhattacharya and Gubbins in Ref.~\citenum{bhattacharya2006fast}. This algorithm is based on the observation that the problem of finding the largest sphere containing $\mathbf r$ and which does not overlap with any monomer can be reformulated as the problem of maximizing the function

\begin{equation}
r(\mathbf r_c) \equiv \min_{i=1,\dots M} \{s_i\} - \sigma/2,
\label{obj_function}
\end{equation}

\noindent subject to the constraint 

\begin{equation}
|\mathbf r_c-\mathbf r| - r(\mathbf r_c) \leq 0,
\end{equation}

\noindent where $\mathbf r_c$ is the position of the sphere's center and $s_i$ is the distance between $\mathbf r_c$ and the centers $\mathbf r_i$ of the monomers: $s_i=|\mathbf r_c - \mathbf r_i|$. If the maximization of the function \eqref{obj_function} is carried out for a large enough number of points $\mathbf r$, the resulting (normalized) histogram of $r$ values will converge to $P_G(r)$.

The problem of calculating $P_G(r)$ reduces therefore to a nonlinear optimization problem, which can be solved with a standard algorithm \cite{press2007numerical}. We have used the open-source Sbplx algorithm of the NLopt library \cite{nlopt}, which is a re-implementation of Subplex \cite{rowan1990functional}. Other choices are possible, although one has to make sure that the chosen algorithm can handle discontinuous objective functions \cite{bhattacharya2006fast}.

\acknowledgements

We thank J. Baschnagel and K. Schweizer for useful discussions, and L. Rovigatti for implementing the code which was used to calculate $P_G$. This work has been supported by LabEx NUMEV (ANR-10-LABX-20) funded by the ‘‘Investissements d'Avenir’’ French Government program, managed by the French National Research Agency (ANR).

\section*{SUPPLEMENTARY INFORMATION}

\setcounter{equation}{0}
\setcounter{figure}{0}
\setcounter{table}{0}
\setcounter{section}{0}

\renewcommand{\theequation}{S\arabic{equation}}
\renewcommand{\thefigure}{S\arabic{figure}}
\renewcommand{\thetable}{S\arabic{table}}
\renewcommand{\thesection}{S\Roman{section}}

\section{Mean-squared displacement} \label{sec:msd}

\begin{figure}
\centering
\includegraphics[width=0.45 \textwidth]{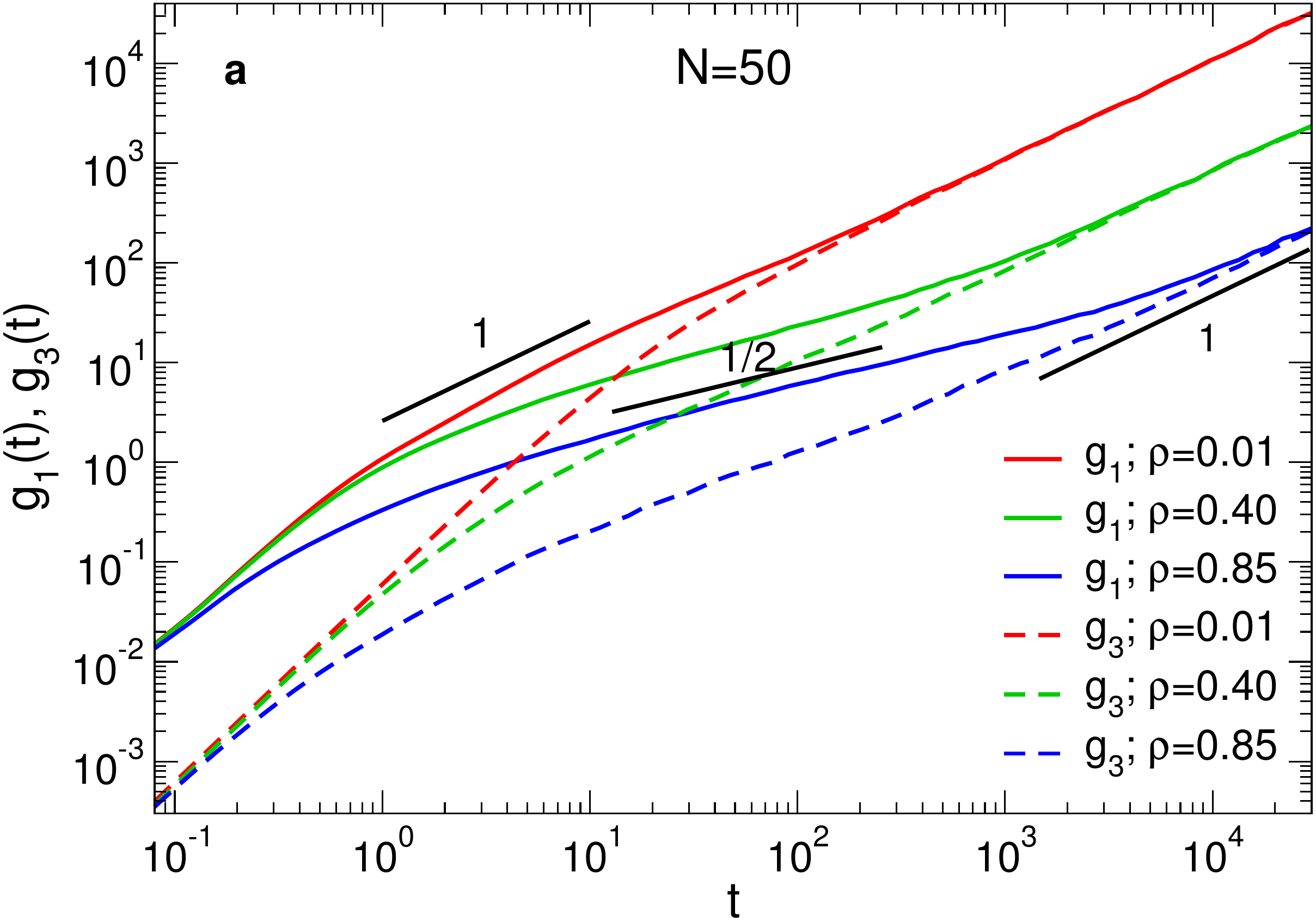}
\includegraphics[width=0.45 \textwidth]{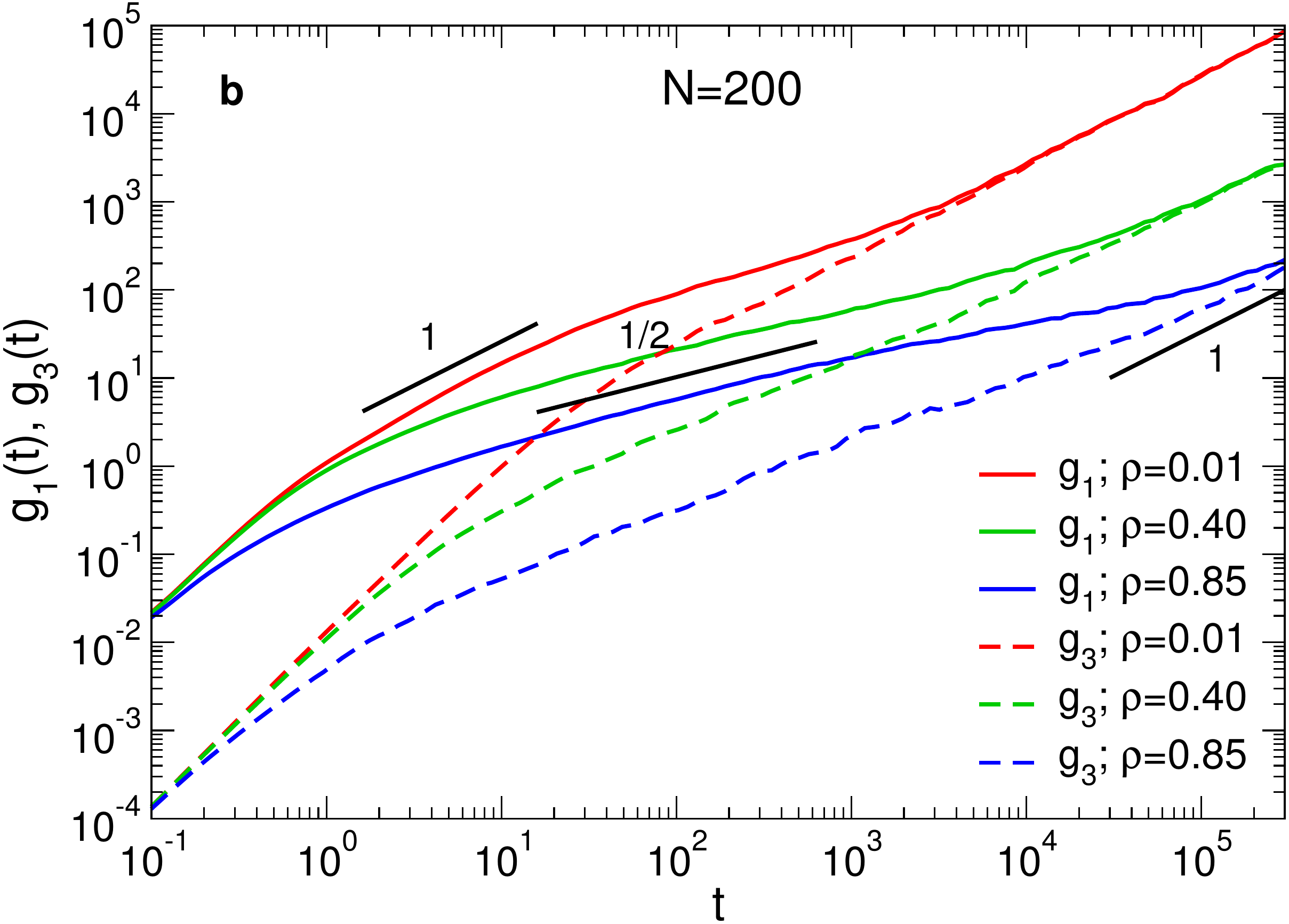}
\includegraphics[width=0.45 \textwidth]{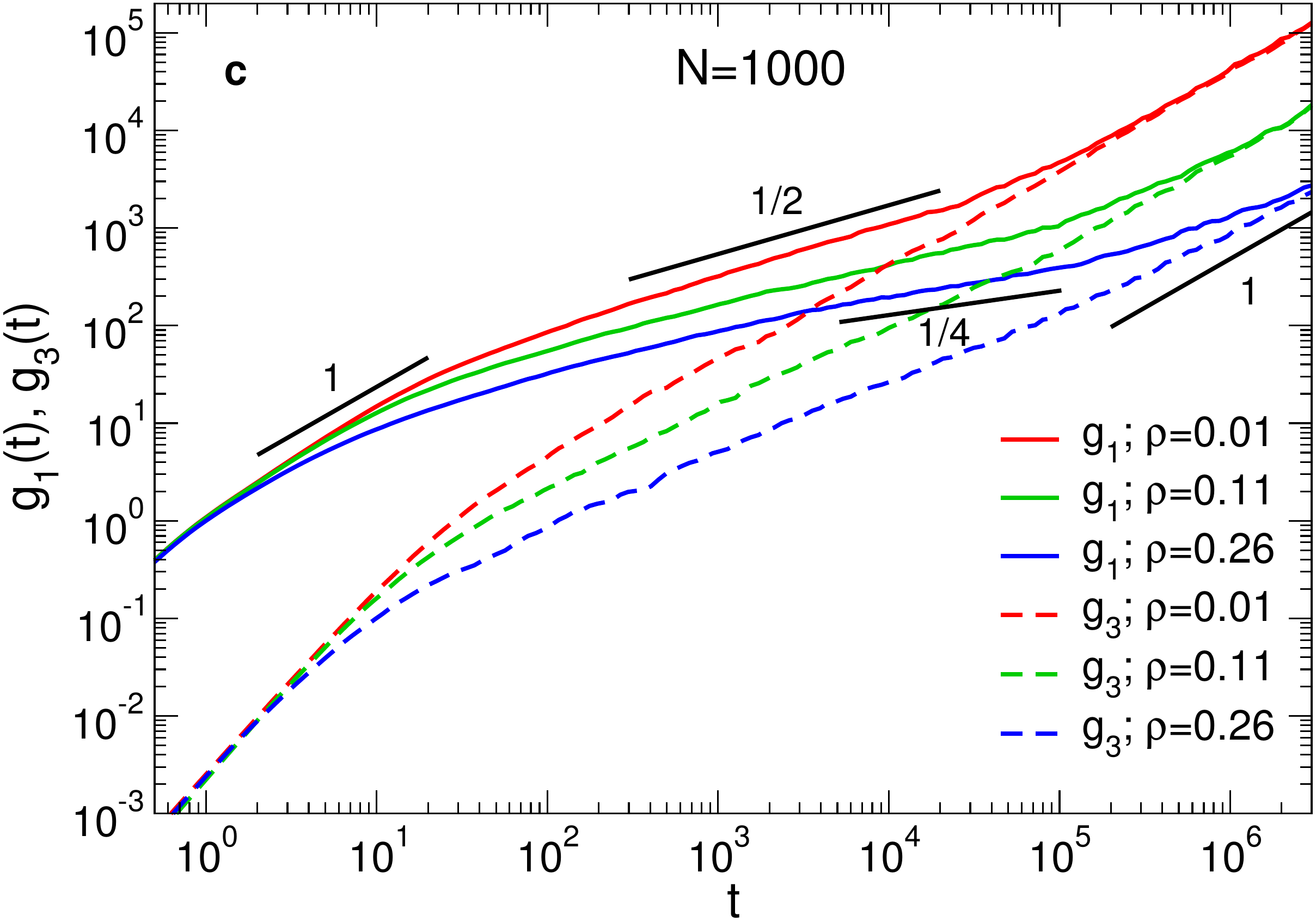}
\caption{MSD of the monomers, $g_1$ (Eq.~\eqref{msd_mono}), and of the centers of mass of the chains, $g_3$ (Eq.~\eqref{msd_chain}), for chain lengths $N=50$ (a), $N=200$ (b) and $N=1000$ (c) and different monomer densities $\rho$. Slopes are reported for comparison with the Rouse model and the reptation model \cite{rubinstein2003polymer,doi1988theory}.}
\label{all_msd}
\end{figure}

In order to check that the time we used to equilibrate the system is sufficiently long, we consider two quantities: the mean-squared displacement (MSD) of the monomers, which we compute as \cite{hsu2016static}

\begin{equation}
g_1(t) \equiv \frac 1 { (N/2+1)}   \sum_{i=N/4}^{3N/4}   \langle [\mathbf r_{i}(t) - \mathbf r_{i}(0)]^2\rangle ,
\label{msd_mono}
\end{equation}

\noindent and the MSD of the centers of mass of the chains,

\begin{equation}
g_3(t) \equiv \langle [\mathbf r_{\text{cm}}(t) - \mathbf  r_{\text{cm}}(0)]^2\rangle,
\label{msd_chain}
\end{equation} 

\noindent where $\mathbf  r_{\text{cm}}(t)$ is the position vector of the center of mass of the chain and $\langle \cdot \rangle$ denotes, as usual, the thermodynamic average. Note that in Eq.~\eqref{msd_mono}, only the central half of the monomers belonging to each chain are considered, in order to suppress the fluctuations caused by chain ends \cite{hsu2016static}.

In the present work, the motion of every monomer is governed by the Langevin equation, and hydrodynamic interactions between monomers are neglected. Therefore, in the absence of entanglements the Rouse model \cite{doi1988theory,rubinstein2003polymer} gives a good approximation of chain dynamics at all densities, whereas in the presence of entanglements, the dynamics is described by the reptation model \cite{doi1988theory,rubinstein2003polymer}. In both cases, when $t>t_\text{rel}$, with $t_\text{rel}$ the longest relaxation time of the system, we expect $g_1(t) = g_3(t) \propto t$  \cite{doi1988theory,hsu2016static}: Therefore, we can check that the system has equilibrated by verifying that this condition is met at long times.

In Fig.~\ref{all_msd}, we show $g_1(t)$ and $g_3(t)$ for different values of $N$ and $\rho$. For the fully-flexible Kremer-Grest model employed here, the entanglement length $N_e$ is  $\simeq 85$ at $\rho=0.85$ ($T=1.0$) \cite{hoy2009topological}, and it decreases with decreasing $\rho$ \cite{fetters1994connection,fetters1999packing,kroger2000rheological}. Therefore, for $N=50$ (Fig.~\ref{all_msd}a) the system is unentangled, for $N=200$ (Fig.~\ref{all_msd}b) it is lightly entangled, and for $N=1000$ (Fig.~\ref{all_msd}c) it is entangled. This is confirmed by the fact that for $N=50$ the behavior of $g_1$ and $g_3$ is in qualitative agreement with the predictions of the Rouse model, whereas for $N=200$ and $1000$ it follows approximately the reptation model  \cite{doi1988theory,rubinstein2003polymer}. This is also evidenced by the comparison with the slopes reported in Fig.~\ref{all_msd}.

For all values of $N$ and $\rho$ considered, we observe that for large times, $g_1(t)\simeq g_3(t)$, and thus we conclude that the equilibration time $t_e$ (see Tab.~\ref{tab:details} in the main text) is larger than the longest relaxation time of the system, $t_\text{rel}$, and therefore all the systems are well equilibrated.

\section{Chain form factor} \label{sec:form_factor}

\begin{figure}
\centering
\includegraphics[width=0.45 \textwidth]{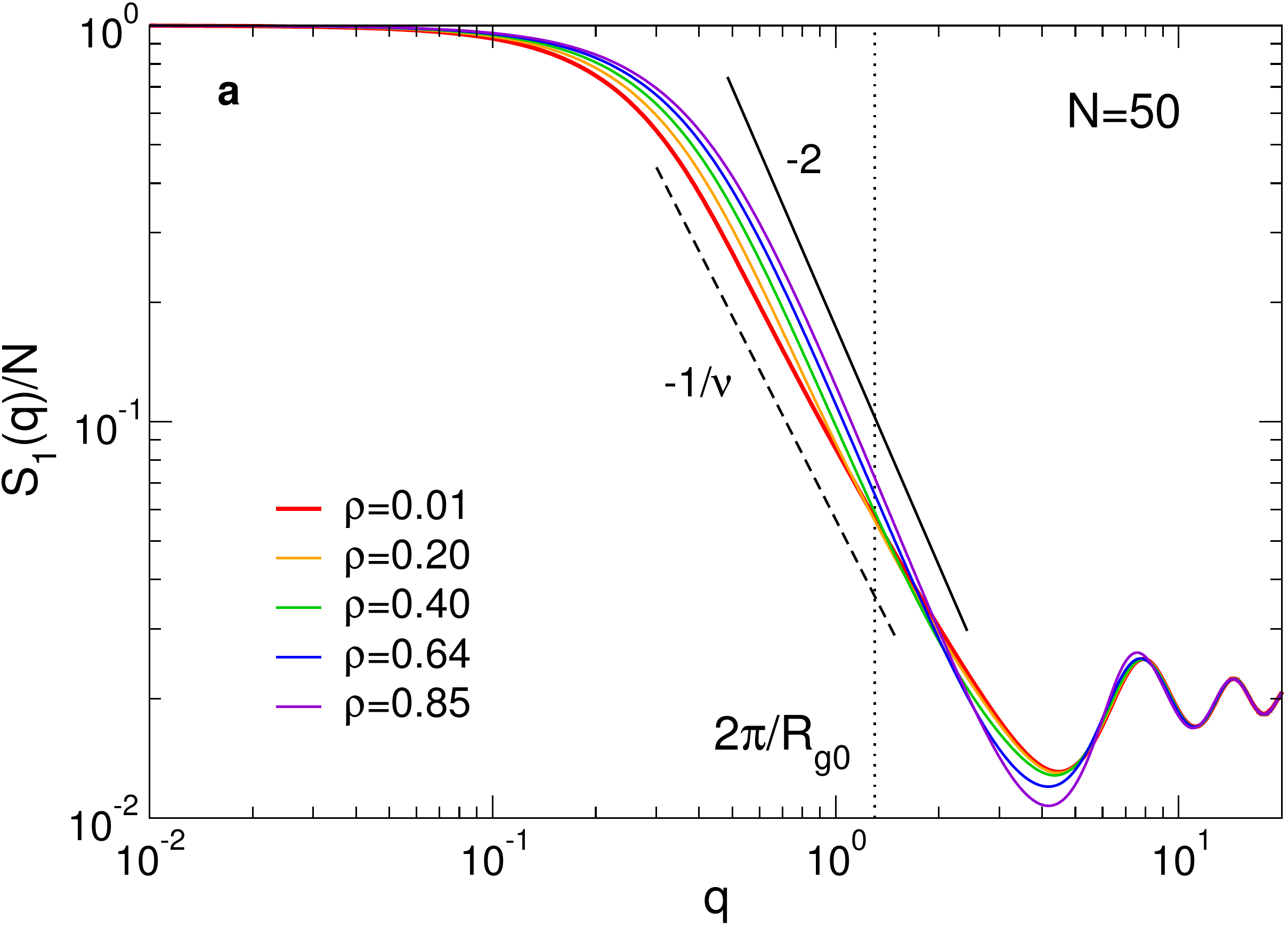}
\includegraphics[width=0.45 \textwidth]{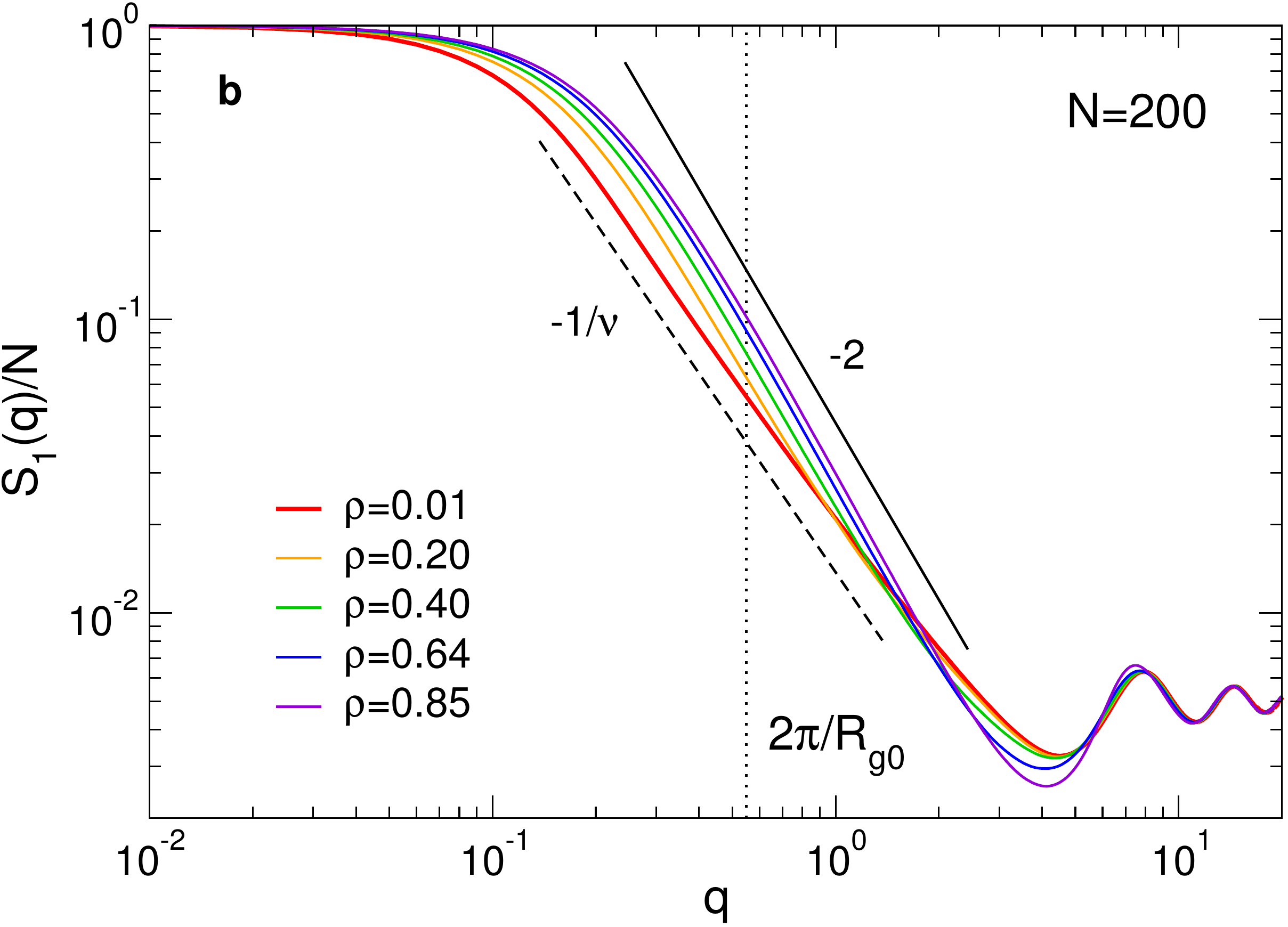}
\includegraphics[width=0.45 \textwidth]{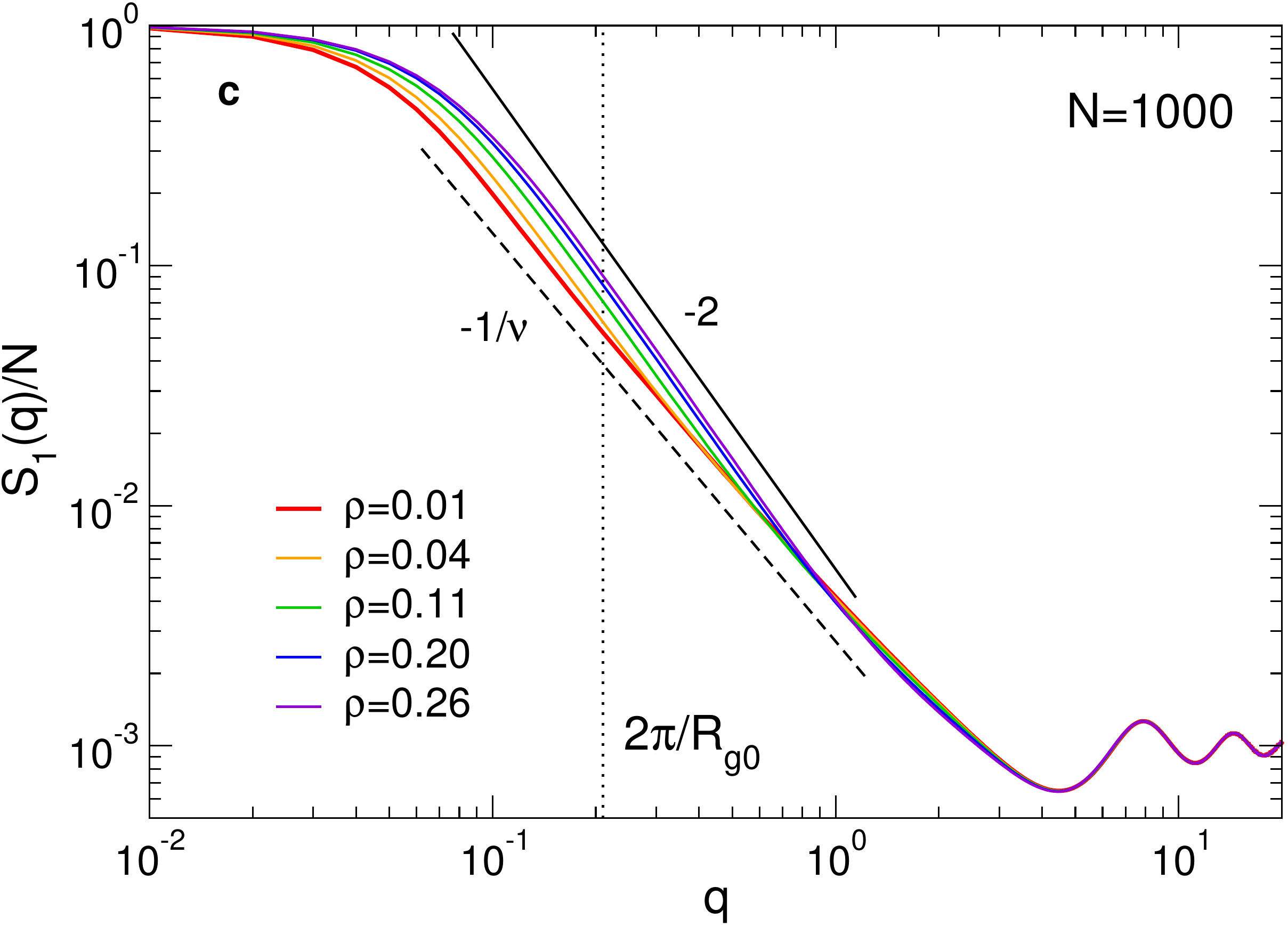}
\caption{Single chain form factor for chain lengths $N=50$ (a), $N=200$ (b) and $N=1000$ (c) and different monomer densities $\rho$. Continuous lines: slope $-2$ (ideal chain). Dashed lines: slope $-1/\nu\simeq -1.70$ (swollen chain).}
\label{form_factor}
\end{figure}

In Fig.~\ref{form_factor}, we show the chain form factor $S_1(q)/N$ for $N=50,200$, and $1000$ and for different values of the monomer density $\rho$.

\noindent We recall that the theoretical prediction for $S_1(q)/N$ is \cite{rubinstein2003polymer,teraoka2002polymer,paul1991crossover}

\begin{equation}
\frac{S_1(q)} N =
\begin{cases}
1/(1+q^2R_g^2/3) & q < 2\pi/ R_g\\
A q^{-2} &  2\pi/R_g<q<2\pi/\xi_c\\
B q^{-1/\nu} &  2\pi/\xi_c<q<2\pi/b\\
\mathcal O(1/N) & q>2\pi/b,\\
\end{cases}
\end{equation}

\noindent where $A,B>0$ are constants. In the dilute regime, $\xi_c\approx R_{g0}$ and therefore the regime  $S_1(q)/N \propto q^{-2}$ disappears; in the concentrated/melt regime, $\xi_c\approx b$ and therefore the regime $S_1(q)/N \propto q^{-1/\nu}$ is not present (here $\nu=0.588$ is the Flory exponent in good solvent). 

Swollen chain behavior $S_1(q)/N \propto q^{-1/\nu}$ and the ideal chain behavior $S_1(q)/N \propto q^{-2}$ are clearly observable in Fig.~\ref{form_factor} respectively at low and high densities for $N=50$ and $N=200$. At intermediate densities, we expect the form factor to transition from a $q^{-2}$ dependence to a $q^{-1/\nu}$ dependence at $q\approx 1/\xi_c$ \cite{degennes1979scaling,rubinstein2003polymer}. Since $S_1(q)/N$ assumes values from $\approx 1/N$ to $1$, this transition is more easily observable for the longer chains: $N=200$ and $N=1000$.

\section{Comparison of $S(q)$ for different values of $N$}

\begin{figure}[b]
\centering
\includegraphics[width=0.45 \textwidth]{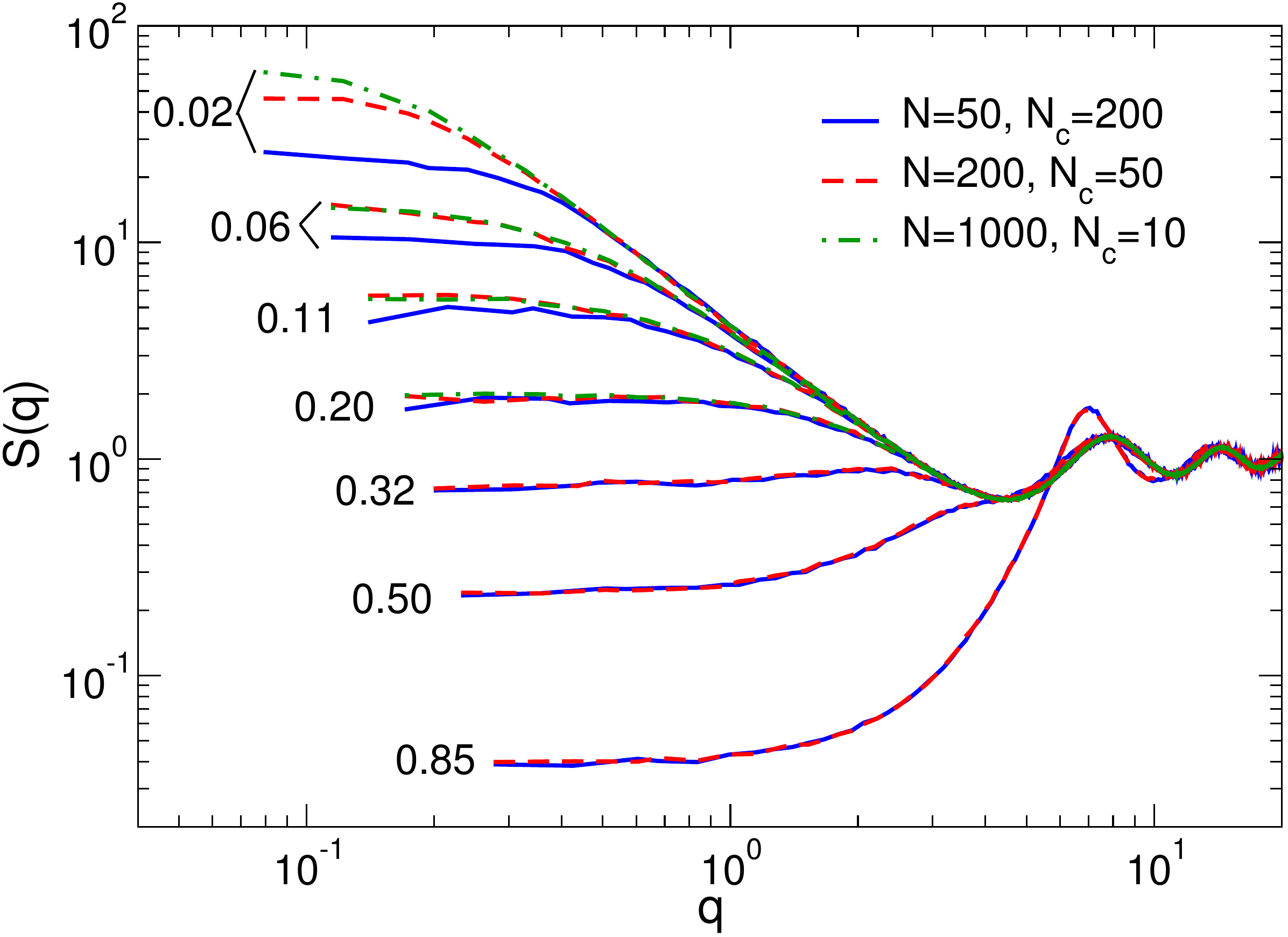}
\caption{Comparison between the monomer structure factors $S(q)$, Eq.~\eqref{sq_def}, of systems with different chain lengths $N$ at the same monomer density $\rho$. For densities $\rho >0.11 \simeq \rho^*(N=50)$, $S(q)$ becomes independent of $N$.}
\label{sq_all_cfr}
\end{figure}

In Fig.~\ref{sq_all_cfr}, we compare the monomer structure factor $S(q)$ for systems with the same density $\rho$ but with different chain lengths $N$. One can see that for densities $\rho\gtrsim0.11$, which is approximately the overlap density of the $N=50$ system, $\rho^{*}(N=50)\simeq0.106$ (see Tab.~\ref{tab:properties} in the Appendix of the main text), $S(q)$ is independent of $N$. This is expected, since for $\rho>\rho^{*}(N=50)$ all the systems here considered are in the semidilute regime, where the global structure is independent of chain length \cite{rubinstein2003polymer}.

\section{Bond angle distribution}

\begin{figure}
\centering
\includegraphics[width=0.45 \textwidth]{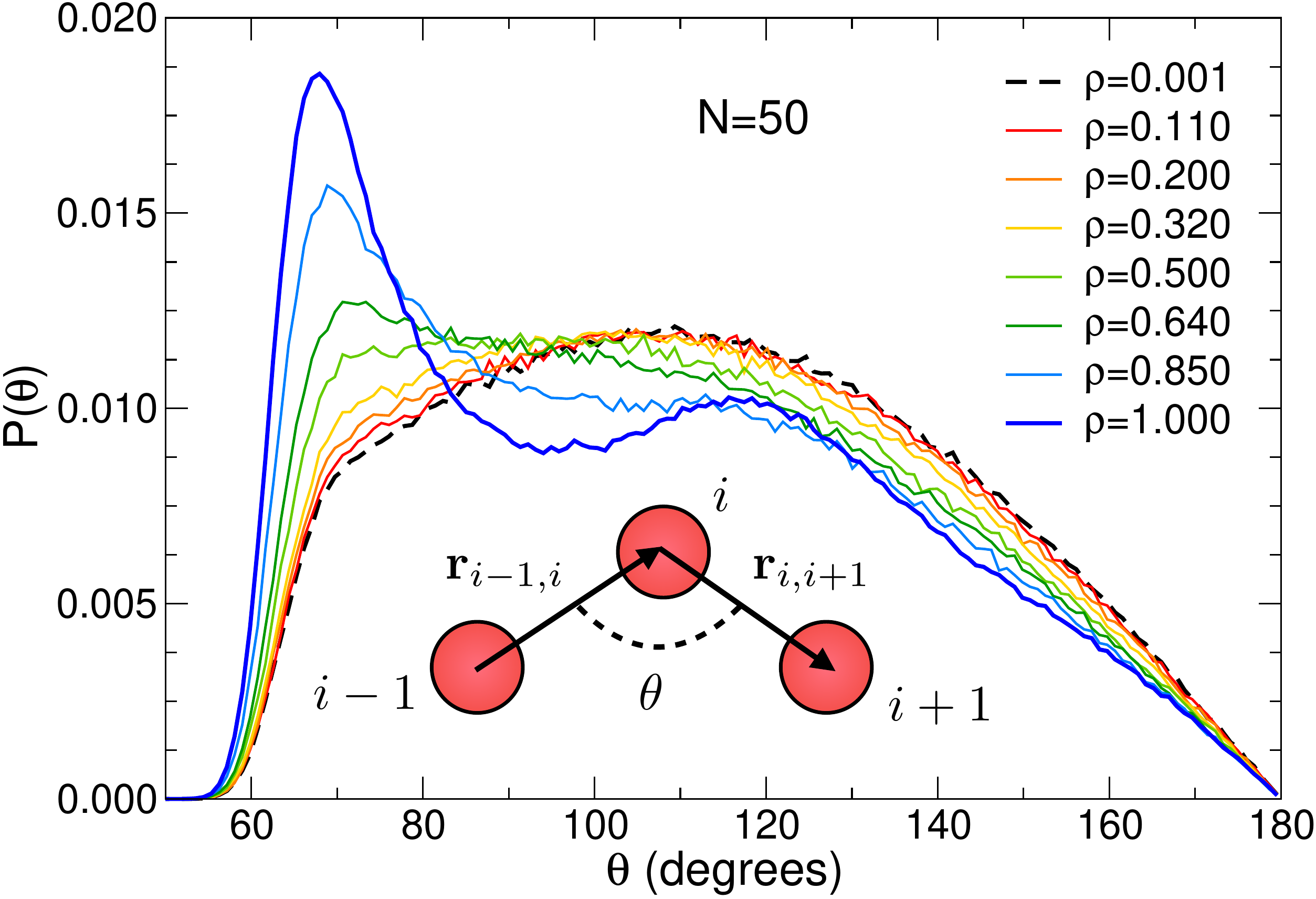}
\caption{Bond angle distribution for $N=50$ and different monomer densities $\rho$. The bond angle $\theta$ is defined by Eq.~\eqref{bond_angle}. Cartoon: schematic representation of how $\theta$ is defined.}
\label{bond_pdf_n50}
\end{figure}

In Fig.~\ref{bond_pdf_n50} we report the bond angle distribution $P(\theta)$ for $N=50$ and different densities (for larger $N$, the results are basically the same). The angle $\theta$ is defined as 

\begin{equation}
\theta \equiv \arccos \left( -\frac{\mathbf r_{i-1,i} \cdot \mathbf r_{i,i+1}}{|\mathbf r_{i-1,i}| \ |\mathbf r_{i,i+1}|}\right),
\label{bond_angle}
\end{equation}

\noindent where $\mathbf r_{i,j}\equiv\mathbf r_i - \mathbf r_j$ and $\mathbf r_i, \ i=2,\dots,N-1$ is the monomer's position vector. We have therefore $\theta=\pi-\theta_1$, where $\theta_s$ is defined in Eq.~\eqref{bond_corr_def} of the main text.

At low density, $P(\theta)$ shows a maximum at $\theta\simeq 112^\circ$ and falls to zero rather sharply at $\theta\simeq 60^\circ$ because of the excluded volume interaction. When $\rho$ is increased past $\rho \simeq 0.3$, the shape of $P(\theta)$ starts to change significantly in that it develops a peak at $\theta\simeq 70^\circ$, signaling that the chains are compressed. Overall, the average bond angle decreases by $\simeq 8\%$ when going from the dilute regime to density $\rho \gtrsim 0.85$. This effect is also observable in the average bond length (not shown), although the decrease is in this case only $\simeq 1\%$. Although it is possible to conceive chain conformations with a small $\langle \theta \rangle$ but a high persistence length (e.g., an ``accordion-like" rigid polymer), the fact that up to very high $\rho$ the form factor is compatible with the scaling predictions for chains in a concentrated solution (Sec.~\ref{sec:form_factor}) suggests that the reduction of $\langle \theta \rangle$ corresponds to a reduction of the persistence length, as also observed from the bond-bond correlation function (Sec.~\ref{sec:gyration} in the main text).

\bibliography{a_main.bib}
\end{document}